\documentclass[a4paper]{article}
\pdfoutput=1 

\usepackage{jheppub} 

\usepackage{graphicx,here,epsfig,wrapfig,amssymb,color,amsmath,bm,breqn}

\usepackage[T1]{fontenc} 
\usepackage{multirow}
\usepackage{amsmath}
\usepackage{pifont}
\usepackage{float}
\allowdisplaybreaks
\usepackage[utf8]{inputenc} 

\usepackage{multirow}

\usepackage{lineno} 

\usepackage{hyperref} 
\hypersetup{
    colorlinks,
    citecolor=black,
    filecolor=black,
    linkcolor=black,
    urlcolor=black
}

\newcommand{\Ds}{\displaystyle}

\newcommand{\nn}{\nonumber}

\newcommand{\ot}{\leftarrow}

\renewcommand{\)}{\right)}

\renewcommand{\vec}[1]{\bm{#1}}

\preprint{JLAB-THY-21-3325}

\title{Extraction of the Sivers function from SIDIS, Drell-Yan, and  $W^\pm/Z$ boson production data with TMD evolution}

\author[a]{Marcin~Bury}
\author[b,c]{Alexei~Prokudin}
\author[a]{Alexey~Vladimirov}

\affiliation[a]{Institut f\"ur Theoretische Physik, Universit\"at Regensburg, D-93040 Regensburg, Germany}
\affiliation[b]{Division of Science, Penn State Berks, Reading,
	PA 19610, USA}
\affiliation[c]{Thomas Jefferson National Accelerator Facility,
	Newport News, VA 23606, U.S.A.}

\emailAdd{marcin.bury@ifj.edu.pl}
\emailAdd{prokudin@jlab.org}
\emailAdd{alexey.vladimirov@ur.de}

\abstract{ 
We perform a global fit of the available polarized Semi-Inclusive Deep Inelastic Scattering (SIDIS), polarized pion-induced Drell-Yan (DY) and  $W^\pm/Z$ boson production data at N$^3$LO and NNLO accuracy of the Transverse Momentum Dependent (TMD) evolution, and extract the Sivers function for $u$, $d$, $s$ and for sea quarks. The Qiu-Sterman function is determined in a model independent way via the operator product expansion from the extracted Sivers function. The analysis is supplemented by additional studies, such as the estimation of applicability region, the impact of the unpolarized distributions' uncertainties, the universality of the Sivers functions, positivity constraints, the significance of the sign-change relation, and the comparison with the existing extractions.}

\begin{document} 

\maketitle 

\section{Introduction}
The asymmetry of the intrinsic nucleon structure is a fascinating phenomenon that demonstrates   the complexity of the underlying theory of strong interactions, Quantum Chromodynamics (QCD). By their very nature, asymmetries are related to a nontrivial three-dimensional (3D) structure of the nucleon and could not be fully understood within the collinear partonic picture. A more adequate approach for studies of such phenomena is the transverse momentum dependent (TMD) factorization theorem~\cite{Collins:1981uk,Collins:1984kg,Meng:1991da,Collins:2011zzd,Ji:2004wu,Ji:2005nu,GarciaEchevarria:2011rb}, which resolves the transverse components of the parton's momentum in addition to the collinear one and is sensitive to the short (perturbative) and long (nonperturbative) distance QCD dynamics. The TMD factorization theorem associates the correlation of the nucleon's spin and the transverse momentum of an unpolarized parton with the Sivers function~\cite{Sivers:1989cc,Sivers:1990fh}. In this paper, we present a global QCD analysis and the extraction of the Sivers function from Semi-Inclusive Deep Inelastic Scattering (SIDIS), Drell-Yan (DY), and $W^\pm/Z$ production data with next-to-next-to-leading order (NNLO) and next-to-next-to-next-to-leading order (N$^3$LO) of TMD evolution and provide a detailed description of results presented in Ref.~\cite{Bury:2020vhj}.

Within the TMD factorization theorem, nonperturbative effects of collinear and transverse parton motion are included in the TMD distribution (TMD PDF) and fragmentation (TMD FF) functions. In its modern form, the TMD factorization theorem has been proven in Refs.~\cite{Collins:2011zzd,GarciaEchevarria:2011rb,Vladimirov:2017ksc}, to which we refer the reader for the theory details. A natural subject of the TMD factorization is inelastic sufficiently inclusive processes with two observed scales, such as SIDIS~\cite{Meng:1991da,Ji:2004wu,Boer:1997nt,Bacchetta:2006tn,Collins:2011zzd}, DY~\cite{Collins:1984kg,Ji:2004xq,Arnold:2008kf,GarciaEchevarria:2011rb} and an almost back-to-back hadron pair production in Semi-Inclusive $e^+e^-$ Annihilation (SIA)~\cite{Collins:1981uk,Boer:1997mf,Collins:2011zzd}. These scales are a large hard scale, $Q$, (that is the virtuality of the photon in SIDIS, the invariant mass of the dilepton pair in DY and SIA) and a smaller scale, $q_T$ (that is the transverse momentum of the produced hadron (SIDIS), the transverse momentum of the dilepton pair (DY), or the transverse momentum disbalance of produced hadrons (SIA)). These processes could be described by several universal TMD PDFs, TMD FFs, and the universal nonperturbative evolution kernel (the so-called Collins-Soper (CS) kernel~\cite{Collins:1984kg}). Over the last few years, it was demonstrated that the TMD factorization framework is indeed universal, and the same TMD functions describe a large body of data for unpolarized SIDIS and DY~\cite{Bacchetta:2017gcc,Scimemi:2019cmh}. The next step of phenomenological verification for the TMD factorization theorem is a demonstration of the universality of polarized distributions and, in particular, the Sivers function -- one of the most prominent TMD distributions. This important step is the goal of our paper.

The Sivers function attracted a lot of attention in the literature~\cite{Efremov:2004tp,Vogelsang:2005cs,Anselmino:2005ea,Anselmino:2008sga,Kang:2009bp,Aybat:2011ta,Gamberg:2013kla,Sun:2013dya,Echevarria:2014xaa,Anselmino:2016uie,Bacchetta:2020gko,Cammarota:2020qcw,Echevarria:2020hpy,Boglione:2021aha} since its invention in the early 90's~\cite{Sivers:1989cc,Sivers:1990fh}, when it was suggested as a possible mechanism for generation of the transverse left-right asymmetry in inclusive pion production in polarized proton-proton scattering~\cite{Anselmino:1994tv}. In the modern generation of polarized experiments, transverse single-spin asymmetries (TSSAs) related to the Sivers function have been measured in SIDIS, DY and $W^\pm/Z$ production processes by HERMES~\cite{Airapetian:2009ae,Airapetian:2020zzo}, COMPASS SIDIS~\cite{Alekseev:2008aa,Adolph:2014zba,Adolph:2012sp,Adolph:2016dvl}, COMPASS DY~\cite{Aghasyan:2017jop}, JLab \cite{Qian:2011py,Zhao:2014qvx} and STAR~\cite{Adamczyk:2015gyk}. The interpretation of these data poses certain difficulties, because in many cases they are collected at a low factorization scale $Q$, and with a relatively low statistics. The situation will improve with the upcoming high-statistics measurements by Jefferson Lab 12 GeV Upgrade~\cite{Dudek:2012vr} which will explore the large-$x$ region. The future Electron-Ion Collider (EIC)~\cite{Boer:2011fh,Accardi:2012qut} has a dedicated program for measurements of TSSAs in the broad kinematic range and with large arms in $Q$, $q_T$, and the collinear momentum fraction $x$. The experimental exploration of the Sivers function is one of the goals of polarized SIDIS and DY experimental programs of  future and existing experimental facilities such as the Electron Ion Collider, Jefferson Lab 12~GeV Upgrade, RHIC~\cite{Aschenauer:2015eha} at BNL, COMPASS~\cite{Gautheron:2010wva,Bradamante:2018ick} at CERN, and SpinQuest at Fermilab~\cite{Chen:2019hhx,Brown:2014sea}.

Theory-wise the Sivers function is the P-odd component of the matrix element of the unpolarized TMD operator. The Sivers function inherits all general properties of TMD distributions. Nonetheless, it is the Sivers function that reveals experimentally the specific feature of TMD operators, namely, the half-infinite Wilson lines that resum the gluon interactions with the spectator particle participating in the 
process~\cite{Belitsky:2002sm}. According to the TMD factorization theorem, Wilson lines in the TMD operator point to the future light-cone time direction in the case of SIDIS (out-going particle), whereas in the case of DY process (in-coming particle), they point to the past. For the majority of TMD distributions, the direction of Wilson lines does not lead to any observable effect due to the T-invariance of strong interaction. However, for the Sivers function (and also for the Boer-Mulders function~\cite{Boer:1997nt}) the change of Wilson line's direction results in the function with the same absolute value and an opposite sign~\cite{Brodsky:2002rv,Brodsky:2002cx,Collins:2002kn}:
\begin{eqnarray}\label{eq:sign}
f_{1T}^\perp(x,k_T)_{\text{[SIDIS]}}=-f_{1T}^\perp(x,k_T)_{\text{[DY]}}.
\end{eqnarray}
Another feature of the half-infinite Wilson lines is revealed in the regime of the small-$b$ (or large transverse momentum), see Ref.~\cite{Ji:2006ub}. For ordinary TMD distributions, half-infinite Wilson lines compensate each other in $b\to0$ limit, and result in spatially compact operators, but not for the Sivers function, as the resulting operator contains gluon fields integrated over all positions along the light-ray~\cite{Boer:2003cm,Ji:2006ub,Koike:2007dg,Kang:2011mr,Scimemi:2018mmi}. The resulting function is the key ingredient of the collinear factorization of TSSAs: a collinear twist-3 function called the Qiu-Sterman (QS) function \cite{Efremov:1981sh,Efremov:1984ip,Qiu:1991pp,Qiu:1991wg}. 
In the present work, we determine the QS function in a model independent way by explicit inversion of the small-$b$ expansion.

An essential aspect of TMD distributions is their scale-dependence (evolution), as predicted by QCD evolution equations. The analyses that described (or predicted) both SIDIS and DY data from Ref.~\cite{Kang:2009bp,Anselmino:2016uie,Cammarota:2020qcw} used the parton model approximation without the TMD evolution and demonstrated a good agreement with the experiment. These analyses assumed suppression of evolution effects in asymmetries and indicated~\cite{Anselmino:2016uie,Cammarota:2020qcw}  the Sivers function's universality, however, they did not capture the full complexity of QCD evolution for TMD distributions. On the other hand, the analyses that included TMD evolution, Refs.~\cite{Echevarria:2014xaa,Echevarria:2020hpy} had problems and did not achieve better agreement with the DY data compared to Refs.~\cite{Kang:2009bp,Anselmino:2016uie,Cammarota:2020qcw}. This situation constitutes a puzzle in establishing the status of the TMD factorization theorem because the scale dependence and the nonperturbative and universal CS kernel are among its principal elements, and the predictive power and precision of the TMD factorization should improve when higher orders are included.

In this work, and Ref.~\cite{Bury:2020vhj}, we demonstrate the universality of the Sivers function and describe simultaneously SIDIS, Drell-Yan, and  $W^\pm/Z$ boson production data using TMD evolution and the universal nonperturbative CS kernel extracted from the unpolarized measurements. 
Our analysis is similar in spirit, but distinct in numerous smaller elements from previous studies \cite{Efremov:2004tp,Vogelsang:2005cs,Anselmino:2005ea,Anselmino:2008sga,Kang:2009bp,Aybat:2011ta,Gamberg:2013kla,Sun:2013dya,Echevarria:2014xaa,Anselmino:2016uie,Bacchetta:2020gko,Cammarota:2020qcw,Echevarria:2020hpy,Boglione:2021aha}. Let us point the most important peculiarities, which altogether lead to the success of our description.
\begin{itemize}
\item We use the \texttt{SV19}~\cite{Scimemi:2019cmh} extraction as an input for unpolarized TMD distributions and nonperturbative TMD evolution (CS kernel). The selection of input unpolarized distributions is of principal importance for the analysis of TSSAs, because they enter both the numerator (convoluted with the Sivers function) and the denominator (the unpolarized cross-section) of the observed asymmetry. \texttt{SV19} extraction is made by the simultaneous fit of the unpolarized DY and SIDIS data on cross-sections differential in the transverse momentum, spanning from relatively low energy experiments such as HERMES, 
to the highest energies of the LHC.

\item The present fit is performed with NNLO and N$^3$LO TMD evolution (together with NNLO matching for unpolarized distribution in \texttt{SV19}). The latest studies \cite{Scimemi:2017etj,Bertone:2019nxa,Scimemi:2019cmh,Bacchetta:2019sam} demonstrated the importance of the perturbative input for a good agreement of theory with the data.

\item In contrast to many of the previous extractions, we do not use any collinear function in the ansatz for the Sivers function. In fact, a parametrization of the Sivers function starting from the collinear PDFs is not a well founded approximation, because the Sivers function generally belongs to a different class of functions, and thus such an approximation only biases the extraction. 

\item The key element of our analysis is the $\zeta$-prescription~\cite{Scimemi:2018xaf}. In the nutshell, the $\zeta$-prescription is a particular selection of the reference scale for the TMD evolution, which totally decouples nonperturbative evolution from the nonperturbative TMD distributions. In this scheme a TMD distribution is defined as a universal function without any specific relation to the collinear distributions. Exactly this property allows us to use NNLO or N$^3$LO TMD evolution (with NNLO or N$^3$LO hard coefficient functions, and NNLO matching for unpolarized distributions inherited from \texttt{SV19}) together with a free parametrization for the Sivers function in a strict theoretical manner.

\item  We use a conservative data cut, which guaranties that the selected data could be described within TMD factorization approach derived in the approximation of a large hard scale $Q$, and a low transverse momentum, $q_T$. Experimental data are also carefully selected to avoid double counting. The resulting data set becomes relatively small in comparison to other extractions,  however the latest 3D binning data set by HERMES \cite{Airapetian:2020zzo} allows us to obtain enough data for the analysis.

\end{itemize}

Let us mention that the \texttt{SV19} extraction~\cite{Scimemi:2019cmh} is also performed using these principles, and demonstrated a perfect agreement with the data. The numerical computations are performed with \texttt{artemide} \cite{artemide} -- the multi-purpose package for phenomenology within the TMD factorization framework. The results of the \texttt{SV19} extraction can be found at \cite{artemide} (and the analysis code can be found at \cite{dataProcessor}).

The paper is organized as follows. In Section~\ref{sec:formalism} we recall aspects of the TMD factorization formalism needed to analyze and interpret the experimental data. The extraction procedure and the uncertainty estimation approach are described in Section~\ref{sec:procedure}. Section~\ref{sec:results} is devoted to the description of results of the global QCD analysis of the data related to the Sivers function, exploration of the properties of extracted Sivers function, and extraction of the QS functions. We finally conclude and discuss further improvements in Section~\ref{sec:conclusions}.

\section{Formalism}
\label{sec:formalism}
We start by summarizing theoretical formalism and presenting expressions that describe DY process and SIDIS cross-sections within TMD factorization. An interested reader can find details about TMD factorization and derivations of cross-sections for these processes in Refs.~\cite{Bacchetta:2006tn,Arnold:2008kf,Kang:2009bp,Boer:2011xd,Collins:2011zzd,Scimemi:2018xaf}.

\subsection{Definition of TMD distributions}

The unpolarized TMD PDF, $f_1$, the unpolarized TMD FF, $D_1$, and the Sivers function, $f_{1T}^\perp$, 
parametrize the matrix element of the vector TMD operator. In the position space, they are defined as follows
\begin{align}
\label{def:TMDPDF}
&\Phi^{[\gamma^+]}_{q\ot h}(x,b;\mu,\zeta)\equiv\int \frac{d\lambda}{2\pi}e^{-ix \lambda P^+}
\langle P,S|\bar q(\lambda n+b)[n\lambda+b,\pm \infty n+b] \frac{\gamma^+}{2} [\pm \infty n,0] q(0)|P,S\rangle
\\\nn&
\qquad\qquad
= f_{1;q\ot h}(x,b;\mu,\zeta)+i\epsilon^{\mu\nu}_T b_\mu S_{\nu} \; M \; f_{1T;q\ot h}^\perp(x,b;\mu,\zeta),
\\\label{def:TMDFF}
&\Delta^{[\gamma^+]}_{q\to h}(z,b;\mu,\zeta)\equiv\frac{1}{2zN_c}\int \frac{d\lambda}{2\pi}e^{-i\lambda P_h^+/z}\sum_X\langle 0|\frac{\gamma^+}{2}[-\infty n +b,n\lambda+b]q(n\lambda+b) |h,X\rangle\\\nn &\qquad\qquad \times \langle h,X|
\bar q(0)[0,-\infty n]|0\rangle
=D_{1,q\to h}(z,b;\mu,\zeta),
\end{align}
where $P^\mu$, $S^\mu$ and $M$ are the momentum, the spin-vector and the mass of the hadron, $P_h^\mu$ is the momentum of the produced hadron $h$, and $b^\mu$ is a transverse vector\footnote{Let us note that the relative position of quark fields is important since it defines the direction of $b^\mu$, and hence the sign convention for the Sivers function. For instance, TMD operators are defined in Ref.~\cite{Gamberg:2011my} as $\sim \bar q(0)...q(\lambda n+b)$, which results in $-i\epsilon^{\mu\nu}_Tb_\mu S_\nu$ prefactor for the Sivers function, compare to Eq.~\eqref{def:TMDPDF}. Taking the direction of $b$ into account, the definition used here coincides with definitions in Refs.~\cite{Gamberg:2011my,Scimemi:2018mmi,Scimemi:2019gge} so that the final expressions for the cross section also coincide. Notice also that Ref.~\cite{Gamberg:2011my} denotes the Sivers function in configuration space as $\tilde f_{1T}^{\perp (1)}$ making explicit the Fourier transform from the momentum space to the position space and a derivative with respect to $b$. This notation corresponds to our notation of $f_{1T}^\perp$ as we start directly from parametrizations of TMD distributions in the position space.}. TMD distributions depend on $x$ and $z$, which are light-cone momentum fractions for incoming quark $q$ and the produced hadron $h$ correspondingly, and the transverse separation $b$. Also, TMD distributions depend on the ultraviolet $\mu$ and the rapidity $\zeta$ renormalization scales, which we discuss below. We adopt the standard notation for components of the light-cone decomposition, $v^\mu=v^+ n^\mu+v^- \bar n^\mu +v^\mu_T$ with $n^2=\bar n^2=0$ and $n\cdot \bar n=1$, and the transverse anti-symmetric tensor $\epsilon_T^{\mu\nu}=\epsilon^{-+\mu\nu}$  with $\epsilon_T^{12}=-\epsilon_T^{21}=1$. The staple gauge-link points to $+\infty$($-\infty$) in the case of TMD distributions measured in SIDIS (DY), and assures the gauge invariance of the TMD operator. The unpolarized distributions are independent of the direction of the staple gauge-link in SIDIS and DY, whereas the Sivers distribution changes   sign while the absolute size remains the same, see Refs.~\cite{Brodsky:2002rv,Brodsky:2002cx,Collins:2002kn}, and Eq.~\eqref{eq:sign} in $b$-space reads
\begin{eqnarray}\label{th:sign-change}
f_{1T}^\perp(x,b;\mu,\zeta)_{\text{[SIDIS]}}=-f_{1T}^\perp(x,b;\mu,\zeta)_{\text{[DY]}}.
\end{eqnarray}
For definiteness, in the formulas for a particular process we use the notation $f_{1T}^\perp$ for the Sivers function without explicit indication of the process, and the sign change between DY and SIDIS is implemented in calculations. All our results of the Sivers function extraction will be presented for the SIDIS definition.

The dependence on the scales $\mu$ and $\zeta$ is given by a pair of TMD evolution equations~\cite{Collins:2011zzd,Scimemi:2018xaf,Chiu:2011qc}
\begin{eqnarray}\label{def:ev1}
\mu^2 \frac{d F(x,b;\mu,\zeta)}{d\mu^2}&=&\frac{\gamma_F(\mu,\zeta)}{2}F(x,b;\mu,\zeta),
\\\label{def:ev2}
\zeta \frac{d F(x,b;\mu,\zeta)}{d\zeta}&=&-\mathcal{D}(b,\mu)F(x,b;\mu,\zeta),
\end{eqnarray}
where $F$ is any TMD distribution ($f_1$, $f_{1T}^\perp$, or $D_1$ in the current context). The first equation is the ordinary renormalization group equation, with $\gamma_F$ being the ultraviolet anomalous dimension for the TMD operator. The second equation is the result of the factorization of rapidity anomalous dimension, with $\mathcal{D}$ being the Collins-Soper kernel\footnote{Our definition of the rapidity anomalous dimension corresponds to $\tilde K$ and $\gamma_\nu$ used in Refs.~\cite{Collins:2011zzd} and \cite{Chiu:2011qc} as $\mathcal{D} = -\tilde K/2=-\gamma_\nu/2$.} (or rapidity anomalous dimension). The Collins-Soper kernel is a fundamental universal function that has explicit operator definition and parametrizes properties of QCD vacuum \cite{Vladimirov:2020umg}.  It is a universal function, nonperturbative at large-$b$ while at small-$b$ it is calculable in terms of the perturbative expansion in the strong coupling constant $\alpha_s$, whereas it has to be extracted from the experimental data. Both quark and rapidity anomalous dimensions are known up to N$^3$LO in the perturbative regime, see Refs.~\cite{Gehrmann:2010ue,Grozin:2014hna,Li:2016ctv,Vladimirov:2016dll}. 

Using the evolution equations one relates measurements performed at different energies. It is convenient to select certain value of the pair $(\mu,\zeta)$ as a reference scale. There are several choices of the reference scale $(\mu,\zeta)$ used in the literature, see Refs.~ \cite{Collins:2011zzd,Bacchetta:2017gcc,Scimemi:2018xaf}. In this work we use the so-called $\zeta$-prescription \cite{Scimemi:2018xaf}. It consists in  selection of the reference scale $(\mu,\zeta)=(\mu,\zeta_\mu(b))$ on the equipotential line (of $(\gamma_F,-\mathcal{D})$-field) that passes through the saddle point. In this case, the reference TMD distribution, called the optimal TMD distribution, is independent on $\mu$ (by definition) and perturbatively finite in the whole range of $\mu$ and $b$.  The solution of the TMD evolution equations from Eqs.~(\ref{def:ev1}, \ref{def:ev2}) can be written in the following simple form
\begin{align}
F(x,b;\mu,\zeta)=\left(\frac{\zeta}{\zeta_\mu(b)}\right)^{-\mathcal{D}(b,\mu)}F(x,b),
\label{eq:optimal}
\end{align}
where $F(x,b)$ on the right-hand side of the equation \eqref{eq:optimal} is the optimal TMD distribution \cite{Scimemi:2017etj}. The functions $\zeta_\mu(b)$ is a known function~\cite{Vladimirov:2019bfa} of the nonperturbative Collins-Soper kernel. In our notations, the optimal TMD distribution $F(x,b)$ has no scaling arguments, which emphasizes its scale independence.

\subsection{Sivers asymmetry in SIDIS}

The differential SIDIS cross section of the inclusive hadron production in the electron scattering off a transversely polarized target ($e(l)+h_1(P,S)\to e(l')+h_2(p_h)+X$) has the following structure \cite{Gourdin:1973qx,Kotzinian:1994dv,Diehl:2005pc,Bacchetta:2006tn}
\begin{eqnarray}
\label{def:SIDIS}
\frac{d\sigma}{dx \, dy\,  \,dz\, d\phi_S d\phi_h\, d P_{hT}^2}
&=&
\frac{\alpha_{\text{em}}^2(Q)}{Q^2}\,
\frac{y}{2(1-\varepsilon)}
\Biggl\{
F_{UU ,T} 
+|S_\perp| \sin(\phi_h-\phi_S) F_{UT ,T}^{\sin(\phi_h -\phi_S)}
+ ... \Bigg\},
\end{eqnarray}
where
\begin{eqnarray}
q^2=-Q^2,\qquad x=\frac{Q^2}{2P\cdot q},\qquad
y=\frac{P\cdot q}{P\cdot l},\qquad z=\frac{P\cdot P_h}{P\cdot q},\qquad
\varepsilon=\frac{1-y}{1-y+\frac{y^2}{4}},
\end{eqnarray}
where $q=l-l'$ is the momentum of the virtual photon. The variables $\phi_h$ and $P_{hT}$ are the angle and the absolute value of transverse component of the produced hadron's momentum, measured in the laboratory frame. The azimuthal angles for transverse components of the produced hadron ($\phi_h$) and the spin of the target hadron  ($\phi_S$) are defined relative to the lepton plane \cite{Bacchetta:2004jz}. The dots denote other angular modulations that are not interesting in the current context, and also the power suppressed structure functions~\cite{Bacchetta:2006tn}, such as $F_{UU ,L}$ and $F_{UT ,L}^{\sin(\phi_h -\phi_S)}$, which do not contribute at our order of accuracy. We define the shorthand notation
\begin{align}
\label{eq:notation}
    \mathcal{B}^{\text{SIDIS}}_n[f D] &\equiv \sum_{q}e_q^2 \int_0^\infty \frac{b db}{2\pi} b^n J_n\left(\frac{b |P_{hT}|}{z}\right) f_{q\ot h_1}(x,b;\mu,\zeta_1)D_{q\to h_2}(z,b;\mu,\zeta_2)
\end{align}
where $f$ and $D$ are TMD PDF and FF, $J_n$ is the Bessel function of the first kind and $e_q$ are electric charges of quarks $q$ and the summation runs over all active quarks and antiquarks. Within the TMD factorization the expressions for structure functions $F_{UU,T}$ and $F_{UT,T}^{\sin(\phi_h-\phi_S)}$ are
\begin{eqnarray}\label{th:FUUT}
&&F_{UU ,T}=\left|C_V(Q^2,\mu^2)\right|^2 \mathcal{B}^{\text{SIDIS}}_0\left[f_{1} D_{1}\right],
\\\label{th:FUTT}
&&F_{UT ,T}^{\sin(\phi_h-\phi_S)}=-M\; \left|C_V(Q^2,\mu^2)\right|^2 \mathcal{B}^{\text{SIDIS}}_1\left[f_{1T}^\perp D_{1}\right]\; ,
\end{eqnarray}
where  $C_V$ is the quark vector form-factor and the hadron mass $M$ is originated from the definition of the Sivers function Eq.~(\ref{def:TMDPDF}).

Let us emphasize the combination $|P_{hT}|/z$ that enters the argument of the Bessel function in Eq.~(\ref{eq:notation}). It is originated from the Lorenz transformation from the factorization frame, where the factorization theorem is derived, to the laboratory frame, where the experimental measurement is performed. This combination serves as a small parameter, and power corrections to Eqs.~(\ref{th:FUUT}) and (\ref{th:FUTT}) have a generic size $\mathcal{O}((P_{hT}/z/Q)^2)$. The accurate transformation between the frames must account for masses of initial and final hadrons. In this case, the argument of the Bessel function is more complicated \cite{Scimemi:2019cmh}. Here, we omit these complications, which is valid in $Q\to\infty$ limit.

The scales of the factorization should be selected such that $\mu\sim Q$, and $\zeta_1\zeta_2=Q^4$~\cite{Collins:2011zzd,Chiu:2011qc,Aybat:2011zv,Aybat:2011ge,Echevarria:2012pw,Vladimirov:2017ksc}. We use
\begin{eqnarray}
\mu^2=Q^2,\qquad \zeta_1=\zeta_2=Q^2.
\end{eqnarray}
The resulting products of TMD distributions are to be evolved to the scale of the experimental measurement. Since the TMD evolution is independent of the flavor and the spin, all structure functions (at the leading TMD twist) have common evolution properties \cite{Idilbi:2004vb}. In the case of the $\zeta$-prescription, using Eq.~\eqref{eq:optimal} one derives that products of TMD distributions in Eq.~\eqref{eq:notation} turn into
\begin{eqnarray}\label{th:evol_stf}
f_{q\ot h_1}(x,b;Q,Q^2) D_{q\to h_2}(z,b;Q,Q^2)&=&
R(b,Q) f_{q\ot h_1}(x,b) D_{q\to h_2}(z,b) \; ,
\end{eqnarray}
where we introduced the evolution factor
\begin{align}\label{eq:evolution_factor} 
R(b,Q) = \left(\frac{Q^2}{\zeta_Q(b)}\right)^{-2\mathcal{D}(b,Q)}
\end{align}
Therefore, in the TMD factorization framework structure functions are Fourier transforms of products of three $b$-dependent universal factors: two TMD distributions $f_{q\ot h}$  and $D_{q\to h}$, and the evolution factor $R$. Each of these factors governs dependence on a particular kinematic variable, $x$ and $z$ for TMD distributions, and $Q$ for evolution factor, and altogether they are integrated over $b$ with a Bessel function.

The single-spin Sivers asymmetry that is measured in SIDIS  experiments, is the ratio of structure functions
\begin{eqnarray}\label{eq:sidisaut}
A_{UT}^{\sin(\phi_h-\phi_S)}&\equiv&\frac{F_{UT ,T}^{\sin(\phi_h -\phi_S)}}{F_{UU ,T}} = -M\frac{\mathcal{B}^{\text{SIDIS}}_1\left[f_{1T}^\perp D_{1}\right]}{\mathcal{B}^{\text{SIDIS}}_0\left[f_{1} D_{1}\right]}\; .
\end{eqnarray}
Combining expressions from Eqs.~(\ref{th:FUUT},~\ref{th:FUTT},~\ref{th:evol_stf}) we obtain the following formula 
\begin{eqnarray}\label{th:AUTsin}
A_{UT}^{\sin(\phi_h-\phi_S)}=
-M\frac{\Ds\sum_{q}e_q^2 \int_0^\infty \frac{b db}{2\pi}\, b\, J_1\left(\frac{b |P_{hT}|}{z}\right) R(b,Q) f^\perp_{1T,q\ot h_1}(x,b)D_{1,q\to h_2}(z,b)}{\Ds\sum_{q}e_q^2 \int_0^\infty \frac{b db}{2\pi}\, J_0\left(\frac{b |P_{hT}|}{z}\right)
R(b,Q) f_{1,q\ot h_1}(x,b)D_{1,q\to h_2}(z,b)}.
\end{eqnarray}

\begin{figure}[t]
\centering
\includegraphics[width=0.7\textwidth]{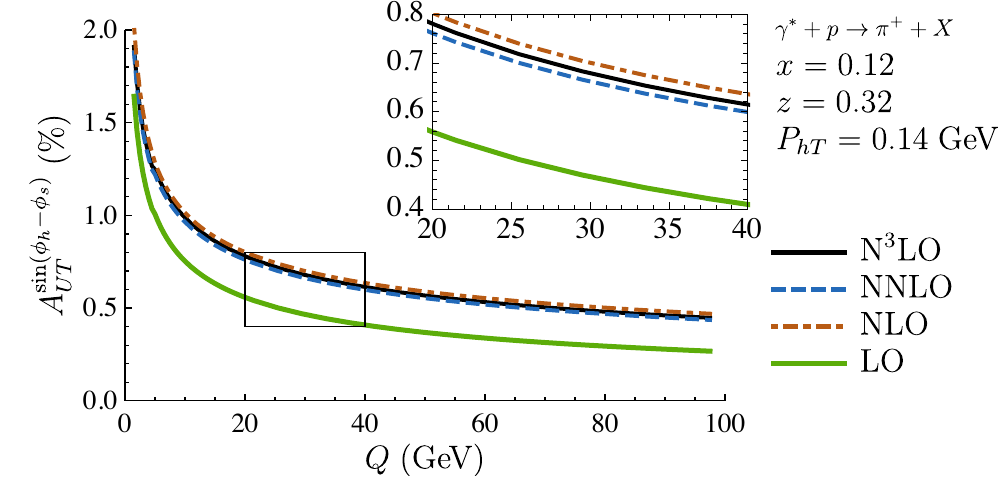}
\caption{\label{fig:evolution} The dependence of the single-spin Sivers asymmetry on $Q$ at fixed values of $x=0.12$, $z=0.32$, and $P_{hT}=0.14$ GeV (these values correspond to a particular bin of HERMES \cite{Airapetian:2020zzo}). Different perturbative orders are compared. In all cases unpolarized TMD  PDF, TMD  FF, the Sivers function and the nonperturbative part of the CS kernel are the same. The change of the perturbative order influences the order of perturbative part of CS kernel, TMD anomalous dimension.}    
\end{figure}

The dependence on $Q$ in (\ref{th:AUTsin}) is enclosed in the factors $R(b,Q)$. They are the only part of our computation that depends on the perturbative input since the hard coefficient functions $|C_V|^2$ exactly cancel in the ratio Eq.~\eqref{th:AUTsin}. The perturbative order is defined by the order of TMD anomalous dimension (\ref{def:ev1}) and by the perturbative part of CS-kernel (\ref{def:ev2}) (see also Eq.~(\ref{def:RAD})). Nowadays, these anomalous dimensions are known up to three-loop order, i.e. up to $\alpha_s^3$ \cite{Gehrmann:2010ue,Grozin:2014hna,Li:2016ctv,Vladimirov:2016dll}. This maximum order (the $\Gamma_{\text{cusp}}$ part is taken with one order higher, i.e. at $\alpha_s^4$ \cite{vonManteuffel:2020vjv}) we refer as N$^3$LO, according to the standard nomenclature (see Ref.~\cite{Scimemi:2019cmh} for extended discussion and references). Currently, one can define four consequent orders of perturbative input, starting from LO, which contains $\Gamma_{\text{cusp}}$ at LO, and null for other anomalous dimensions. In Fig.~\ref{fig:evolution}  we demonstrate\footnote{ We anticipate and use in Fig.~\ref{fig:evolution} our results of extraction of the Sivers function that we will perform in Sec.~\ref{sec:fitdata}. The $Q$ dependence of the asymmetry depends mainly on the evolution factor $R$ that is known from the analysis of unpolarized data. The dependence on the parameters describing nonperturbative TMD functions is quite weak therefore a similar $Q$ behavior is anticipated for all TSSAs that include $J_1\left(\frac{b |P_{hT}|}{z}\right)$.} the comparison of different orders and the general behavior of asymmetry as a function of $Q$. The convergence of the series is good. The difference between orders is almost homogeneous at different $Q$ and $\sim 50\%$ at LO$\to$NLO, $\sim -7\%$ at NLO$\to$NNLO, and $\sim 3\%$ at NNLO$\to$N$^3$LO. Also, we notice a very rapid behavior of asymmetry at small values of $Q$. Fig.~\ref{fig:evolution} is given for SIDIS and values of $x=0.12$, $z=0.32$, and $P_{hT}=0.14$ GeV are taken from a particular bin in HERMES kinematics \cite{Airapetian:2020zzo}. For other values of $x$, $z$, and $P_{hT}$, and for DY measurement, the behavior is similar.

\subsection{Transverse single-spin asymmetry in DY process}

Using notations of Ref.~\cite{Arnold:2008kf}, the differential cross-section for DY reaction $(h_1(P_1,S)+h_2(P_2)\to l^+(l)+l^-(l')+X$) can be written as
\begin{eqnarray}\label{def:DY}
&& \frac{d\sigma}{dQ^2 \, dy \, d\varphi \, dq_T^2}=\frac{\alpha_{\text{em}}^2(Q)}{9 s Q^2}\Big\{
F_{UU}^1+|S_T|\sin(\varphi-\phi_S) F_{TU}^1
 +...\Big\},
\end{eqnarray}
where
\begin{eqnarray}
q^2=Q^2,\qquad
s=(P_1+P_2)^2,\qquad
x_1=\sqrt{\frac{Q^2}{s}}e^y,\qquad x_2=\sqrt{\frac{Q^2}{s}}e^{-y},
\end{eqnarray}
where $q=l+l'$ is the momentum of the electroweak boson, and $y$ is its rapidity. The variables $\varphi$ and $q_T$ are the angle and the absolute value of the transverse component of the momentum of  the electroweak boson  measured in the center-of-mass frame. The ellipsis denotes other DY structure functions.

The shorthand notation from Eq.~\eqref{eq:notation} for DY reads
\begin{align}
\label{eq:notationdy}
\mathcal{B}^{\text{DY}}_n[f_1 \, f_2] &\equiv \sum_{q}e_q^2 \int_0^\infty \frac{b db}{2\pi} b^n J_n\left(b |q_T|\right) f_{1; q\ot h_1}(x_1,b;\mu,\zeta_1)f_{2; \bar q\ot h_2}(x_2,b;\mu,\zeta_2)\,,
\end{align}
where $f_1$ and $f_2$ are TMD distributions, $J_n$ is the Bessel function of first kind and $e_q$ are electric charges of quarks $q$ and the summation runs over all active quarks and anti-quarks. The TMD factorization gives the following expressions for structure functions (cf.~(\ref{th:FUUT},~\ref{th:FUTT}))
\begin{eqnarray}\label{th:FUU1}
&&F_{UU}^1=\left|C_V(-Q^2,\mu^2)\right|^2 \mathcal{B}^{\text{DY}}_0[f_1\, f_1]\,,
\\\label{th:FUT1}
&&F_{TU}^1=-M\left|C_V(-Q^2,\mu^2)\right|^2
\mathcal{B}^{\text{DY}}_1[f_{1T}^\perp \, f_1]\,,
\end{eqnarray}
where $C_V(-Q^2,\mu^2)$ is the quark vector form factor for the space-like momentum transfer. Further steps are analogous to the SIDIS case (\ref{th:evol_stf}) and the final expression for the transverse spin-asymmetry in the $\zeta$-prescription reads
\begin{eqnarray}\label{th:AUT}
A_{TU}\equiv\frac{F_{TU}^1}{F_{UU}^1}=
-M\frac{\Ds\sum_{q}e_q^2 \int_0^\infty \frac{b db}{2\pi}b\, J_1\(b |q_T|\) R(b,Q) f^\perp_{1T,q\ot h_1}(x_1,b)f_{1,\bar q\ot h_2}(x_2,b)}{\Ds\sum_{q}e_q^2 \int_0^\infty \frac{b db}{2\pi}J_0\(b |q_T|\)
R(b,Q) f_{1,q\ot h_1}(x_1,b)f_{1,\bar q\ot h_2}(x_2,b)}.
\end{eqnarray}
In some cases, discussed also in the following Sections, experimental measurements provide asymmetries which are related to $A_{TU}$. In particular, the asymmetry $A_N$~\cite{Adamczyk:2015gyk} measured by STAR Collaboration is defined as the asymmetry relative to $\cos\varphi$ with $\vec S_T$ oriented along $\vec y$. In this case, $\cos \varphi=-\sin(\varphi-\phi_s)$ and thus 
\begin{eqnarray}\label{def:AN}
A_N=-A_{TU}.
\end{eqnarray}
Another important case is the process $h_1(P_1)+h_2(P_2,S)\to l^+l^-+X$ (\textit{i.e.} with the polarized hadron $h_2$) measured by COMPASS~\cite{Aghasyan:2017jop}. In this case
\begin{eqnarray}\label{def:AUT}
A_{UT}=-A_{TU}(f_{1T}^\perp\leftrightarrow f_1),
\end{eqnarray}
where the exchange of Sivers and unpolarized TMD PDFs takes place in the numerator of (\ref{th:AUT}). In order to explain the origin of the minus sign in \eqref{def:AUT} let us recall that the definition in Eq.~(\ref{def:TMDPDF}) is written for the operator with $\gamma^+$. The directions $n$ and $\bar n$ are associated with the large components of $P_1$ and $P_2$, correspondingly. Therefore, the Sivers function for the operator with $\gamma^-$, corresponding to the large momentum $P_2$, has a prefactor $-i\epsilon_T^{\mu\nu}b_\mu S_\nu$, since $\epsilon^{+-\mu\nu}=-\epsilon_T^{\mu\nu}$. The detailed discussion on different asymmetries related to $A_{UT}$ and relation between them can be found in Refs.~\cite{Kang:2009sm,Anselmino:2009st}.

At high $Q$ the Drell-Yan pair production should account also for weak boson channels. To account for various channels in Eq.~\eqref{eq:notationdy} one should replace
\begin{eqnarray}
\label{eq:ew}
\sum_q e_q^2~\to~\sum_{l,q,\text{ch.}}z_l^{\text{ch.}}z_q^{\text{ch.}}\Delta^{\text{ch.}},
\end{eqnarray}
where $z_l$ and $z_q$ are combinations of couplings associated with lepton and quark vertices, and $\Delta$ is the product of propagators multiplied by $Q^4$. In the case of the neutral boson production, expressions for $z_f$ (where $f=q$ or $l$) and $\Delta$,  for the channels in Eq.~\eqref{eq:ew}, $\text{ch.}\in \{\gamma\gamma,\gamma Z,ZZ\}$ are
\begin{align}\nn
&\text{ch.}=\gamma\gamma : \qquad &&z_f^{\gamma\gamma}=e_f^2,\qquad &&\Delta^{\gamma\gamma}=1,
\\\label{th:zz-Z}
&\text{ch.}=\gamma Z : \qquad &&z_f^{\gamma Z}=\frac{T_3-2e_fs_W^2}{2s_W^2c_W^2},\qquad &&\Delta^{\gamma Z}=\frac{2Q^2(Q^2-M_Z^2)}{(Q^2-M_Z^2)^2+\Gamma_Z^2 M_Z^2},
\\\nn
&\text{ch.}=ZZ : \qquad &&z_f^{ZZ}=e_f^2,\qquad &&\Delta^{ZZ}=\frac{Q^4}{(Q^2-M_Z^2)^2+\Gamma_Z^2M_Z^2},
\end{align}
where $f=q,l$. While in the case of charged-boson production there exists a single channel
\begin{eqnarray}\label{th:zz-W}
\text{ch.}=WW &:& \qquad z_f^{WW}=\frac{|V_{ff'}|^2}{4s_W^2},\qquad \Delta^{WW}(Q^2)=\frac{Q^4}{(Q^2-M_W^2)^2+\Gamma_W^2M_W^2}.
\end{eqnarray}
In these expressions $e_f$ is the electric charge, $T_3$ is the third component of the iso-spin, $s_W$ and $c_W$ are the sine and the cosine of the Weinberg angle, $V_{ff'}$ is the element of the Cabibbo-Kobayashi-Maskawa matrix, $\Gamma_{Z}$ and $M_Z$ are the decay-width and the mass of the Z boson, and $\Gamma_W$ and $M_W$ are the decay-width and the mass of the W boson.

Let us emphasize that in the presented expressions, we omitted the target- and product-mass corrections. Accounting for these corrections modifies many factors in formulas (\ref{def:SIDIS}, \ref{th:FUUT}, \ref{th:FUTT}, \ref{def:DY}, \ref{th:FUT1}, \ref{th:FUU1}), including the values of collinear variables $x$ and $z$, values for $P_{hT}$ and $\varepsilon$ (the corresponding expressions can be found in Refs.~\cite{Scimemi:2019cmh,Boglione:2019nwk}). It was shown that the accounting for these power corrections improves the quality of extraction of unpolarized TMD distributions, see Ref.~\cite{Scimemi:2019cmh}. At small-$Q$ or large $q_T/Q$, which is a typical situation for the kinematical region of many experimental measurements, these corrections push values of kinematic variables closer to the phase space's border, and theory predictions become unstable. To stabilize predictions with respect to these corrections, we decided to neglect mass corrections in this analysis and leave a more profound analysis of the influence of target and produced hadron mass corrections to a future publication. We also make sure that neglecting mass corrections does not spoil our results by restricting the data used in our analysis to small values of $q_T/Q$, where the influence of target and product mass corrections is relatively small.

\subsection{Nonperturbative parametrization for the Sivers function}

The optimal TMD distributions in the expressions (\ref{th:AUTsin}, \ref{th:AUT}) are universal nonperturbative functions which should be determined from the comparison with experimental data. In principle, at small values of $b$ these functions could be related to collinear distributions through the operator product expansion. Generally,  accounting of the small-$b$ relation significantly reduces the parametric freedom since any modification of small-$b$ behavior appears as power corrections in $b^2$. Thus, the typical ansatz for a TMD distribution $F(x,b)$ has the following generic form
\begin{eqnarray}\label{th:gen_ansatz}
F_{f\ot h}(x,b)=\sum_{f'}[C_{f\ot f'}(\ln(b\mu))\otimes f_{f'\ot h}(\mu)](x)\cdot f_{NP}(x,b),
\end{eqnarray}
where $C$ is the perturbative coefficient function, $f_{f'\ot h}(x,\mu)$ is a collinear parton distribution, $\otimes$ is an integral convolution in variable $x$ (for twist-2 collinear distributions it is the Mellin convolution), and $f_{NP}$ is some function that parametrizes nonperturbative $b$-shape of the TMD distribution such that $f_{NP}(x,b\to 0)= \mathcal{O}(b^2)$. The expressions for perturbative matching coefficient for unpolarized distributions were obtained at NNLO in Refs.~\cite{Scimemi:2019gge,Echevarria:2016scs}, and for Sivers function at NLO in Ref.~\cite{Scimemi:2019gge}. 

The ansatz (\ref{th:gen_ansatz}) works well  for the unpolarized TMD distributions, where corresponding collinear distributions are known very precisely from different experiments. In particular, the  \texttt{SV19} global fit of DY and SIDIS data made in Ref.~\cite{Scimemi:2019cmh}  is based on this ansatz with NNLO coefficient functions. The \texttt{SV19} fit was performed with NNLO and N$^3$LO TMD evolution, and with the Collins-Soper kernel parameterized as
\begin{eqnarray}\label{def:RAD}
\mathcal{D}(b,\mu)=\mathcal{D}_{\text{resum}}(b^*,\mu)+c_0 bb^*,
\end{eqnarray}
where $b^*=b/\sqrt{1+(b/(2\; \text{GeV}^{-1})^{2}}$, $\mathcal{D}_{\text{resum}}$ is the resummed expression for the perturbative part, and $c_0$ is a fitting parameter. 
The model (\ref{def:RAD}) has linear asymptotic at large-$b$. This behavior is in agreement with the model calculations and analysis of large-$b$ behavior of the Collins-Soper kernel performed in Refs.~\cite{Collins:2014jpa,Vladimirov:2020umg}. The coefficient $c_0$ can be related to the gluon condensate and its extracted values agrees~\cite{Vladimirov:2020umg} with its known values. The \texttt{SV19} fit demonstrates perfect agreement of the theory with the data. In particular, $\chi^2/N_{pt}$ is 1.1 for DY (with $N_{pt}=457$) and 0.95 for SIDIS (with $N_{pt}=582$). 

The pion unpolarized TMD PDF was extracted in the same framework in Ref.~\cite{Vladimirov:2019bfa}, what we refer to as \texttt{Vpion19}. The values of nonperturbative parameters for \texttt{SV19} and \texttt{Vpion19} fits together with Monte-Carlo replicas are publicly available via \texttt{artemide} repository\footnote{The original \texttt{SV19} fit accounted for the target-mass corrections in kinematics. In the current fit we omit mass-correction. For consistency, we have rerun the \texttt{SV19} code without target-mass corrections. The resulting $\chi^2/N_{pt}$ is around 1.3 for DY and 1.05 for SIDIS, see also discussion in Ref.~\cite{Scimemi:2019cmh}. The updated sets of TMD distributions marked as \texttt{SV19\_all=0} can be found in \cite{artemide}. The same procedure is applied to \texttt{Vpion19} TMD set for the pion TMD distributions.} \cite{artemide}.

In the case of the Sivers function the perturbative NLO matching has been  derived in  Refs.~\cite{Scimemi:2018mmi,Scimemi:2019gge}. It has the the following expression (see also Sec. \ref{sec:QS})
\begin{eqnarray}\label{th:Sivers=small-b}
f_{1T}^\perp(x,b)=-\pi T(-x,0,x;\mu)+\pi a_s(\mu) C(\mu^2b^2)\otimes \mathcal{T}(x;\mu)+\mathcal{O}(a_s^2)\; ,
\end{eqnarray}
where $a_s=g^2/(4\pi)^2$ is the strong coupling constant and the function $T(x_1,x_2,x_3)$ parametrizes twist-3 quark-gluon-quark correlator $\sim \bar q(z_1) \gamma^+ F_{\mu+}(z_2)q(z_3)$. This function mixes with functions $\Delta T$, $G_\pm$ that parametrizes another twist-3 correlations, and in (\ref{th:Sivers=small-b}) are collectively denoted as $\mathcal{T}$. The twist-3 distributions have support $-1<x_{1,2,3}<1$ with $x_1+x_2+x_3=0$, and the integral convolution $\otimes$ in (\ref{th:Sivers=small-b}) projects it to a single variable $x$. The tree order term in Eq.~\eqref{th:Sivers=small-b}, $T(-x,0,x)$, is the Qiu-Sterman (QS) function ~\cite{Efremov:1981sh,Efremov:1983eb,Qiu:1991pp,Qiu:1998ia}. The QS function is not an autonomous function, in the sense that its evolution involves the values of arguments outside of the line $(-x,0,x)$, and mixes with functions $\Delta T$ and $G_{\pm}$~\cite{Braun:2009mi}. None of these functions is known, and thus accounting for small-$b$ asymptotic in the sense of Eq.~(\ref{th:gen_ansatz}) is not a feasible way of constructing the expression to be used in phenomenology. Therefore, 
in the present work, we will not use the small-$b$ matching for Sivers function. 

Instead, we  consider the optimal Sivers function as a generic nonperturbative function that we will extract directly from the data. We do not put any special restrictions on the parametrization, apart from the usual constraints. We require $f_{1T}^\perp(x\to 1,b)\lesssim (1-x)$, $f_{1T}^\perp(x\to 0,b)\lesssim x^{-1}$ to ensure integrability and vanishing of the Sivers function at $x=0$ and $x=1$. Also, we require that $f_{1T}^\perp(x,b)$ is a function of $x$ and $b^2$ to mimic the operator product expansion structure. We have explored many parametric forms and selected the following one, which is flexible enough to reveal the Sivers function, but at the same time is not overwhelmed with free parameters:
\begin{eqnarray}\label{def:model}
f_{1T;q\ot h}^\perp(x,b)=N_q \frac{(1-x) x^{\beta_q} (1+\epsilon_q x)}{n(\beta_q,\epsilon_q)}\exp\left(-\frac{r_0+x r_1}{\sqrt{1+r_2 x^2 b^2}}b^2\right),
\end{eqnarray}
where $n(\beta,\epsilon)= (3+\beta+\epsilon+\epsilon \beta)\Gamma(\beta+1)/\Gamma(\beta+4)$, such that
\begin{eqnarray}
\int_0^1 dx f_{1T;q\ot h}^\perp(x,0)=N_q.
\end{eqnarray}
The $b$-dependent factor mimics $f_{NP}(x,b)$ used in \texttt{SV19} fit, with a reduced number of parameters. Notice that $b$ and $x$ dependencies do not factorize in our parametrization. The   experimental data on Sivers asymmetries is available for various final states, including charged pions and kaons. The quark composition of those final states allows access to $u$, $d$, $s$ quark flavors but is not sufficient to distinguish other sea quarks, such as $\bar u$, $\bar d$, and $\bar s$. The Sivers function for heavy quark flavors $b$ and $c$ cannot be extracted with the current data either. Thus, we will distinguish separate functions for $u$, $d$, $s$ quarks, and a single \textit{sea} Sivers function for $\bar u$, $\bar d$ and $\bar s$ quarks. We nullify the Sivers function for $b$ and $c$ flavors. We also set $\beta_s=\beta_{sea}$ and $\epsilon_s=\epsilon_{sea}=0$, since they are not restricted by the existing experimental data.
Large-$x$ region of the data is also limited at the moment to $x \lesssim 0.5$ and we therefore are using a general $(1-x)$ factor in our parametrization. In total we have 12 free parameters: $N_u$, $N_d$, $N_s$ and $N_{sea}$ that dictates the general scale, $\beta_u$, $\beta_d$ and $\beta_{sea}$ that gives small-$x$ asymptotic ($\beta_i>-1$), $\epsilon_u$ and $\epsilon_d$ to fine-tune of valence distributions, and $r_0$, $r_1$ and $r_3$ for $x$-dependence in parameterization of transverse momentum behavior ($r_i>0$).

Let us emphasize that the absence of small-$b$ matching in the optimal Sivers function is not in contradiction with the perturbative order of TMD evolution (NNLO and N$^3$LO in the current case) or the perturbative order of matching to other distributions (NNLO for unpolarized distributions). The utilization of different orders for components in TMD factorization is consistent within the $\zeta$-prescription, as well as, in other schemes with fixed reference scale for TMD distributions, discussed e.g. in Ref.~\cite{Collins:2014jpa}, but is not consistent in the resummation-like schemes e.g. used in Refs~ \cite{Echevarria:2014xaa,Bacchetta:2020gko,Echevarria:2020hpy}. In the latter scheme, one would need to use the matching function for Sivers function at N$^3$LO, which is not yet available~\cite{Scimemi:2019gge}. For resummation-like schemes of scale-fixation, where the scales of TMD distributions depend on $b$ in an arbitrary manner, such an approach is inconsistent. In this case, the orders of TMD evolution and matching coefficients must be adjusted to guarantee the compensation of scaling logarithms. 

\section{Global analysis procedure}
\label{sec:procedure}

In this Section we discuss basic principles of the global QCD analysis, data selection, fit procedure, and the study of the limits of TMD factorization. 

\subsection{Data selection}
\label{sec:data}

\begin{figure}
\begin{center}
\includegraphics[width=0.7\textwidth]{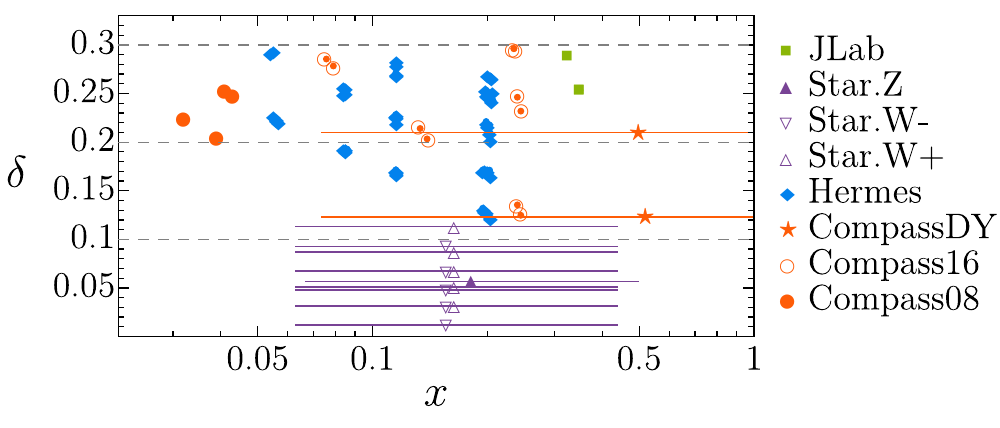}
\caption{\label{fig:points} Distribution of the experimental data over the values of $x$ and $\delta$ Eq.~(\ref{def:delta}).}
\end{center}
\end{figure}

The TMD factorization theorem is derived in the limit of large-$Q$ and a small relative transverse momentum $\delta$, defined as
\begin{eqnarray}\label{def:delta}
\delta=\frac{|P_{hT}|}{zQ} (\text{in SIDIS}),\qquad \delta=\frac{|q_T|}{Q} (\text{in DY}).
\end{eqnarray}
The large-$Q$ requirement is needed to suppress the   power corrections $\sim M^2/Q^2$ and $\sim \Lambda^2/Q^2$, where $\Lambda$ is a general nonperturbative scale of QCD. Since $M$ and $\Lambda$ are $\sim 1$ GeV, we impose the restriction $\langle Q\rangle >2$ GeV, which limits possible power corrections to around $10-20$\% for the lowest energy data points. The optimal values of $\delta$ for applicability of TMD factorization were studied in Ref.~\cite{Scimemi:2017etj} (and were further confirmed by independent studies in Refs.~\cite{Scimemi:2019cmh,Bacchetta:2019sam}), where it was shown that phenomenologically TMD factorization is valid for $\delta<0.2 - 0.3$, and is strongly violated for large values of $\delta$. In the current study we impose $\delta<0.3$, assuring that we accommodate data points from as many experiments as possible, still preserving applicability of TMD factorization, see Fig.~\ref{fig:points}. Summarizing our data selection cuts, we apply the following selection criteria 
\begin{eqnarray}\label{def:cuts}
\langle Q\rangle >2\; \text{GeV}\qquad \text{and}\qquad \delta<0.3.
\end{eqnarray}
These restrictions are consistent with the applicability of the TMD factorization theorem as discussed in Ref.~\cite{Scimemi:2017etj}. However, we hope that a part of power corrections cancels in the ratio of structure functions measured experimentally (\ref{th:AUTsin}, \ref{th:AUT}). The more stringent conditions (say $\delta<0.2$) would secure the TMD approach, but they are hardly applicable to the modern data, which is dominated by the low-energy measurements. Our data selection cuts (\ref{def:cuts}) are the most stringent among all other extractions of Sivers function, compare to Refs.~\cite{Efremov:2004tp,Vogelsang:2005cs,Anselmino:2005ea,Anselmino:2008sga,Gamberg:2013kla,Sun:2013dya,Echevarria:2014xaa,Anselmino:2016uie,Bacchetta:2020gko,Cammarota:2020qcw}.

The Sivers asymmetry in SIDIS has been measured by HERMES \cite{Airapetian:2009ae,Airapetian:2020zzo}, COMPASS \cite{Alekseev:2008aa,Adolph:2016dvl}\footnote{We do not include COMPASS measurements \cite{Adolph:2012sp,Adolph:2014zba} because we are interested in multi-dimensional binning of \cite{Adolph:2016dvl} and these two measurements overlap substantially in their experimental sample with \cite{Adolph:2016dvl}.} and JLab Hall A \cite{Qian:2011py} collaborations. DY measurements of the transverse spin-asymmetry were performed by the COMPASS  Collaboration~\cite{Aghasyan:2017jop} in the pion-induced DY process and by the STAR  Collaboration~\cite{Adamczyk:2015gyk} in $W^\pm/Z$ production. After application of our data selection cuts (\ref{def:cuts}) we have 76 data points in total (63 for SIDIS, and 13 for DY).  The distribution of the points in the $(x,\delta)$-plane is shown in Fig.~\ref{fig:points}. The synopsis of data is presented in Table~\ref{tab:data}.

\begin{table}
\begin{center}
\begin{tabular}{l||c| c|c|c}
Dataset name &Ref. & Reaction & \# Points & Av.Uncertainty
\\\hline
\multirow{4}{*}{Compass08} & \multirow{4}{*}{\cite{Alekseev:2008aa}} 
   & $d^{\uparrow}+\gamma^*\to \pi^+$ & 1~/~9 & 1.2\%\\
&  & $d^{\uparrow}+\gamma^*\to \pi^-$ & 1~/~9 & 1.1\%\\
&  & $d^{\uparrow}+\gamma^*\to K^+$ & 1~/~9 &   3.4\%\\
&  & $d^{\uparrow}+\gamma^*\to K^-$ & 1~/~9 &   5.1\%\\
\hline
\multirow{2}{*}{Compass16} & \multirow{2}{*}{\cite{Adolph:2016dvl}} 
   & $p^{\uparrow}+\gamma^*\to h^+$ & 5~/~40 & 1.6\%\\
&  & $p^{\uparrow}+\gamma^*\to h^-$ & 5~/~40 & 2.0\%\\
\hline
\multirow{4}{*}{Hermes} & \multirow{4}{*}{\cite{Airapetian:2020zzo}} 
   & $p^{\uparrow}+\gamma^*\to \pi^+$ & 11~/~64 & 2.6\%\\
&  & $p^{\uparrow}+\gamma^*\to \pi^-$ & 11~/~64 & 3.1\%\\
&  & $p^{\uparrow}+\gamma^*\to K^+$ & 12~/~64 & 6.1\%\\
&  & $p^{\uparrow}+\gamma^*\to K^-$ & 12~/~64 & 10.8\%\\
\hline
\multirow{4}{*}{JLab} & \multirow{4}{*}{\cite{Qian:2011py,Zhao:2014qvx}} 
   & $p^{\uparrow}+\gamma^*\to \pi^+$ & 1~/~4 & 13.9\%\\
&  & $p^{\uparrow}+\gamma^*\to \pi^-$ & 1~/~4 & 8.0\%\\
&  & $p^{\uparrow}+\gamma^*\to K^+$ & 1~/~4 & 7.0\%\\
&  & $p^{\uparrow}+\gamma^*\to K^-$ & 0~/~4 & -- \\
\hline\hline
SIDIS total & & & 63 & 
\\\hline\hline
CompassDY & \cite{Aghasyan:2017jop} & $\pi^-+d^{\uparrow}\to \gamma^*$ & 2~/~3 & 12.2\%\\\hline
Star.W+ & \multirow{3}{*}{\cite{Adamczyk:2015gyk}} & $p^{\uparrow}+p\to W^+$ & 5~/~5 & 16.1\%\\
Star.W- &  & $p^{\uparrow}+p\to W^-$ & 5~/~5 & 32.2\%\\
Star.Z  &  & $p^{\uparrow}+p\to \gamma^*/Z$ & 1~/~1 & 33.\%\\
\hline\hline
DY total & & & 13 & \\
\hline\hline
Total & & & 76 & \\\hline
\end{tabular}
\caption{\label{tab:data} Synopsis of the data sets used in the analysis. The fourth column ``\# Points'' shows the number of data points selected after application of cuts from Eq.~(\ref{def:cuts}) and the total number of available data points. The last column shows the average uncorrelated error for points in the data set (after application of (\ref{def:cuts})).}
\end{center}
\end{table}

A large portion of the SIDIS data comes from a recent HERMES analysis~\cite{Airapetian:2020zzo} that uses a three-dimensional kinematic binning and enlarged phase space. It is the three-dimensional binning that allows a clean separation of the TMD factorization region. On the contrary, the Compass and JLab measurements provide effectively ``one-dimensional binning'', i.e., only one of the kinematic variables has narrow binning, while the rest are integrated over a wide range. Only the  $P_{hT}$-differential measurements could be studied in such cases. The $z$-differential and $x$-differential measurements have $P_{hT}$ integrated over the full kinematic range and thus could not be fully described by the TMD factorization theorem. Even for the $P_{hT}$-differential binning, the TMD factorization is hard to apply due to the presence of $z^{-1}$ in the data selection rules (\ref{def:cuts}). Almost every bin of COMPASS and JLab measurements borders with a region of the phase space where the TMD factorization is strongly violated ($P_{hT}/z\sim Q$). Consequently, we were forced to use the average kinematics to include these data points into the fit. The ignorance of the bin integration effects is compensated by large uncertainties of these measurements but could lead to a systematic error in our extraction. We also use the averaged kinematics for HERMES measurement as it is suggested by the HERMES collaboration, because effects of the bin-integration are already included in the systematic uncertainty of the data\footnote{We thank Gunar Schnell for clarification of this point.}.

In the case of DY measurements, the bin integration effects are larger due to the larger bin sizes. These effects are especially significant for electroweak boson production, where the cross-section changes rapidly. Thus, we perform the integration over the bin size separately for the numerator and denominator of Eq.~(\ref{th:AUT}).

\subsection{Fit procedure and estimation of uncertainties}
\label{sec:fit}

To estimate the goodness of theory prediction against the experimental measurements we use the $\chi^2$-test function defined as
\begin{eqnarray}
\chi^2=\sum_{i,j=1}^n(m_i-t_i)V_{ij}^{-1}(m_j-t_j),
\end{eqnarray}
where $m_i$ is the central value of the $i$'th measurement, $t_i$ is the theory prediction for it, and $V_{ij}$ is the covariance matrix. The covariance matrix is build as usual
\begin{eqnarray}
V_{ij}=\delta_{ij}\sum_{l=1}^{N_{\text{uncor.}}}(\sigma^{(l),\text{uncor.}}_{i})^2+\sum_{l=1}^{N_{\text{cor.}}}\sigma^{(l),\text{cor.}}_i\sigma^{(l),\text{cor.}}_j,
\end{eqnarray}
where $\sigma^{(l),\text{uncor.}}_{i}$ ($\sigma^{(l),\text{cor.}}_i$) are uncorrelated (correlated) uncertainties of $i$'th measurement. In the case of asymmetries the main source of the correlated uncertainties is the beam/target polarization dilution. This definition takes into account the nature of experimental uncertainties, and gives a faithful estimate of the agreement between the experimental data and the theory prediction.

The evaluation of the theory prediction for a given set of model parameters is made by \texttt{artemide}. \texttt{Artemide} \cite{artemide} is the fortran library for numerical computations  within TMD factorization approach. It allows for a flexible implementation of any nonperturbative model for TMD distributions alongside with NNLO and N$^3$LO perturbative order precision~\cite{Gehrmann:2010ue,Echevarria:2015byo,Echevarria:2016scs,Vladimirov:2016dll}. The evaluation of $\chi^2$ function and the data analysis is performed via  the \texttt{Python} interface for \texttt{artemide}, which is (together with all programs used for the current fit) available in Ref.~\cite{dataProcessor}. The minimization of $\chi^2$ is made with \texttt{iminuit} package (MINUIT2) \cite{iminuit}.

The measured polarized asymmetries are weighted by unpolarized cross-section, see (\ref{th:AUTsin},~\ref{th:AUT}). The unpolarized cross-section is also evaluated by \texttt{artemide} with \texttt{SV19} and \texttt{Vpion19} sets. In contrast to other extractions, the \texttt{SV19} set of unpolarized distribution is extracted from the fit to the large set of DY and SIDIS data without application of additional normalization conditions.

\begin{figure}[htb]
\includegraphics[width=0.49\textwidth]{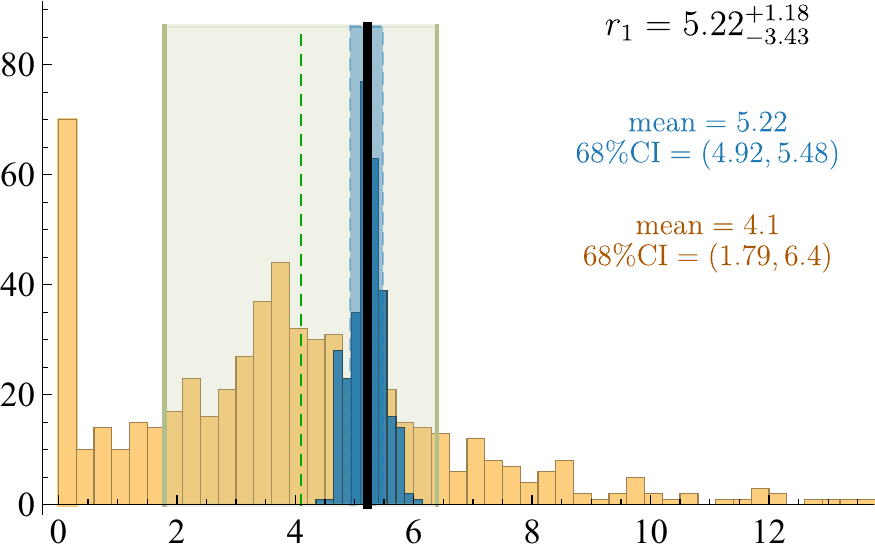}
~
\includegraphics[width=0.49\textwidth]{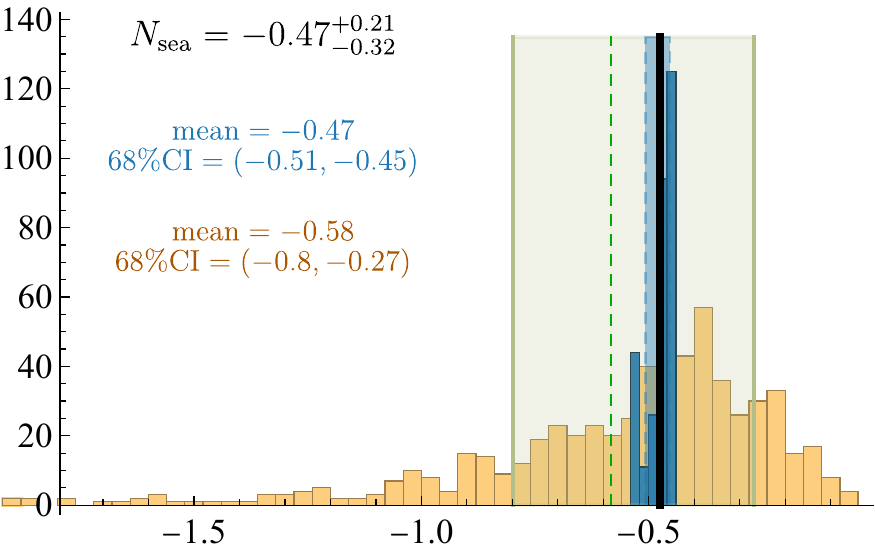}
\caption{\label{fig:params} Examples of histograms of parameter distribution for $r_1$ and $N_{\text{sea}}$ obtained in the joined fit of SIDIS and DY data. The orange histogram is the distribution due to experimental uncertainties (500 replicas). The blue histogram is the distribution due to \texttt{SV19} extraction (300 replicas). The green dashed line and green band (the black line and blue band) show positions of the mean value and 68\%CI for distribution due to experimental uncertainties (due to \texttt{SV19} fit).}
\end{figure}

There are two sources of uncertainties in the determination of the Sivers function. The main one is the uncertainties of the experimental data. The other important source is the uncertainties in the determination of unpolarized TMD distributions and the nonperturbative part of TMD evolution. We have estimated both uncertainties using the replica method~\cite{Ball:2008by}. The present precision of the experimental data does not restrict values of parameters for the Sivers function strongly, and the $\chi^2$-test function does not have an isolated global minimum. The obtained uncertainty bands and underlying distributions are very broad and asymmetric, see Fig.~\ref{fig:params}. Therefore, we use the notations from Eq.~(\ref{def:uncer}) to faithfully represent the uncertainties of our determination. However, we stress that only the whole set of obtained replicas is meaningful and provides complete information.

To estimate the uncertainties due to the experimental data, we follow the procedure described in Ref.~\cite{Ball:2008by}. The method's essence is to generate pseudo-data replicas derived from the experimental data with central values randomly distributed according to the experimental uncertainties. In this procedure, one accounts for the origin of the uncertainties and properly re-scales the error-bands in the pseudo-data. For each replica, we find the values of the nonperturbative parameters that minimize the value of $\chi^2$-test function. The resulting set of parameter-vectors samples the empirical probability density distribution of parameters. Therefore, with a sufficiently large number of replicas, we can reliably estimate uncertainties for the Sivers function and related observables. We use 500 replicas, which give a reliable precision (2-3 significant digits) that we have verified by computing and comparing to 1000 replicas for several cases.

The uncertainty for \texttt{SV19} extraction of unpolarized TMD distributions is accounted for by including the distribution of 300 replicas from \texttt{SV19} analysis (both for NNLO and N$^3$LO). To estimate the uncertainties due to unpolarized TMD PDFs, we have minimized $\chi^2$ for each replica of \texttt{SV19} set with central values of experimental data points. Note that this computation requires a re-evaluation of the normalization factors for asymmetries for each replica.

\begin{figure}[htb]
\includegraphics[width=0.475\textwidth]{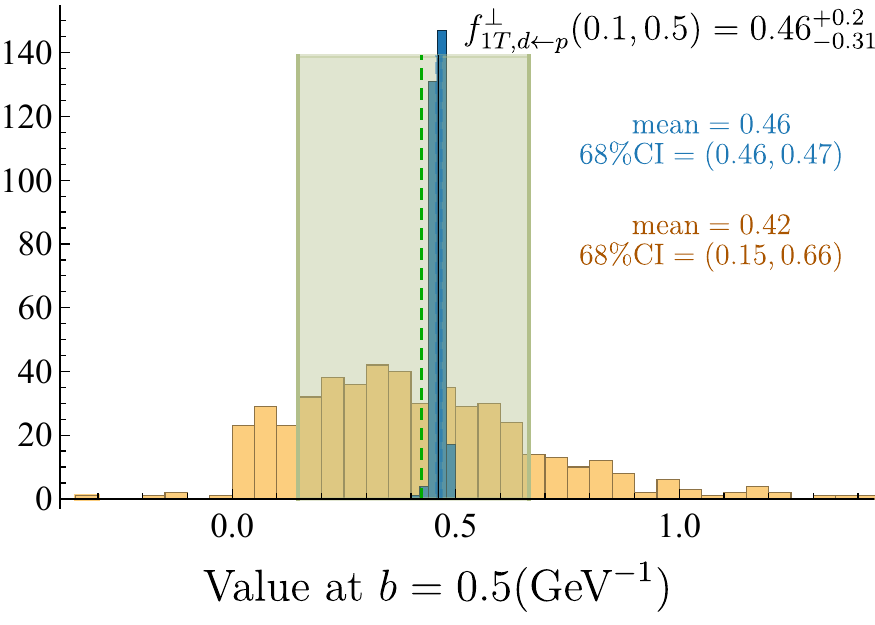}
\includegraphics[width=0.475\textwidth]{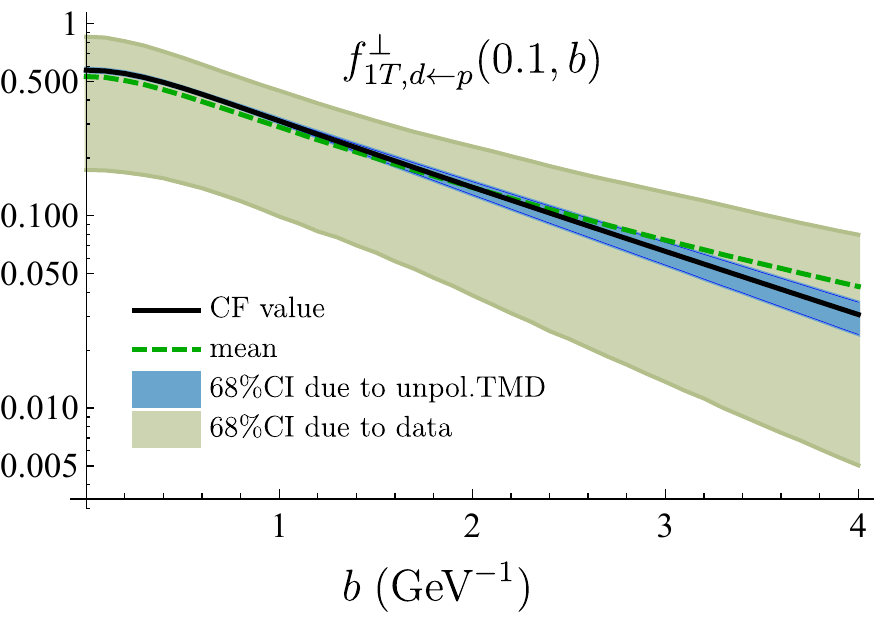}
\caption{\label{fig:params2} An example of evaluation of an observable (here it is the optimal Sivers function for $d$-quark at $x=0.1$ as a function of $b$). The observable is computed for all sets of replicas for each point (an example of the distribution at $b=0.5$ (GeV$^{-1}$) is shown in the left panel). The color notation is the same as in Fig.~\ref{fig:params}}
\end{figure}

After the implementation of these two procedures, we have two distributions of model parameters -- due to the data uncertainty and due to the uncertainties of unpolarized TMD distributions. For the ideal data/theory input, almost Gaussian statistics for distributions of parameters and observables should be observed. In our case, we have found the distributions that deviate significantly from the Gaussian shape (especially the ones due to the experimental uncertainties). Typically, they are skewed and have long, power-like tails. Two examples of replica distributions (we select parameters $r_1$ and $N_{\text{sea}}$ as examples with the widest distribution due to \texttt{SV19} uncertainty) and their parameters determination are shown in Fig.~\ref{fig:params}. We also observed that the distribution due to \texttt{SV19} uncertainty is much narrower (in most cases by order of magnitude) and less skewed compared to the distribution due to the uncertainty of the experimental data. Therefore, we use the mean value of the distribution due to \texttt{SV19} uncertainty as to the central fit value (CF value). CF value is the value of our best estimate of the true values for the parameters. The uncertainty is given by 68\% confidence interval (68\%CI) computed with distribution due to the data uncertainty using the bootstrap method, see Ref.~\cite{bootstrapbook}. The results for observables are presented in the form
\begin{eqnarray}\label{def:uncer}
\text{observable}=\text{CF value}^{+\delta_1}_{-\delta_2},
\end{eqnarray}
where $\delta_{1,2}$ distances to boundaries of 68\%CI. For continuous functions such as the Sivers function, we use the same method for each point; see an example in Fig.~\ref{fig:params2}. The resulting distributions of replicas are available as a part of \texttt{artemide} distribution \cite{artemide_sivers}. 

Let us explain our choice of CF values. There are several strategies for determination of CF value used in the literature, compare e.g. with \cite{Bacchetta:2020gko,Echevarria:2020hpy,Anselmino:2008sga}. Usually, one uses the fit to central values of data, and thus it is close to the true minimum of $\chi^2$ test. However, such a choice is also problematic because resulting parameters could lie outside of 68\%CI (see e.g.~\cite{Bacchetta:2020gko}). Such a situation could happen due to the over-fitting or due to the skewness of distributions. The usage of CF value avoids these problems because averaging over \texttt{SV19} replicas washes out possible over-fitted cases and remains close to the global minimum of $\chi^2$, and therefore is highly probable.

In the plots that represent our results, we do not show the uncertainty due to the unpolarized \texttt{SV19} input. The main reason is that it is small in comparison to the data-related uncertainty. Another reason is that two uncertainty bands could not be combined to the total uncertainty as a quadrature due to the essential non-Gaussianity of distributions. The accurate determination of total uncertainty requires the generation of multiple replicas for each replica of \texttt{SV19} and thus is very computationally costly. Provided that the uncertainty due to the data is dominating in all cases compared to the uncertainty due to the unpolarized distributions, we have decided to showcase only the former. However, generally speaking, the uncertainty band due to the unpolarized input is not negligible, contrary to the commonly used assumption. To our best knowledge, this is the first estimation of such uncertainty in the analysis of polarized TMD distributions. 

\subsection{Test of the factorization region limit}
\label{sec:delta-test}
\begin{figure}[htb]
\begin{center}
\includegraphics[width=0.65\textwidth]{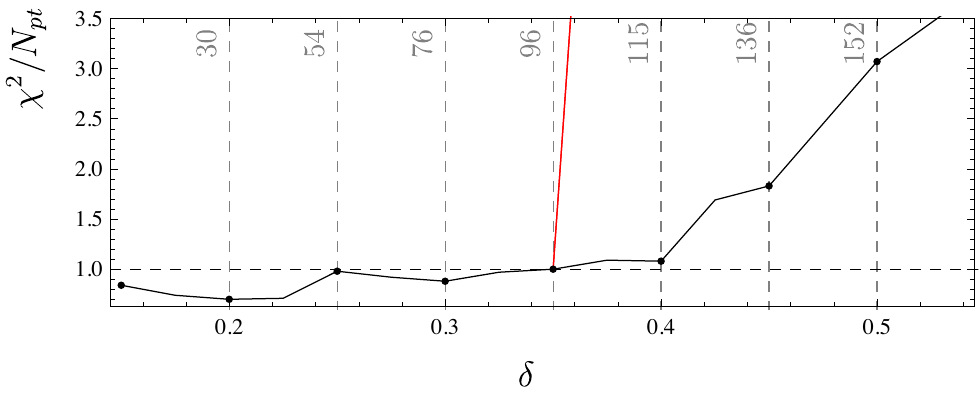}
\end{center}
\vskip -0.7cm
\caption{\label{fig:chiVSdelta} Comparison of $\chi^2/N_{pt}$ for different values of $\delta$. The fits are made for SIDIS+DY at N$^3$LO setup. Gray numbers at the top of the figure show the number of points. At $\delta=0.375$ the additional point of $\pi$-induced DY contributes with a very large $\chi^2$ (red line). For $\delta>0.375$ only SIDIS points are added.}
\end{figure}
The TMD factorization works at small values of $\delta$, see Eq.~(\ref{def:delta}), see also discussion in Ref.~\cite{Boglione:2019nwk}. A priori, the size of power corrections, which violate the factorization approach, is not known. For that reason,  one should implement a data selection cut, Eq.~(\ref{def:cuts}), and exclude the data with large values of $\delta$. In this section, we perform a survey of different values of $\delta$ and thus test the boundary for the TMD factorization for asymmetries.

To test TMD factorization's applicability, we perform fits of the data selected with different values of $\delta$ in Eq.~(\ref{def:cuts}). The fits are executed at NNLO accuracy. In the region where the factorization theorem is applicable, one expects the value of $\chi^2/N_{pt}\simeq 1$, and grows to larger values outside of the applicability region. In other words, the plot of $\chi^2/N_{pt}$ versus $\delta$ should have a plateau in the validity range of the factorization theorem. Such a test has been suggested in Ref.~\cite{Scimemi:2017etj} and successfully applied for unpolarized SIDIS and DY data analysis, where it was found that the optimal data cut is $\delta\sim 0.2-0.25$, see Refs.~\cite{Scimemi:2019cmh,Bacchetta:2019sam}. It has been shown that for DY processes, the data with $\delta>0.3$ are poorly described by the TMD factorization formula. In contrast,  the situation for SIDIS is better, and one could go up to $\delta\sim 0.3-0.35$ \cite{Scimemi:2019cmh} without significant loss of quality of the fit. These numbers served as the rationale for the initial estimation of our current data selection cut in Eq.~(\ref{def:cuts}).

The results of our current test are shown in Fig.~\ref{fig:chiVSdelta}, which has a clear plateau $\chi^2/N_{pt}\simeq 1$ for $\delta<0.4$. The quality of the fit drops drastically for $\delta>0.4$ for SIDIS. This result agrees with the general expectations. Indeed, one could expect that power corrections partially cancel in asymmetry, and thus the kinematic range for the applicability of factorization theorems becomes slightly wider. Since, in the SIDIS unpolarized case $\delta<0.3-0.35$ the observation of rough agreement for $\delta\lesssim0.4$ is anticipated.

The situation for DY is less certain because the total number of points is small. All points included into the fit have $\delta<0.22$ (see Fig.~\ref{fig:points}). There is only one additional point to include. This point is measured in pion-induced DY at COMPASS~\cite{Aghasyan:2017jop}, and it has $\delta=0.36$ and a wide $q_T$-bin  up to values $q_T\simeq Q$. This point is outside of the applicability range, and the prediction strongly disagrees with the measurement ($\chi^2\sim8$). The main source of the disagreement is the denominator in Eq.~(\ref{th:AUT}), which becomes negative. The negative values for cross-section are typical for TMD factorization formula in the region beyond its validity. To get the positive cross-section valid in the full range of $q_T$ one should match it to the collinear picture via the so-called $Y$-term~\cite{Collins:2011zzd}. This goes far beyond the present study.

We conclude that even though the region of TMD factorization widens slightly for asymmetries, one has to be cautious when including the data outside of the TMD factorization region. In the following sections, we analyze only the data with $\delta<0.3$. This value corresponds to our best estimate of the region of data appropriate for the TMD factorization approach description. Future work that will include matching to the collinear factorization is needed to widen the region of the data used in the global analysis.

\section{Results of extraction}
\label{sec:results}

This is the main Section of our work. We describe in detail results of N$^3$LO extraction of the Sivers function, also presented in Ref.~\cite{Bury:2020vhj}. We discuss the Sivers function in momentum and position spaces, discuss positivity constraints, show the 3D tomography of the nucleon via the Sivers function, extract the Qiu-Sterman functions, and study the significance of the sign change of the Sivers function between SIDIS and DY.

\subsection{Fit of the data}
\label{sec:fitdata}
\begin{figure}[htb]
\begin{center}
\includegraphics[width=0.85\textwidth]{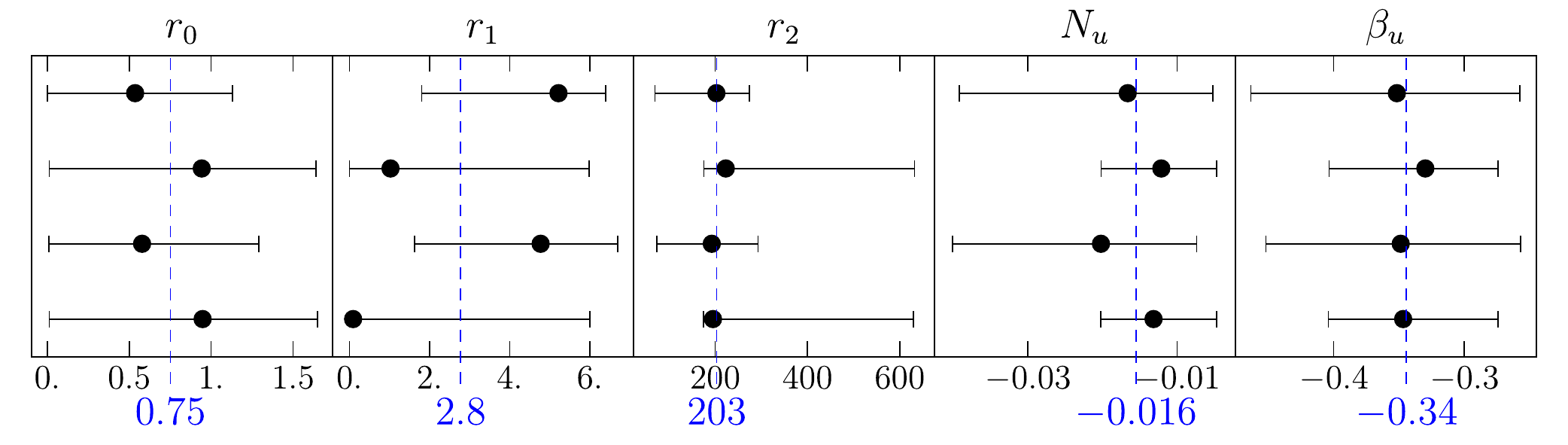}
\includegraphics[width=0.85\textwidth]{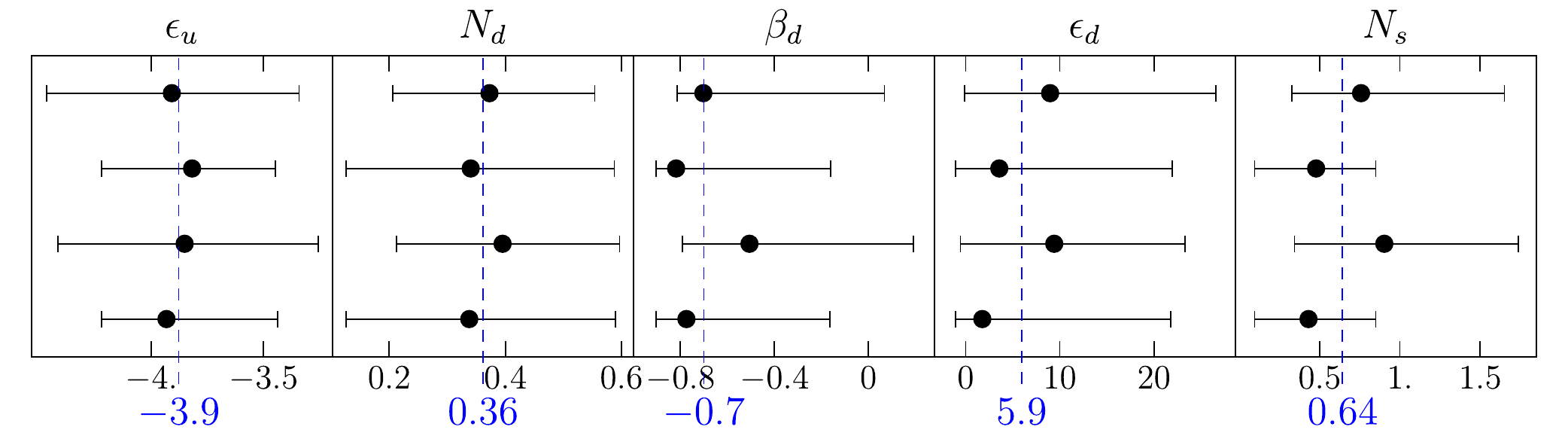}
\includegraphics[width=0.5\textwidth]{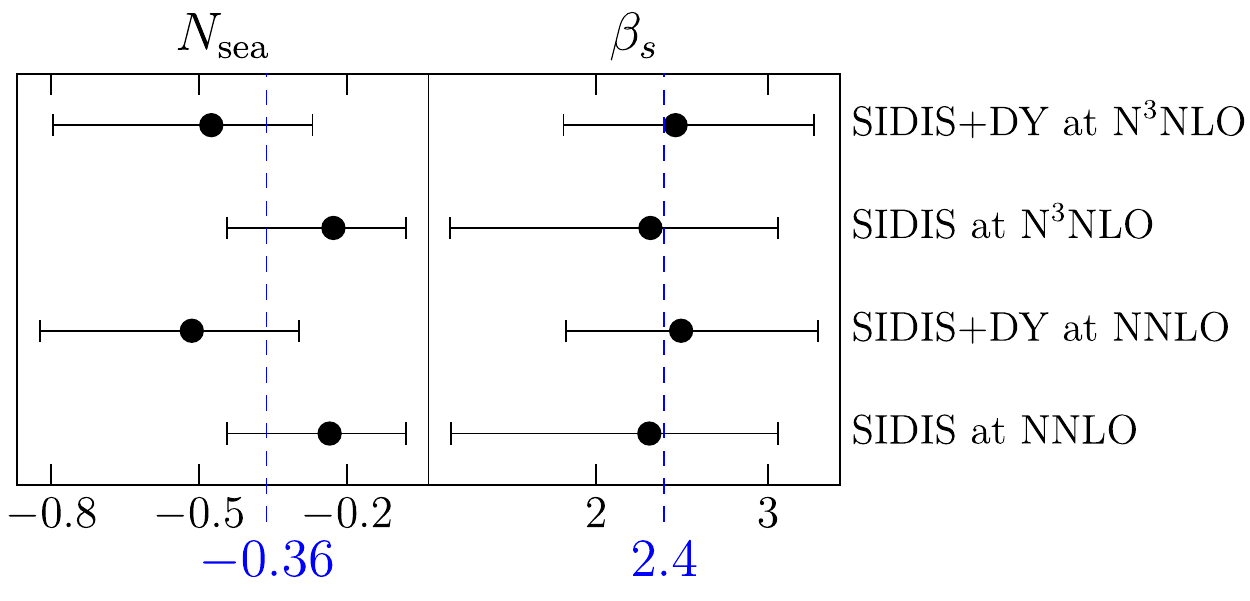}    
\end{center}

\caption{\label{fig:parameters-plot} The   comparison of values for model parameters obtained in fits. Numerical values of parameters are given in Table \ref{tab:params}. The blue dashed line and blue numbers show the average over all fits, in order to estimate the stability of fit results.}
\end{figure}

Using the approach described in the previous sections, we performed several fits with different setups. In particular, we distinguish the fits with and without inclusion of DY data, with a purpose to estimate the  universality of the Sivers function. Also we performed separate fits at NNLO and N$^3$LO perturbative precision for the TMD evolution. The synopsis of $\chi^2$ values is presented in Table~\ref{tab:chi2}. The distribution of contributions to $\chi^2$ per experiments is shown in Table~\ref{tab:chi2-drop-down}. The values of nonperturbative parameters extracted in these fits are given in Table~\ref{tab:params} and in Fig.~\ref{fig:parameters-plot}.
\begin{table}[b]
\begin{center}
\begin{tabular}{l||c|c||c}
Name & $\chi^2/N_{pt}$[SIDIS] & $\chi^2/N_{pt}$[DY] &  $\chi^2/N_{pt}$[total]
\\\hline
SIDIS at NNLO & $0.88_{+0.03}^{+0.13}$ & $1.29_{-0.30}^{+0.45}$ no fit&  $0.95_{+0.00}^{+0.16}$
\\
SIDIS+DY at NNLO & $0.90_{+0.02}^{+0.13}$ & $0.94_{-0.01}^{+0.25}$ \phantom{no fit}&  $0.91_{+0.04}^{+0.13}$
\\\hline
SIDIS at N$^3$LO & $0.87_{+0.03}^{+0.13}$ & $1.23_{-0.24}^{+0.50}$ no fit&  $0.93_{+0.01}^{+0.16}$
\\
SIDIS+DY at N$^3$LO & $0.88_{+0.04}^{+0.15}$ & $0.90_{+0.00}^{+0.31}$ \phantom{no fit} & $0.88_{+0.05}^{+0.15}$
\end{tabular}
\end{center}
\caption{\label{tab:chi2} Values for $\chi^2/N_{pt}$ in different fits. Note, that for the cases included in the fit the CF value of $\chi^2$ lies outside the 68\%CI. This is because CF realizes the minimum of $\chi^2$ distribution, whereas the 68\%CI (roughly) excludes 16\% of boundary replicas.}
\end{table}

The main observation that follows from Table~\ref{tab:chi2} is that the Sivers function extracted in SIDIS data only nicely describes the DY data, even without extra tuning. Indeed, the Sivers function extracted at NNLO from SIDIS data results in $\chi^2/N_{pt} = 1.29$ for DY data, and the agreement improves slightly at N$^3$LO. The overall quality of the description of SIDIS and DY data from the SIDIS data only fit with $\chi^2/N_{pt} = 0.93$ indicates the consistency of the data with the universality of the Sivers function. The inclusion of DY data into the fit reduces the $\chi^2$ associated with DY. It modifies some nonperturbative parameters, especially those related to the high-$x$ behavior of the Sivers function, such as $\beta$ and $\epsilon$, see Fig.~\ref{fig:parameters-plot}.  Let us mention that this is the first consistent description of polarized Drell-Yan data in the TMD formalism  with TMD evolution. The inclusion of DY data in the fit does not result in the worsening of a description of SIDIS data, and thus the fit demonstrates the absence of the tension between SIDIS and DY data. The inclusion of both SIDIS and DY data sets results in a better overall description of the data $\chi^2/N_{pt} = 0.88$ (at N$^3$LO) and the most precise determination of the Sivers function. We, therefore, believe that our extraction demonstrates the universality of the Sivers function.

\begin{table}[htb]
\renewcommand{\arraystretch}{1.2}
\centering
\begin{tabular}{l|c|c || l|c|c}
SIDIS & $N_{\text{pt}}$ & $\chi^2/N_{pt}$ & DY & $N_{\text{pt}}$ & $\chi^2/N_{pt}$
\\\hline
Compass08 & 4 & $0.29_{-0.01}^{+0.30}$ & CompassDY & 2 & $0.19_{-0.15}^{+0.49}$
\\
Compass16 & 10 & $0.34_{+0.01}^{+0.36}$ & STAR.W+ & 5 & $0.72_{+0.40}^{+0.69}$
\\
Hermes $\pi^+$ & 11 & $0.79_{-0.05}^{+0.16}$ & STAR.W- & 5 & $0.92_{-0.14}^{+0.31}$
\\
Hermes $\pi^-$ & 11 & $0.49_{+0.01}^{+0.08}$ & STAR.Z & 1 & $2.04_{-1.15}^{+0.10}$
\\
Hermes $K^+$ & 12 & $1.36_{-0.11}^{+0.15}$ & & & 
\\
Hermes $K^-$ & 12 & $1.62_{-0.01}^{+0.12}$ & & & 
\\
Jlab & 3 & $0.26_{-0.04}^{+1.13}$ & & & 
\\\hline
SIDIS total & 63 & $0.88_{+0.05}^{+0.15}$ & DY total & 13 & $0.90_{+0.00}^{+0.31}$
\\\hline\hline
\multicolumn{4}{c|}{Total SIDIS and DY}  & 76 & $0.88_{+0.05}^{+0.15}$
\\\hline
\end{tabular}
\caption{\label{tab:chi2-drop-down}
Distribution of $\chi^2$  values over the data sets. The results are for the fit SIDIS+DY at N$^3$LO. For other fits the distributions are analogous.}
\end{table}

The values of model parameters extracted with and without DY data agree within 68\%CI, see Fig.~\ref{fig:parameters-plot}. The main impact of the inclusion of DY data into fit happens on CF values of parameters $(r_1,N_d,N_{\text{sea}},\beta_s)$ and to a lesser degree on $(\beta_u,\epsilon_u)$. Other parameters are almost unaffected by the inclusion of DY data due to the bigger experimental uncertainty of DY data, see Table.~\ref{tab:data}.
\begin{table}[htb]
\renewcommand{\arraystretch}{1.2}
\begin{center}
\begin{tabular}{l||c|c||c|c|}
\multirow{2}{*}{Parameter} & SIDIS & DY+SIDIS & SIDIS &DY+SIDIS
\\ & at NNLO & at NNLO & at N$^3$LO & at N$^3$LO
\\\hline
$r_0$  	& $0.95_{-0.94}^{+0.70}$ 	& $0.58_{-0.57}^{+0.71}$ 	& $0.94_{-0.93}^{+0.71}$ 	& $0.54_{-0.53}^{+0.60}$
\\\hline
$r_1$ 	& $0.09_{-0.09}^{+5.90}$  	& $4.8_{-3.1}^{+1.9}$ 		& $1.02_{-1.02}^{+4.96}$	& $5.22_{-3.43}^{+1.18}$
\\\hline	
$r_2$ 	& $195._{-20.}^{+434.}$		& $192._{-119.}^{+101.}$ 		& $223._{-47.}^{+409.}$		& $203._{-133.}^{+71.}$
\\\hline
$N_u$ 	& $-0.013_{-0.007}^{+0.008}$& $-0.020_{-0.020}^{+0.013}$ & $-0.012_{-0.008}^{+0.007}$ &$-0.017_{-0.023}^{+0.011}$
\\\hline
$\beta_u$ & $-0.35_{-0.06}^{+0.07}$	& $-0.35_{-0.10}^{+0.09}$ 	& $-0.33_{-0.07}^{+0.06}$	& $-0.36_{-0.11}^{+0.09}$
\\\hline
$\epsilon_u$ & $-3.9_{-0.3}^{+0.5}$	& $-3.9_{-0.6}^{+0.6}$ 		& $-3.8_{-0.4}^{+0.4}$		& $-3.9_{-0.6}^{+0.6}$
\\\hline
$N_d$ 	& $0.34_{-0.21}^{+0.25}$	& $0.40_{-0.18}^{+0.20}$ 	& $0.34_{-0.21}^{+0.25}$	& $0.37_{-0.17}^{+0.18}$
\\\hline
$\beta_d$& $-0.77_{-0.13}^{+0.61}$	& $-0.51_{-0.29}^{+0.70}$ 	& $-0.82_{-0.08}^{+0.66}$	& $-0.7_{-0.11}^{+0.77}$
\\\hline
$\epsilon_d$& $1.8_{-2.8}^{+20.0}$	& $9.4_{-9.9}^{+13.9}$  	& $3.6_{-4.7}^{+18.4}$		& $9.0_{-9.1}^{+17.6}$
\\\hline
$N_s$ 	& $0.43_{-0.34}^{+0.42}$	& $0.90_{-0.56}^{+0.83}$ 	& $0.48_{-0.38}^{+0.37}$	& $0.76_{-0.43}^{+0.89}$
\\\hline
$N_{\text{sea}}$& $-0.23_{-0.21}^{+0.15}$& $-0.51_{-0.31}^{+0.22}$ & $-0.23_{-0.22}^{+0.15}$& $-0.47_{-0.32}^{+0.21}$
\\\hline
$\beta_{s}=\beta_{\text{sea}}$& $2.3_{-1.2}^{+0.7}$& $2.5_{-0.7}^{+0.8}$ & $2.3_{-1.2}^{+0.7}$& $2.5_{-0.7}^{+0.8}$
\end{tabular}
\end{center}
\caption{\label{tab:params} Values of parameters obtained in various fits. The visual representation of the table is given in Fig.~\ref{fig:parameters-plot}.}
\end{table}

A previous attempt to describe DY $W$ and $Z$ data with TMD evolution was made in Ref.~\cite{Echevarria:2020hpy}. It faced serious problems giving the best $\chi^2/N_{pt}\sim 1.5 - 1.9$ for $W^\pm/Z$ data (depending on additional assumptions on the evolution of the Sivers function). In our case, the $\chi^2/N_{pt}$ for DY  and $W^\pm/Z$ data close to 1 even without the inclusion of these data into the fit. Unfortunately, it is not possible to directly compare our and \cite{Echevarria:2020hpy} approaches and find the origin of the disagreement. Although the main spirit of both works is similar, the number of smaller differences is significant. Let us mention the differences between these works that, in our opinion, influence the results. First, in \texttt{SV19} and in the present extraction, we operate with the data that definitely belong to the TMD factorization region. In contrast, in Ref.~\cite{Echevarria:2020hpy} the cut is much wider, $\delta<0.75$, assuming possible cancellation of power corrections to TMD factorization (see Sec.~\ref{sec:delta-test}). Second, we use the unpolarized TMD distributions extracted in \texttt{SV19}. So far, \texttt{SV19} is the only extraction of unpolarized TMD distributions that describes both SIDIS and DY data without additional normalization conditions. The other differences are the TMD evolution implementation ($\zeta$-prescription vs. CSS-like ansatz) and the nonperturbative model for the Sivers function. In particular, our parametrization for the Sivers function is more flexible compared to Ref.~\cite{Echevarria:2020hpy} and allows sea quark contributions to be large in the large-$x$ region.

In the remainder of the Section, we will discuss details of the description of various SIDIS and DY data sets coming from various experiments considered in this analysis (see also Table~\ref{tab:chi2-drop-down}). We present results of the description and discuss the data.

\begin{figure}[t]
\begin{center}
\includegraphics[width=0.495\textwidth]{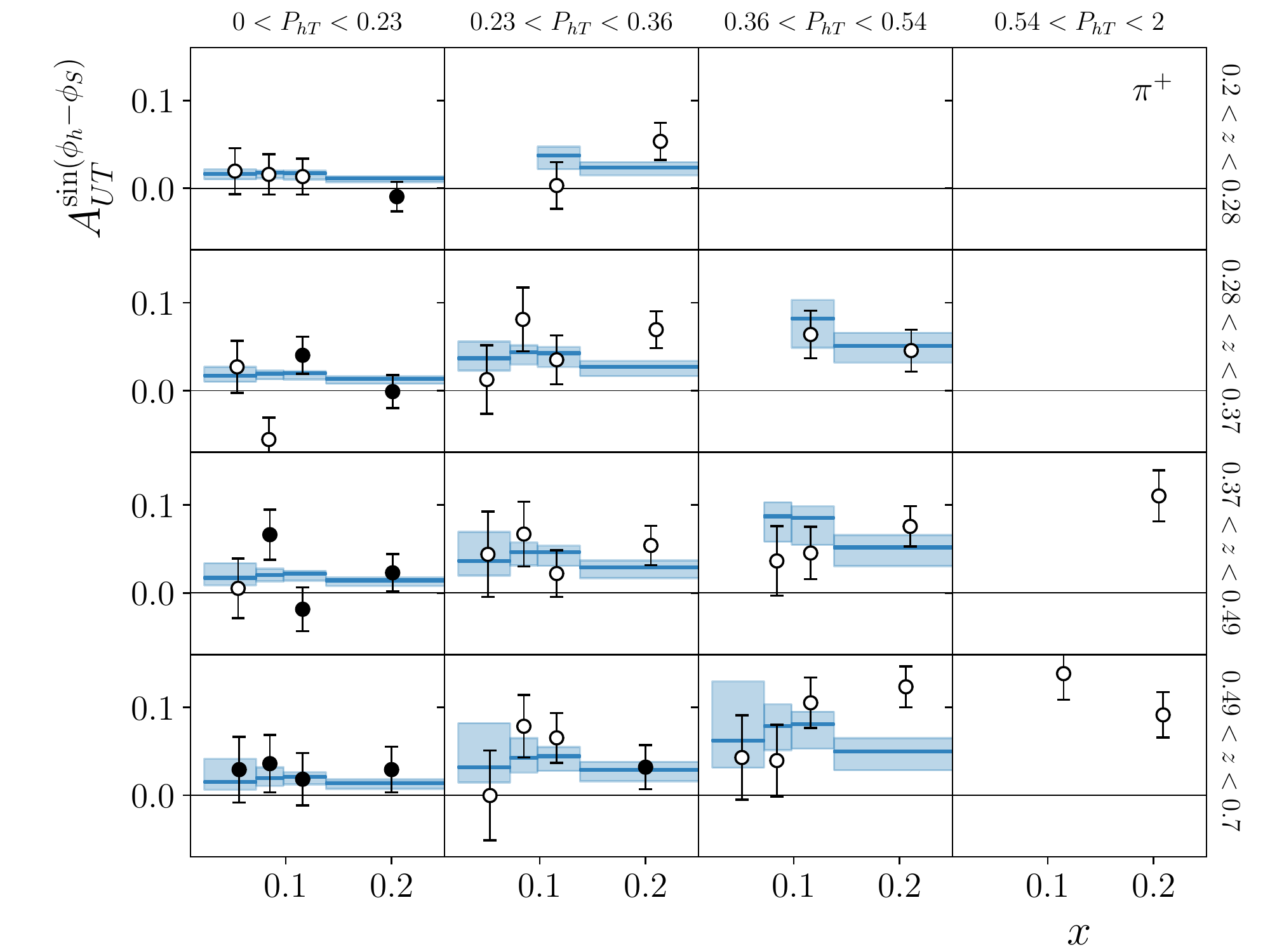}
\includegraphics[width=0.495\textwidth]{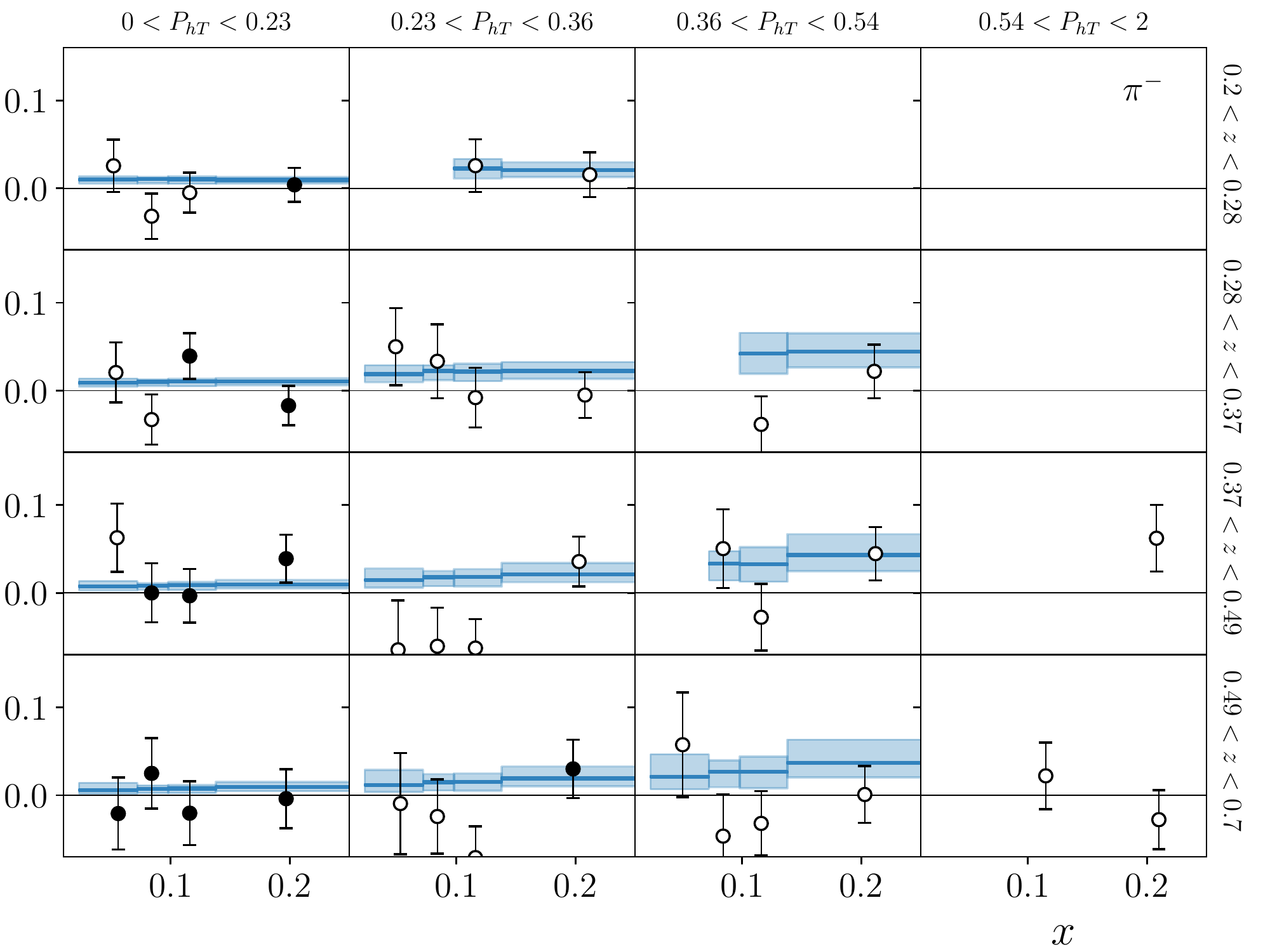}
\includegraphics[width=0.495\textwidth]{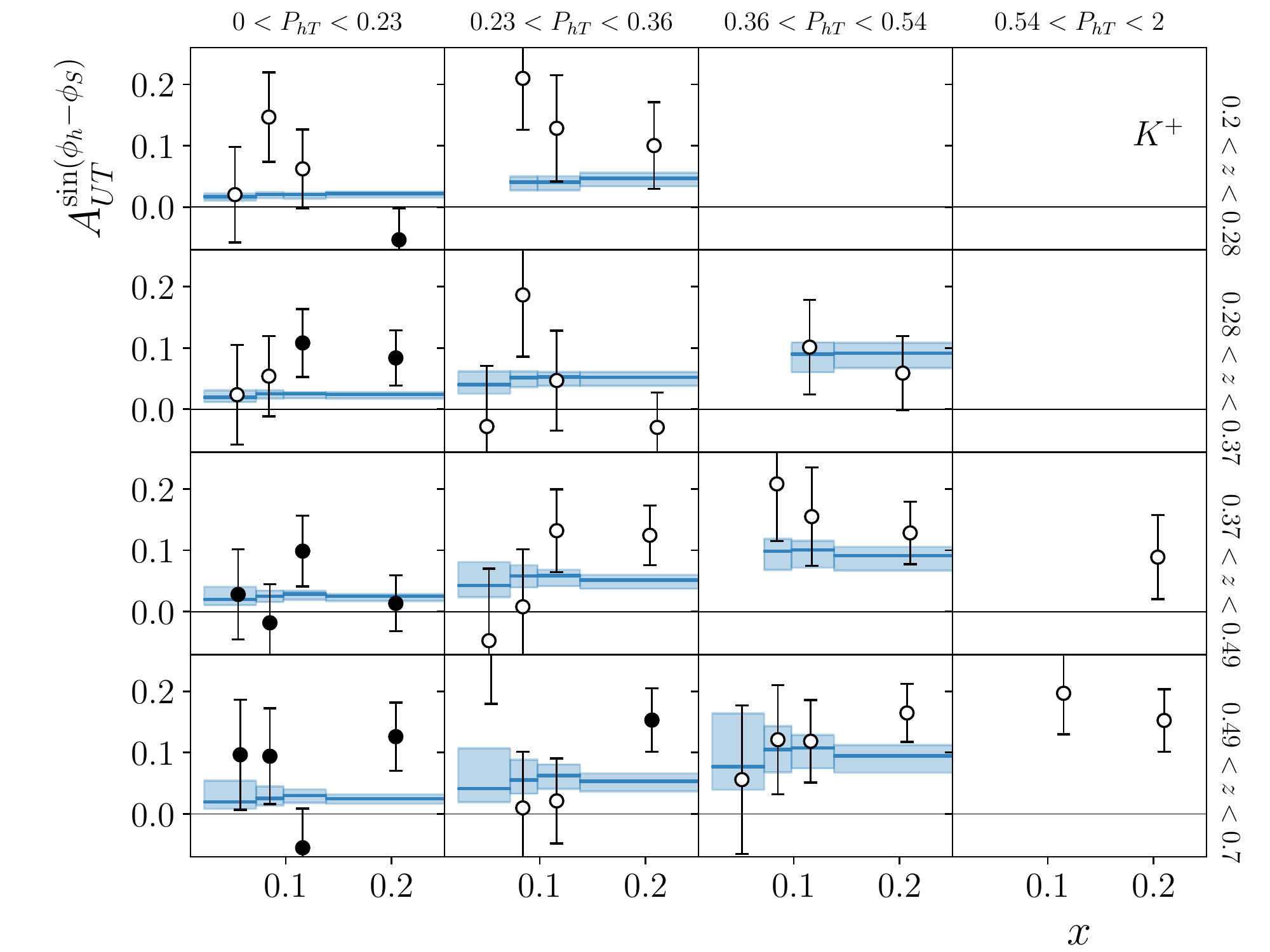}
\includegraphics[width=0.495\textwidth]{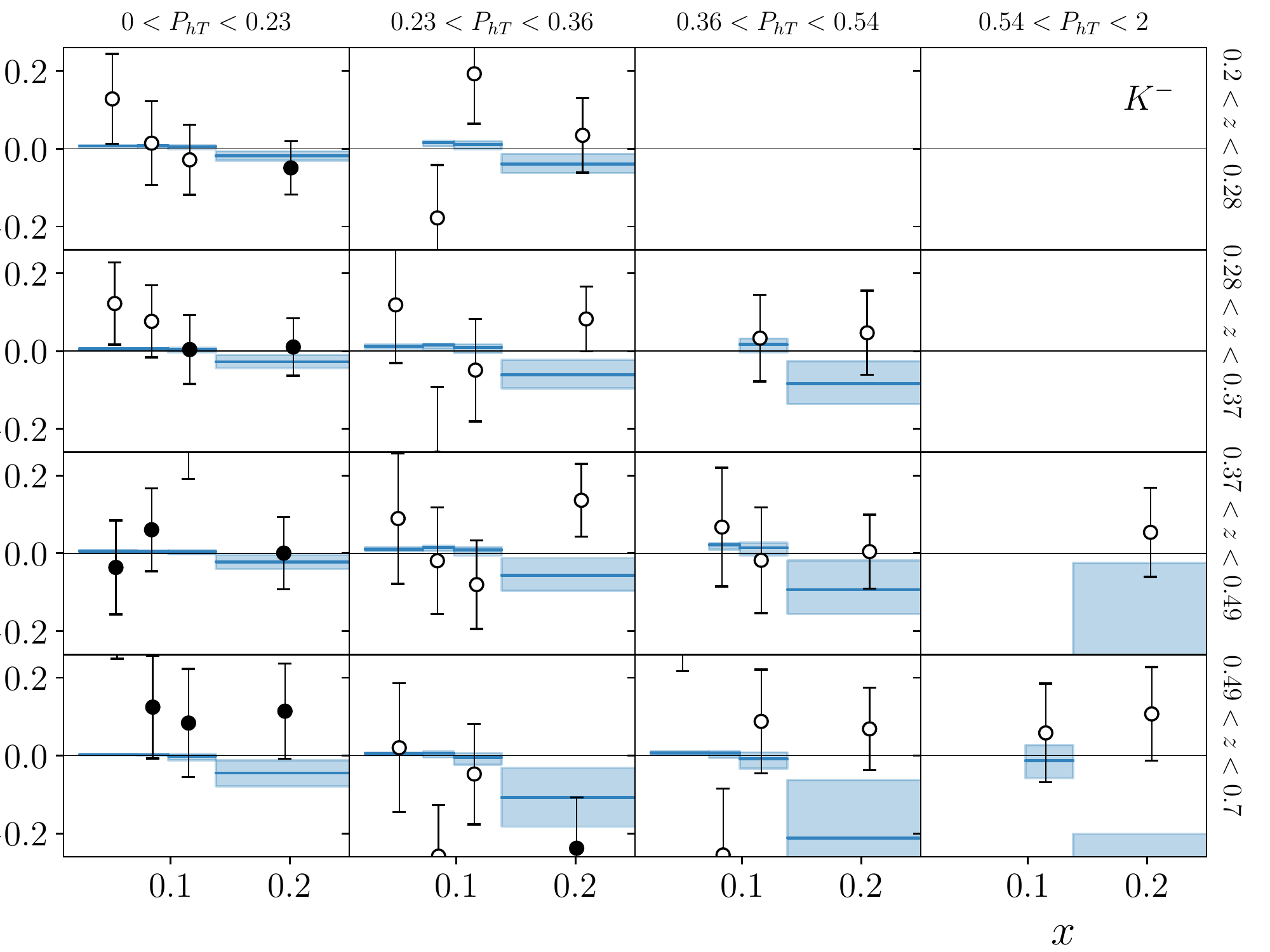}
\end{center}
\vskip -0.5cm
\caption{\label{fig:hermes} Description of HERMES data~\cite{Airapetian:2020zzo} for $\pi^\pm$ and $K^\pm$, only data with $\delta<0.5$ are shown. The data are presented as the function of $x$ and the 3D binning of the data is indicated by the bin sizes in $P_{hT}$ (GeV) and $z$.  Solid (open) symbols data used (not used) in the fit. Blue line is the CF and the blue box  is 68\%CI of the fit of the data and prediction for the data not used in the fit.}
\end{figure}

\textit{HERMES data set~\cite{Airapetian:2020zzo}}. In our fit we use the latest updated data on Sivers asymmetry in SIDIS by the HERMES Collaboration~\cite{Airapetian:2020zzo} on the proton target for $\pi^\pm$, $K^\pm$. The incident electron energy is $P_{lab} = 27.5$ GeV. Events were selected subject to the requirements $Q^2$ > 1 GeV$^2$, $W^2$ > 10 GeV$^2$, $0.1 < y < 0.95$, and $0.023<x<0.6$. Hadrons were accepted if $0.2 < z < 0.7$. The data are presented in a three-dimensional binning in $x$, $z$, and $P_{hT}$ (GeV). The correlated uncertainty of the data is 7.3\% due to the accuracy of the target polarization determination. Importantly, the systematic uncertainty of HERMES data already includes possible effects of the bin-integration, and thus the theory prediction for this data set must be evaluated using the average bin kinematics.
For the SIDIS subset, the largest $\chi^2/N_{pt}$ is for $K^-$ production measured at HERMES (typical values $\sim 1.6$ for 12 points). The next-to-the-largest $\chi^2/N_{pt}$ is for $K^+$-production measured at Hermes (typical values $\sim 1.3$ for 12 points). The rest of the SIDIS data, for $\pi^\pm$ and $h^\pm$, have partial $\chi^2/N_{pt}$ which are smaller than 1. These relatively large contributions may be related to either poor knowledge of Kaon fragmentation functions or sea quark Sivers function. Description of HERMES data is presented in Fig.~\ref{fig:hermes} where we plot only the data with $\delta < 0.5$ and show description for both the data used in the fit (solid points) and the data not used in the fit (open points).

\begin{figure}[t]
\begin{center}
\includegraphics[width=0.59\textwidth]{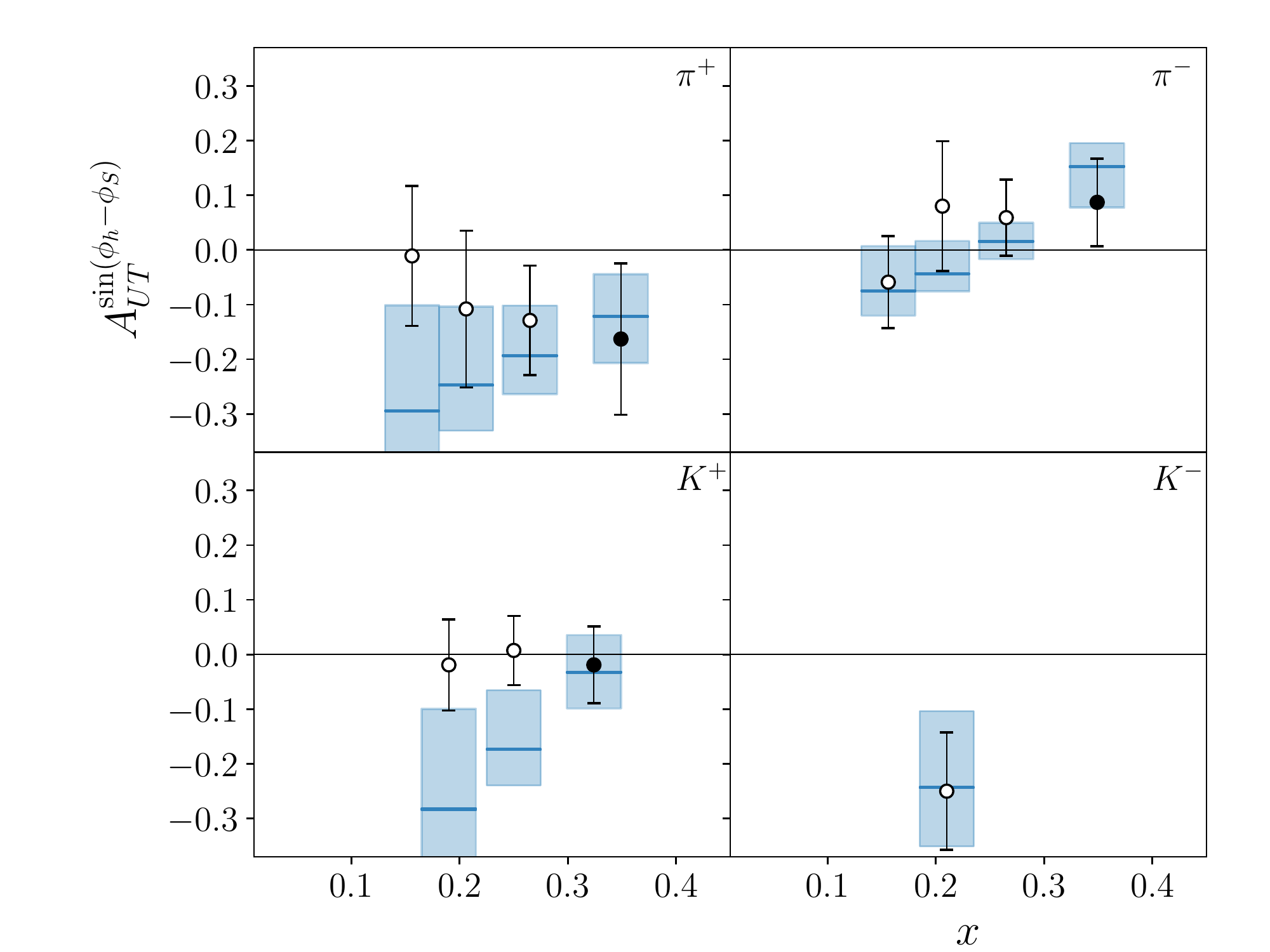}
\end{center}
\vskip -0.5cm
\caption{\label{fig:jlab} Description of JLab HALL~A data~\cite{Qian:2011py,Zhao:2014qvx} for $\pi^\pm$ and $K^\pm$. Solid (open) symbols data used (not used) in the fit. Blue line is the CF and the blue box  is 68\%CI of the fit of the data and prediction for the data not used in the fit.}
\end{figure}
\textit{JLab data set~\cite{Qian:2011py,Zhao:2014qvx}}. Jefferson Lab experiments in HALL A measured Sivers asymmetry on $^3$He target for $\pi^\pm$ \cite{Qian:2011py}  and $K^\pm$ \cite{Zhao:2014qvx}. The experiment, conducted at Jefferson Lab using a 5.9 GeV electron beam, covers a range of $0.14 <x< 0.34$ with $1.3 <Q^2< 2.7$ GeV$^2$. SIDIS events were selected using cuts on the four-momentum transfer squared $Q^2 > 1$ GeV$^2$, the hadronic final-state invariant mass $W > 2.3$ GeV. The data were presented as Bjorken-$x$ projection. We show the description of JLab HALL~A data~\cite{Qian:2011py,Zhao:2014qvx} ~in Fig.~\ref{fig:jlab}.

\begin{figure}[b]
\begin{center}
\includegraphics[width=0.59\textwidth]{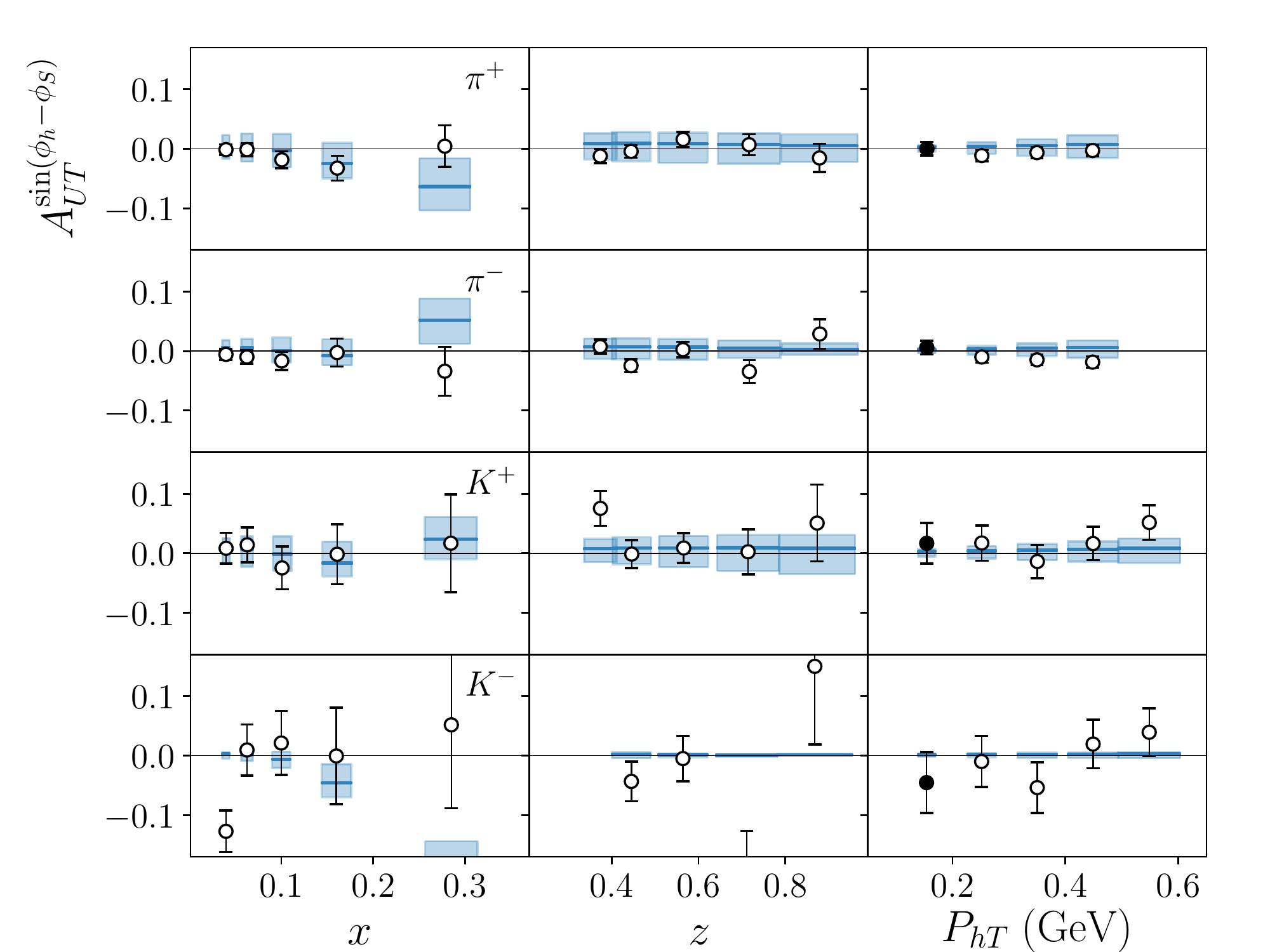}
\end{center}
\vskip -0.5cm
\caption{\label{fig:compass} Description of COMPASS SIDIS data~\cite{Alekseev:2008aa} for $\pi^\pm$ and $K^\pm$, only data with $\delta<0.5$ are shown. Solid (open) symbols data used (not used) in the fit. Blue line is the CF and the blue box  is 68\%CI of the fit of the data and prediction for the data not used in the fit.}
\end{figure}

\begin{figure}[t]
\begin{center}
\includegraphics[width=0.495\textwidth]{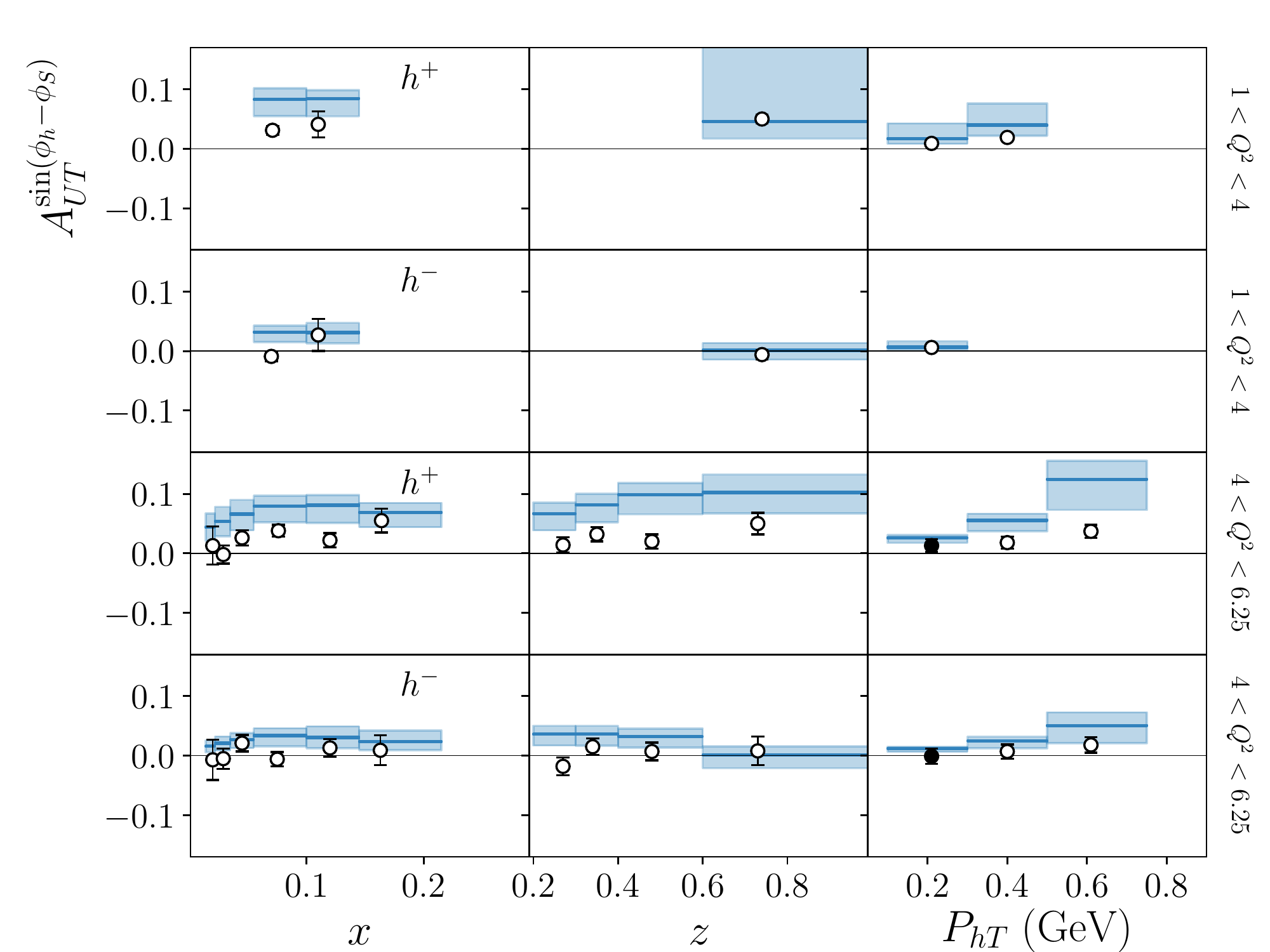}
\includegraphics[width=0.495\textwidth]{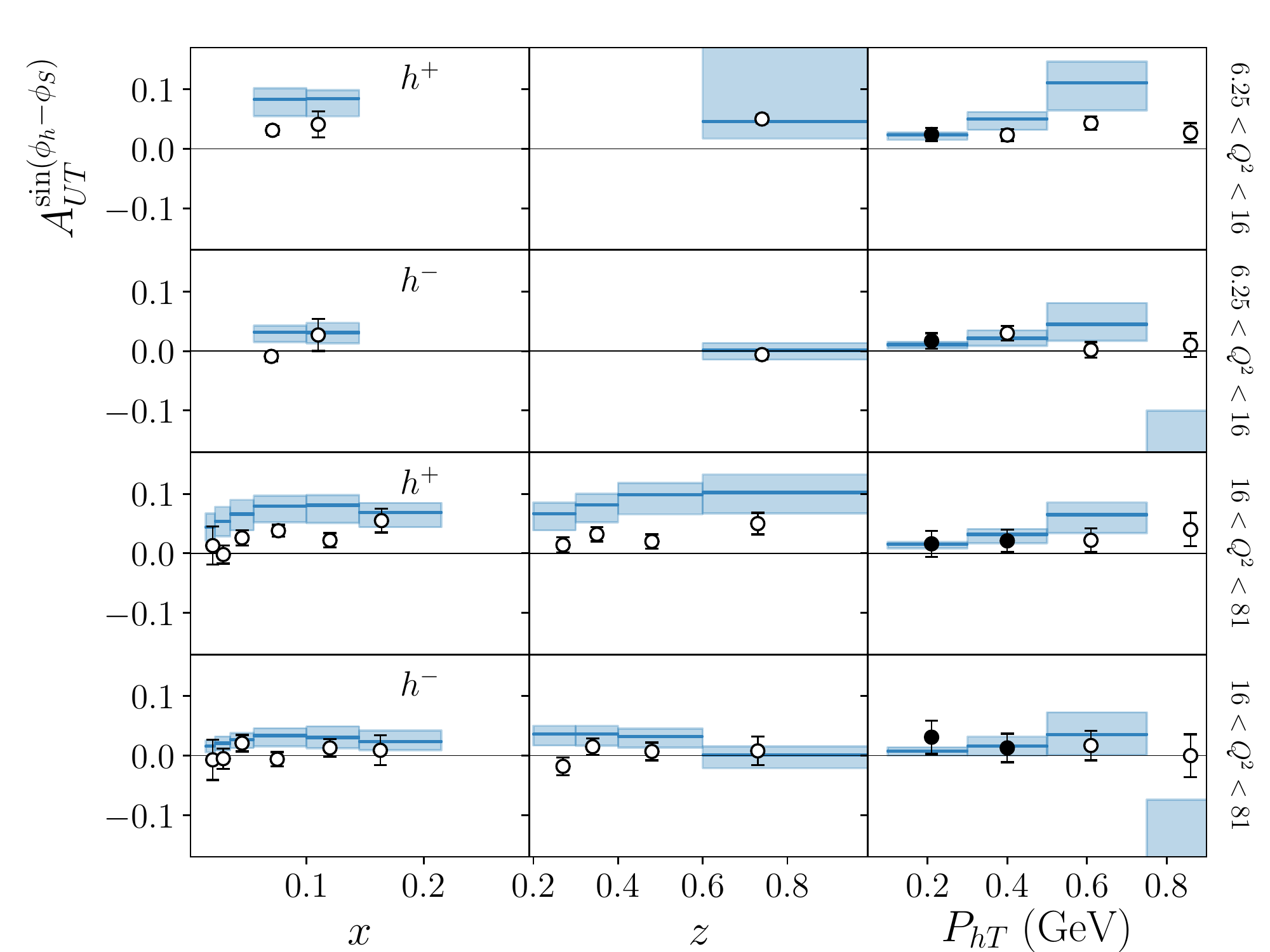}
\end{center}
\vskip -0.5cm
\caption{\label{fig:compass16} Description of multi-dimensional COMPASS SIDIS proton data~\cite{Adolph:2016dvl}. Sivers asymmetry for $z > 0.1$ in the four $Q^2$-ranges as a function of $x$, $z$ and $P_{hT}$ for unidentified charged hadrons $h^\pm$, only data with $\delta<0.5$ are shown. Solid (open) symbols data used (not used) in the fit. Blue line is the CF and the blue box  is 68\%CI of the fit of the data and prediction for the data not used in the fit.}
\end{figure}
\textit{Compass08~\cite{Alekseev:2008aa} and Compass16~\cite{Adolph:2016dvl}  data sets}. COMPASS measured the Sivers asymmetry using different targets (iso-scalar samples from 2003-2004 data \cite{Alekseev:2008aa} and proton sample for unidentified charged hadrons from 2010, multi-dimensional data~\cite{Adolph:2016dvl})  with incident muon energy $P_{lab} = 160$ GeV. In these measurements, the cuts on the photon virtuality $Q^2$ > 1 GeV$^2$ and the mass of the hadronic final state $W^2 > 25$ GeV$^2$ were applied, as well as $0.1 < y < 0.9$.  To simulate the isospin target (deuteron), we make the iso-spin rotation for components of the Sivers function
\begin{eqnarray}
f^\perp_{1T,u\ot d}=f^\perp_{1T,d\ot d}=\frac{f^\perp_{1T,u\ot p}+f^\perp_{1T,d\ot p}}{2}.
\end{eqnarray}
The measurement Compass08 is made for $\pi^\pm$ and $K^\pm$ fragmenting hadrons (we omit the $\pi^0$ and $K^0$ measurements because \texttt{SV19} extraction does not have these fragmentation functions). The Compass16 measurements is made for charged hadrons $h^\pm$, which in \texttt{SV19} are approximated as sum of pion and kaon components $h^\pm=\pi^\pm+K^\pm$ ignoring the higher-mass contribution.
We show the description of COMPASS SIDIS data~\cite{Alekseev:2008aa} in Fig.~\ref{fig:compass} and~\cite{Adolph:2016dvl} in Fig.~\ref{fig:compass16}. One can see that, as in previous cases, the data description is good even for the data not used in the fit.

\begin{figure}[b]
\begin{center}
\includegraphics[width=0.6\textwidth]{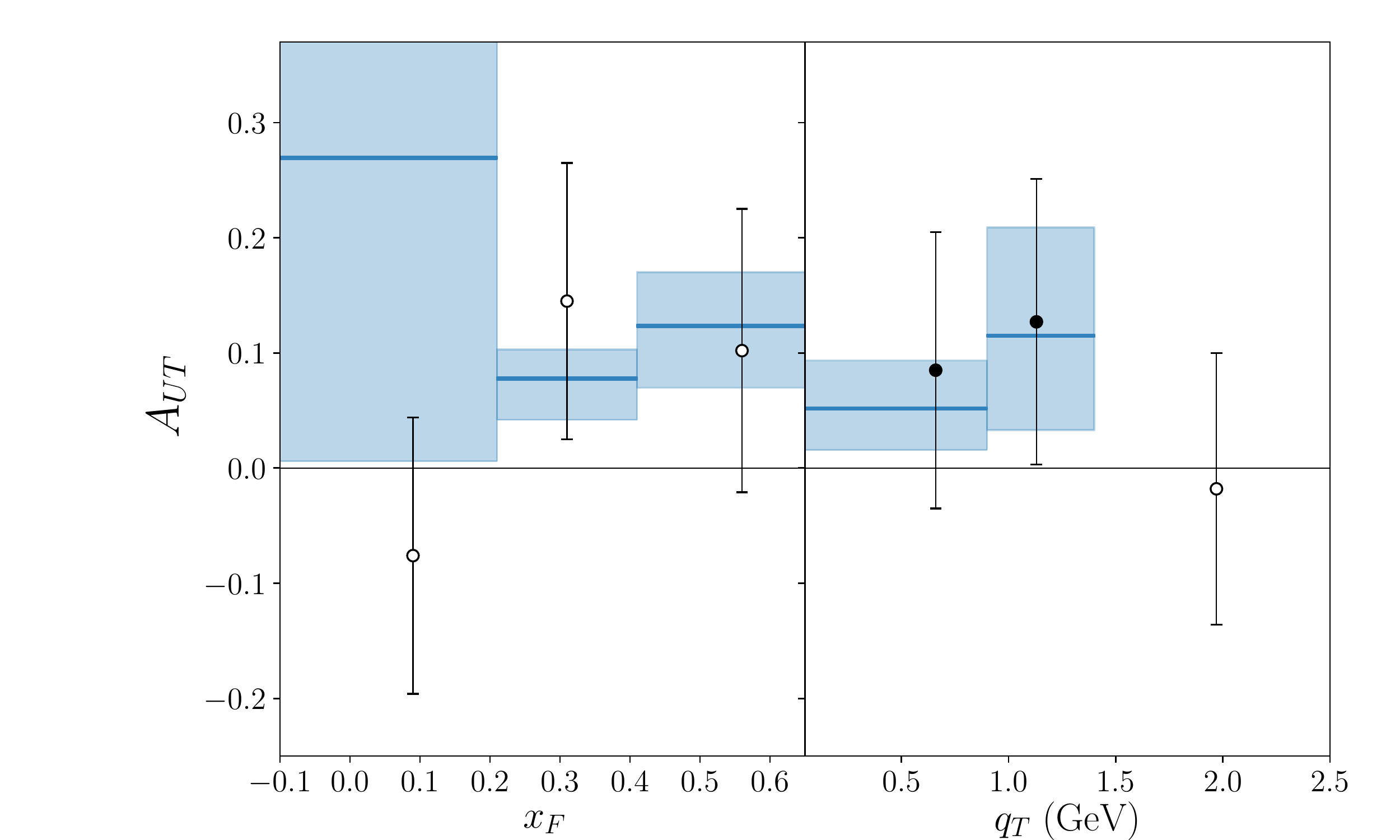}
\end{center}
\vskip -0.5cm
\caption{\label{fig:compassDY} Description of Compass DY data~\cite{Aghasyan:2017jop} as a function of $x_F$ and $q_T$ (GeV). Solid (open) symbols data used (not used) in the fit. Blue line is the CF and the blue box  is 68\%CI of the fit of the data and prediction for the data not used in the fit.}
\end{figure}
\textit{CompassDY~\cite{Aghasyan:2017jop} data set}. The data were taken using a high-intensity $\pi^-$ beam of 190 GeV and the transversely polarized isoscalar NH$_3$ target.  Sivers asymmetry was extracted using di-muon events with the invariant mass between 4.3 GeV and 8.5 GeV.  The measured asymmetry, $A_{UT}$, is given in (\ref{def:AUT}). Notice that our definition of $A_{UT}$ from Eq.~\eqref{def:AUT} corresponds to the definition from  Ref.~\cite{Aghasyan:2017jop} $A_{UT}=A_T^{\sin{\phi_S}}$. The data is presented in the one-dimensional binnings over $x_\pi$, $x_N$, $x_F$, $q_T$. In non-$q_T$ binning, the integration over $q_T$ spans up to 5 GeV, i.e., includes the domain with $q_T>Q$. Therefore, only the $q_T$-binned data could be analyzed within TMD factorization. We show a description of the data in Fig.~\ref{fig:compassDY}. One can see that the resulting Sivers function describes well the data on $q_T$-dependence that we use in the fit and predict the data on $x_F$-dependence not used in the fit. 

\begin{figure}[t]
\begin{center}
\includegraphics[width=0.6\textwidth]{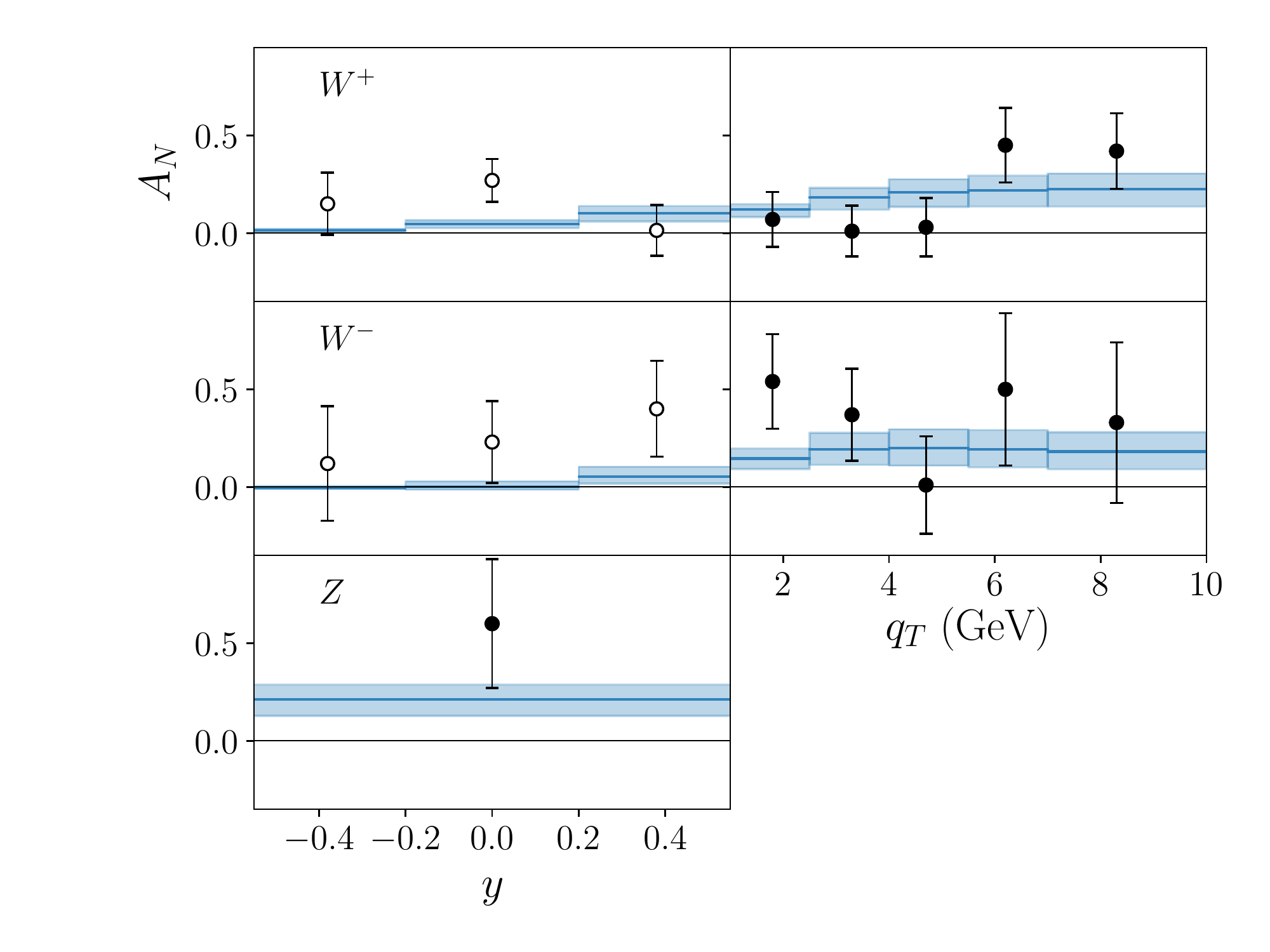}
\end{center}
\vskip -0.5cm
\caption{\label{fig:star15} Description of  the transverse single-spin asymmetry data~\cite{Adamczyk:2015gyk} for $W^\pm$ and $Z$ boson production measured by STAR in polarized proton-proton collisions at $\sqrt{s} = 500$ GeV. Left column, the data as a function of $y$ for $W^\pm$ and $Z$, the right column, the data as a functions of $q_T$ GeV for $W^\pm$. Solid (open) symbols data used (not used) in the fit. Blue line is the CF and the blue box  is 68\%CI of the fit of the data and prediction for the data not used in the fit.}
\end{figure}
\textit{STAR~\cite{Adamczyk:2015gyk} data set}. The STAR Collaboration at RHIC measured the transverse single-spin asymmetry of weak boson ( charged ($W^\pm$) and neutral ($Z/\gamma$)) production in polarized proton-proton collisions at $\sqrt{s}$=500 GeV. It is described by $A_N$ (\ref{def:AN}) with inclusion of modified factors (\ref{th:zz-Z},~\ref{th:zz-W}). The results were presented as a function of rapidity, $y$, and the bosons' transverse momentum, $q_T$. The measured values of asymmetry are much higher (up 60\%) than typical asymmetries in SIDIS, which present a certain problem in their description. We show the description of STAR data~\cite{Adamczyk:2015gyk}   in Fig.~\ref{fig:star15}. One can see that our global analysis gives a good description of $q_T$ dependent data for $W^\pm$ production.   We also describe well $y$ dependent data that is not used directly in the fit for $W^\pm$ and a single point for $Z$-boson production that we use in the fit. It is the first agreement with the data of extraction of the Sivers function with TMD evolution to our best knowledge.  For the DY subset, the main contribution to the $\chi^2/N_{pt}$ is due to a single $Z-$boson production point ($A_N=0.6\pm0.33$) measured at RHIC. Despite the large error, this single point contributes significantly with $\Delta \chi^2=(2.9,1.6,2.8,1.6)$ into fits (SIDIS at NNLO, SIDIS+DY at NNLO, SIDIS at N$^3$LO, SIDIS+DY at N$^3$LO). Let us notice that for $W$ and $Z$ bosons productions, one should also account for contributions of $c$ and $b$ quarks, which are currently neglected.

N$^3$LO fit does not essentially change the result of the fit compared to NNLO. It is expected because the difference between NNLO and N$^3$LO evolution is relatively marginal, see Fig.~\ref{fig:evolution}, especially in comparisons to the large uncertainties of experimental measurements of asymmetries. The values of $\chi^2$ are practically unchanged. As for the values of parameters, we observe that they agree within the error-bands, thus corroborating the stability of evolution effects and the fit results.

\subsection{Sivers function in the position and the momentum spaces}

\begin{figure}[t]
\begin{center}
\includegraphics[width=0.46\textwidth]{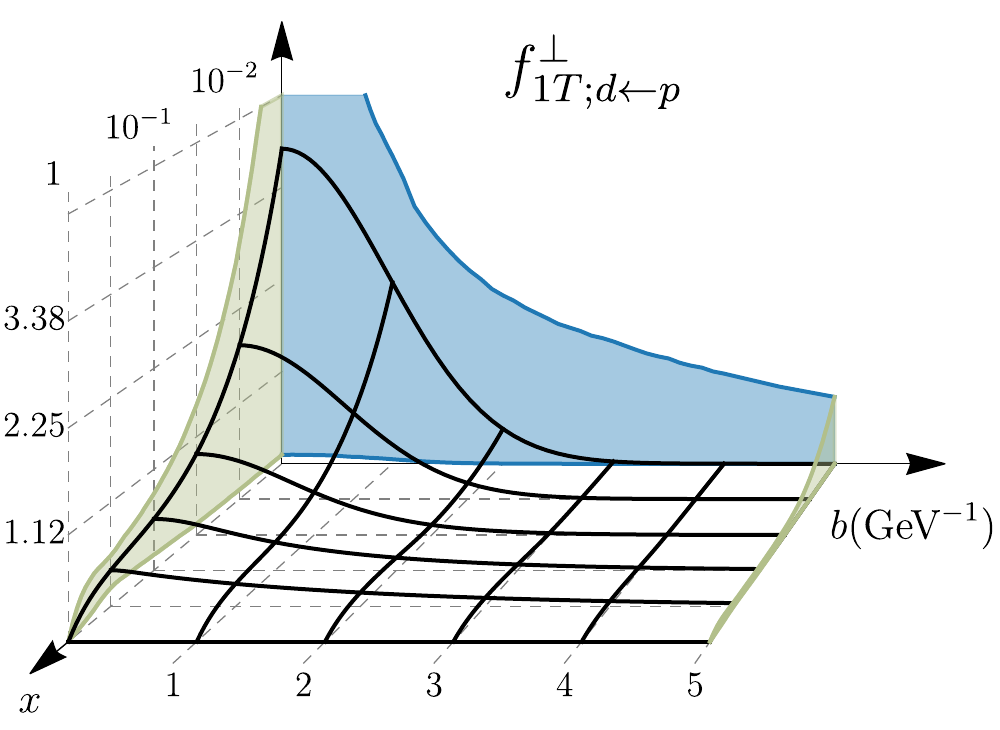}
\includegraphics[width=0.46\textwidth]{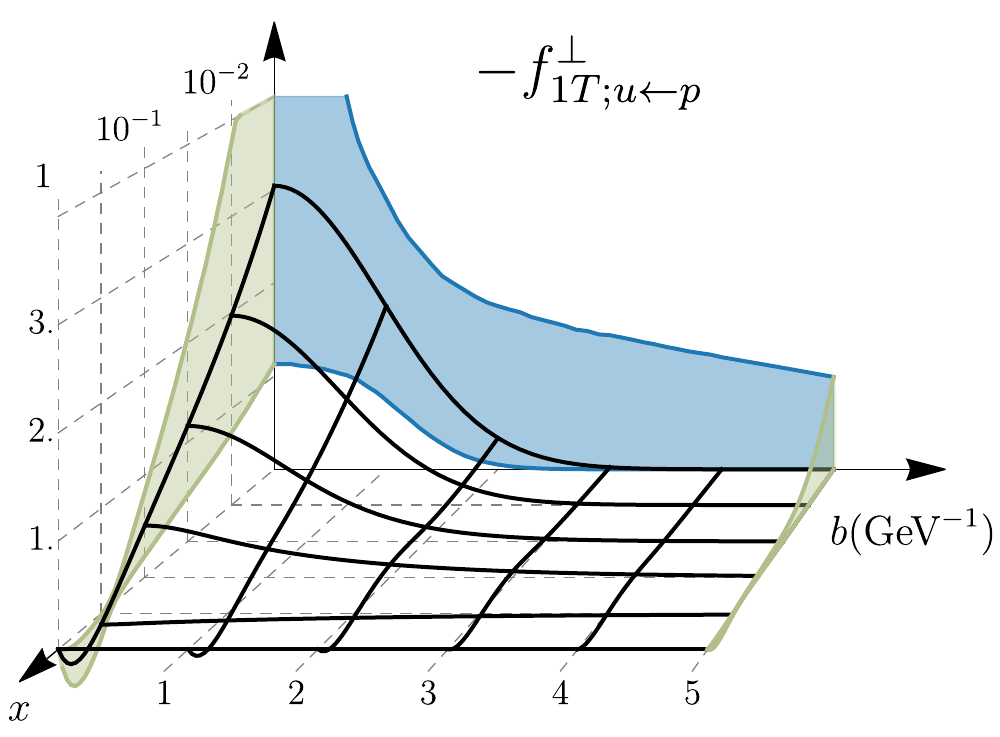}
\end{center}
\caption{\label{fig:sofaPlot} The   $(b,x)$-landscape of the optimal Sivers function $f_{1T}^\perp(x,b)$ for $d$-quark (the left panel) and $u$-quark (the right panel). The grid shows the CF value, whereas the shaded (blue and green) regions on the boundaries demonstrate the 68\%CI.}
\end{figure}

The extracted Sivers function in position space for $u$ and $d$ quarks is shown in Fig.~\ref{fig:sofaPlot}. Its values have notably large uncertainties, which we demonstrate by shaded areas. Another distinctive feature of our extraction is a non-positive definiteness of the Sivers function. The Sivers function does not have the probabilistic interpretation and can have nodes~\cite{Boer:2011fx,Kang:2011hk}, which is realized by the parameter $\epsilon$. Moreover, the presence of a node is predicted by various models~\cite{Lu:2004au,Courtoy:2008vi,Bacchetta:2008af,Boer:2011fx}. The Sivers function for $u$ quark in our extraction, see Fig.~\ref{fig:sofaPlot}, turns positive at large-$x$. However, it can stay negative within 68\%CI. Although such behavior looks unusual, it does not contradict any known properties of the Sivers function. 

\begin{figure}[t]
\begin{center}
\includegraphics[width=0.45\textwidth]{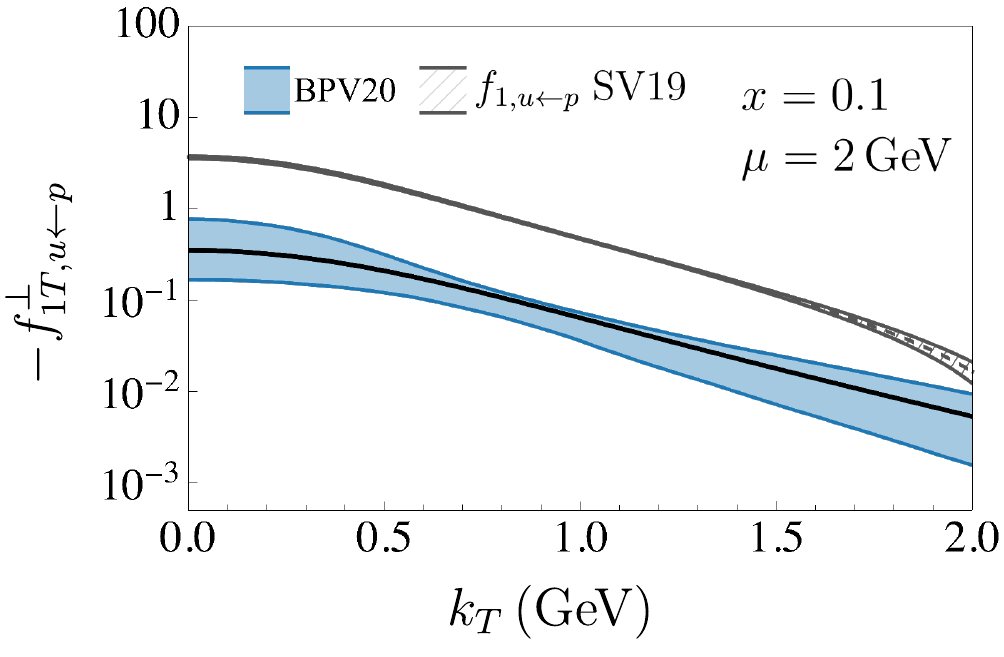}(a)
\includegraphics[width=0.45\textwidth]{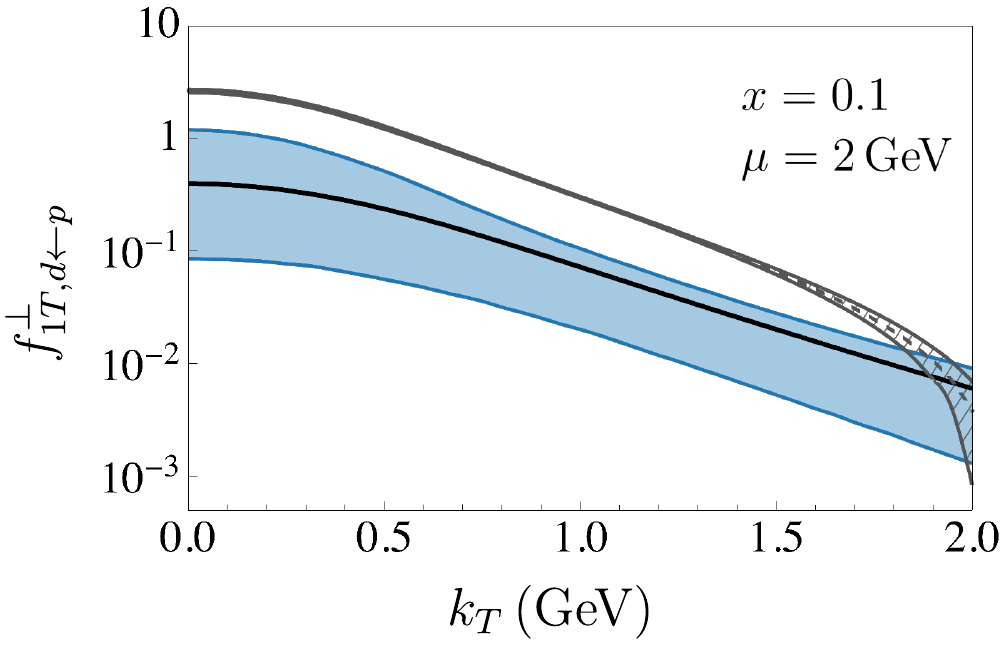}(b)
\includegraphics[width=0.45\textwidth]{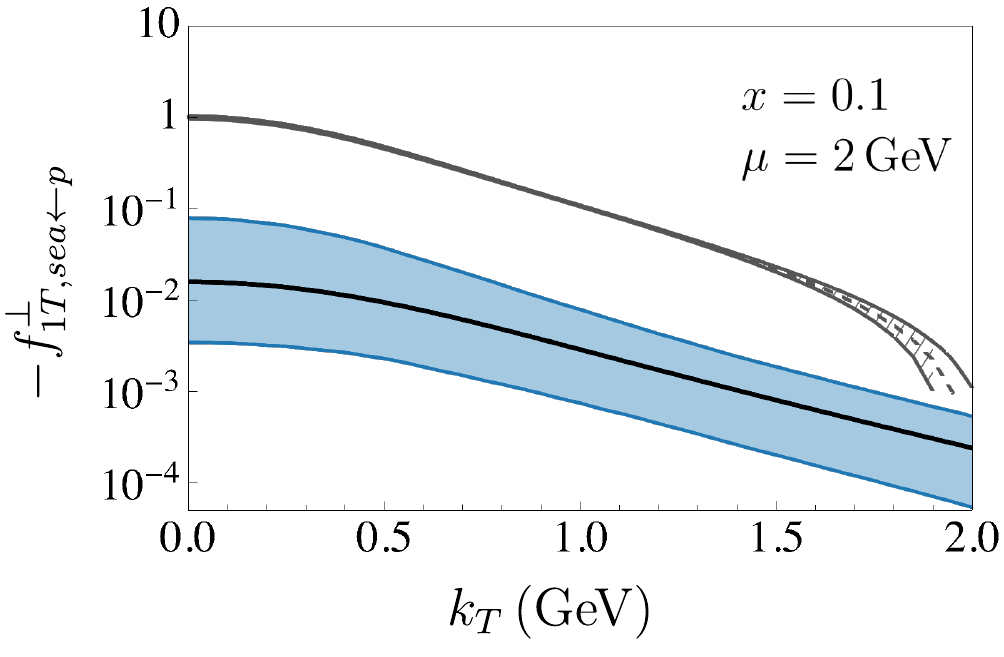}(c)
\includegraphics[width=0.45\textwidth]{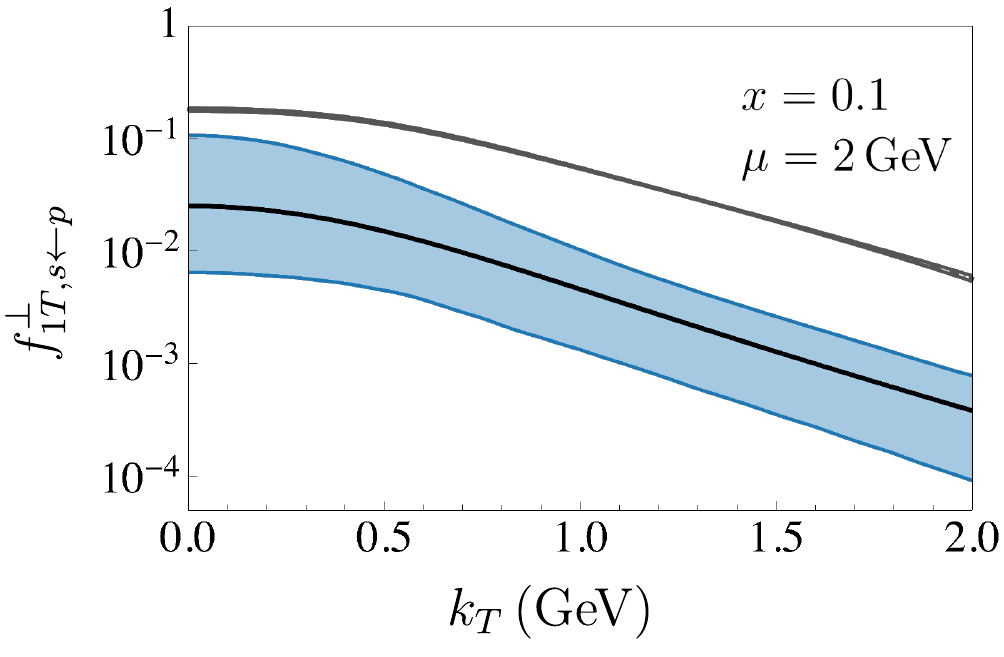}(d)
\end{center}
\caption{\label{fig:kT-profile} Sivers function in the momentum space (black solid line) for $u$, $d$, $sea$, and $s$ quarks at $x=0.1$ and $\mu=2$ GeV. The blue band is the 68\%CI. The gray dashed line is the unpolarized TMD PDF extracted in SV19 shown for the comparison (for $u$ and $sea$-quark the Sivers function is multiplied by $-1$ and $sea$-quark the Sivers function is compared to $\bar{u}$ unpolarized TMD PDF).}
\end{figure}

In the momentum representation the TMD distributions for unpolarized quarks are defined as\footnote{Notice that we do not distinguish the symbols for the Sivers functions in the position and the momentum spaces, they are related by the Fourier transform of  Eq.~(\ref{eq:ft}). It is intended by the functional arguments, $b$ or $k_T$, which function we use.} 
\begin{eqnarray}\label{eq:momspace}
\int \frac{d^2b}{(2\pi)^2}e^{i(bk_T)} \Phi^{[\gamma^+]}_{q\ot h}(x,b;\mu,\zeta)=
 f_{1;q\ot h}(x,k_T;\mu,\zeta)-\frac{\epsilon_T^{\mu\nu}k_{T\mu} S_{T\nu}}{M} f_{1T;q\ot h}^\perp(x,k_T;\mu,\zeta),
\end{eqnarray}
where $k_T$ is the two-component Euclidean vector of traverse momentum, and $\Phi^{[\gamma^+]}_{q\ot h}$ is given by the left-hand-side of Eqn.~(\ref{def:TMDPDF}). Performing the angular integration in Eq.~(\ref{eq:momspace}) we find
\begin{eqnarray}
f_{1;q\ot h}(x,k_T;\mu,\zeta)&=&\int_0^\infty \frac{b db}{2\pi}  J_0(b|k_T|)f_{1;q\ot h}(x,b;\mu,\zeta),
\\
f_{1T;q\ot h}^\perp(x,k_T;\mu,\zeta)&=&M^2\int_0^\infty \frac{b db}{2\pi} \frac{b}{|k_T|} J_1(b|k_T|)f^\perp_{1T;q\ot h}(x,b;\mu,\zeta).\label{eq:ft}
\end{eqnarray}
The momentum space representation has complicated evolution properties since the TMD evolution factor is multiplicative in the position space. The notion of the optimal TMD distribution is less useful in the momentum space because it involves the integration over all scales. For that reason, we only show the  TMD distributions in the momentum space at a fixed scale.

The extracted Sivers function is shown in Fig.~\ref{fig:kT-profile}. The Fourier transformation,  Eq.~(\ref{eq:ft}), effectively inverses the ranges of variables. Therefore, a large uncertainty at large-$b$ (given by parameters $r_{0,1,2}$) transforms to a large uncertainty at small-$k_T$. For comparison, we also show the values and uncertainties of the unpolarized TMD PDFs extracted in SV19 fit. We observe that the Sivers function's typical size is about 4-5 times as small as the corresponding unpolarized distribution. Figure~\ref{fig:kT-profile} shows the functions at $x=0.1$, for other values of $x$ of the data used in our fit $x\sim 0.01 - 0.25$ profiles are similar.

\begin{figure}[htb]
\begin{center}
\includegraphics[width=0.5\textwidth]{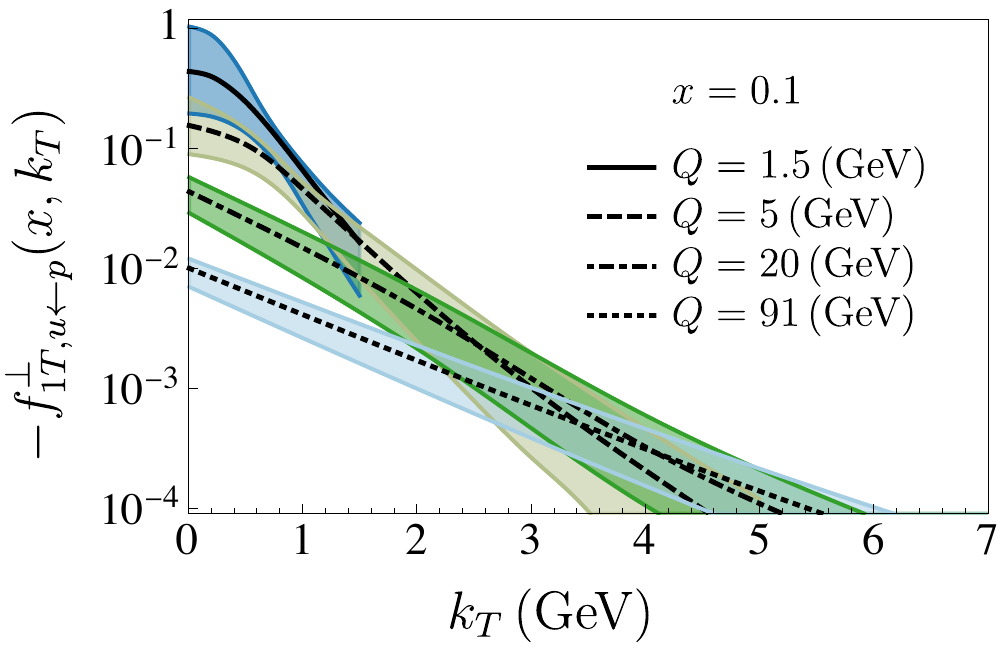}
\end{center}
\caption{\label{fig:kT-profile-Q} Sivers function in the momentum space  for $u$ quark at $x=0.1$  as a function of $k_T$ (GeV). The bands are the 68\%CI. The calculations are performed at four different values of $Q$.}
\end{figure}

In Fig.~\ref{fig:kT-profile-Q}, we demonstrate the impact of QCD evolution in the momentum space. We show $u$ quark Sivers function calculated by Eq.~\eqref{eq:ft} at four different scales $Q=1.5$, 5, 20, 91 GeV. As one can see, the evolution modifies the shape and the amplitude of the Sivers function.

\subsection{Positivity constraints for the Sivers function}

In Ref.~\cite{Bacchetta:1999kz} the positivity constraints for TMD distributions were derived assuming the positive-definiteness of the polarization matrix due to its probabilistic interpretation in the parton model. In particular, the positivity constraint involving the Sivers function is
\begin{eqnarray}\label{def:positivity}
\frac{k_T^2}{M^2}\left(g_{1T}(x,k_T)^2+f_{1T}^\perp(x,k_T)^2\right)\leqslant f_1(x,k_T)^2,
\end{eqnarray}
where $g_{1T}$ is the worm-gear T or Kotzinian-Mulders~\cite{Tangerman:1994eh,Kotzinian:1995cz} function. Generally, such positivity constraints are not respected in the quantum field theory due to renormalization effects, which are only enhanced in the TMD case by renormalizing rapidity divergences. Recall in particular that even cross-sections become negative in the region outside of the TMD factorization validity. In some cases the violation of positivity constraints is very significant, e.g., for linearly polarized gluon TMD PDF discussed in Ref.~\cite{Gutierrez-Reyes:2019rug}. As far as our analysis includes the TMD evolution, we expect that the positivity constraint is not applicable, given that it is based on the tree order approximation argument. Nonetheless, it is instructive to check the constraint from Eq.~(\ref{def:positivity}).

\begin{figure}[htb]
\centering
\begin{tabular}{ccccc}
\rotatebox[origin=c]{90}{{$k_T$ (GeV)}} &  
\hspace{-2.5mm}\raisebox{-.5\height}{\includegraphics[width=3.95cm]{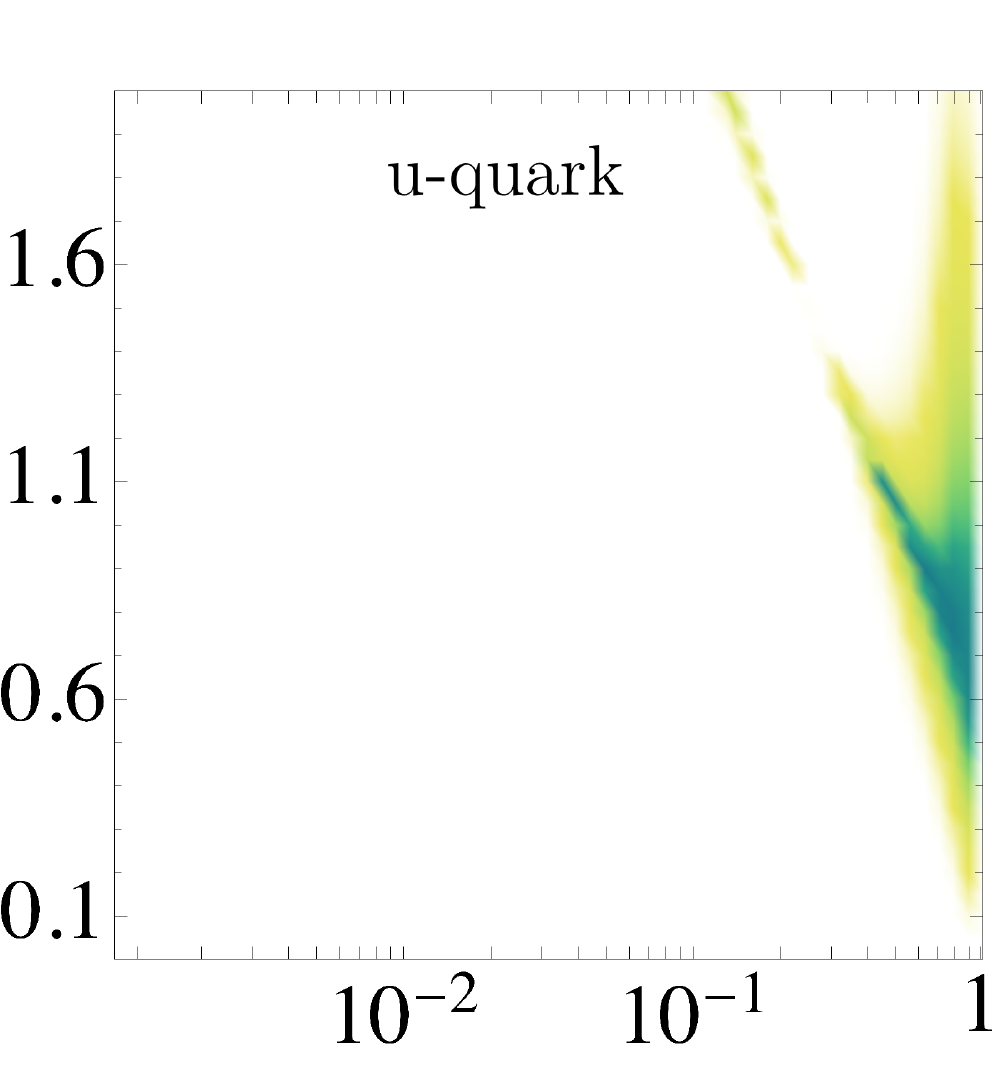}} &  
\hspace{-4.5mm}\raisebox{-.5\height}{\includegraphics[width=3.5cm]{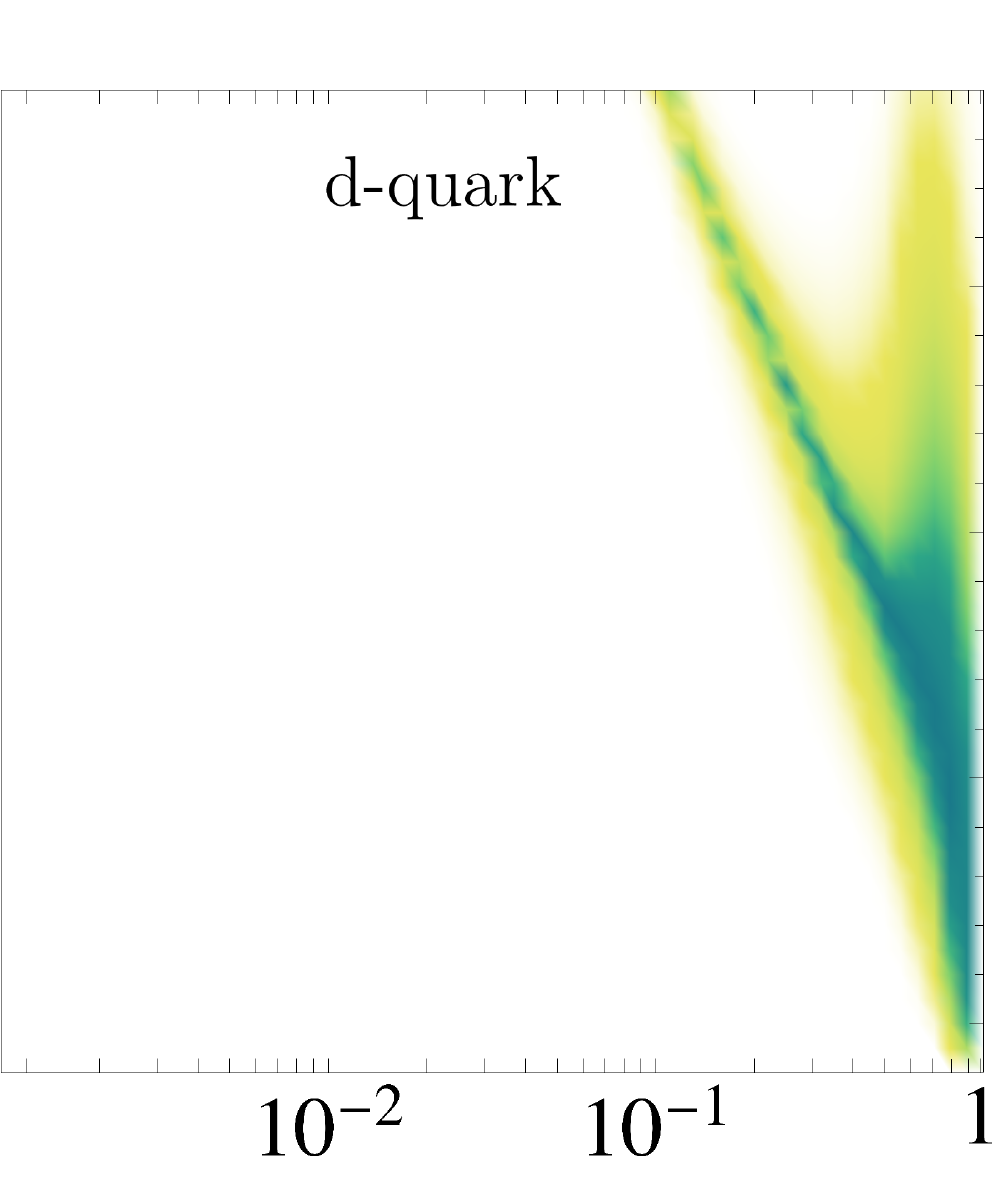}} & 
\hspace{-4.5mm}\raisebox{-.5\height}{\includegraphics[width=3.5cm]{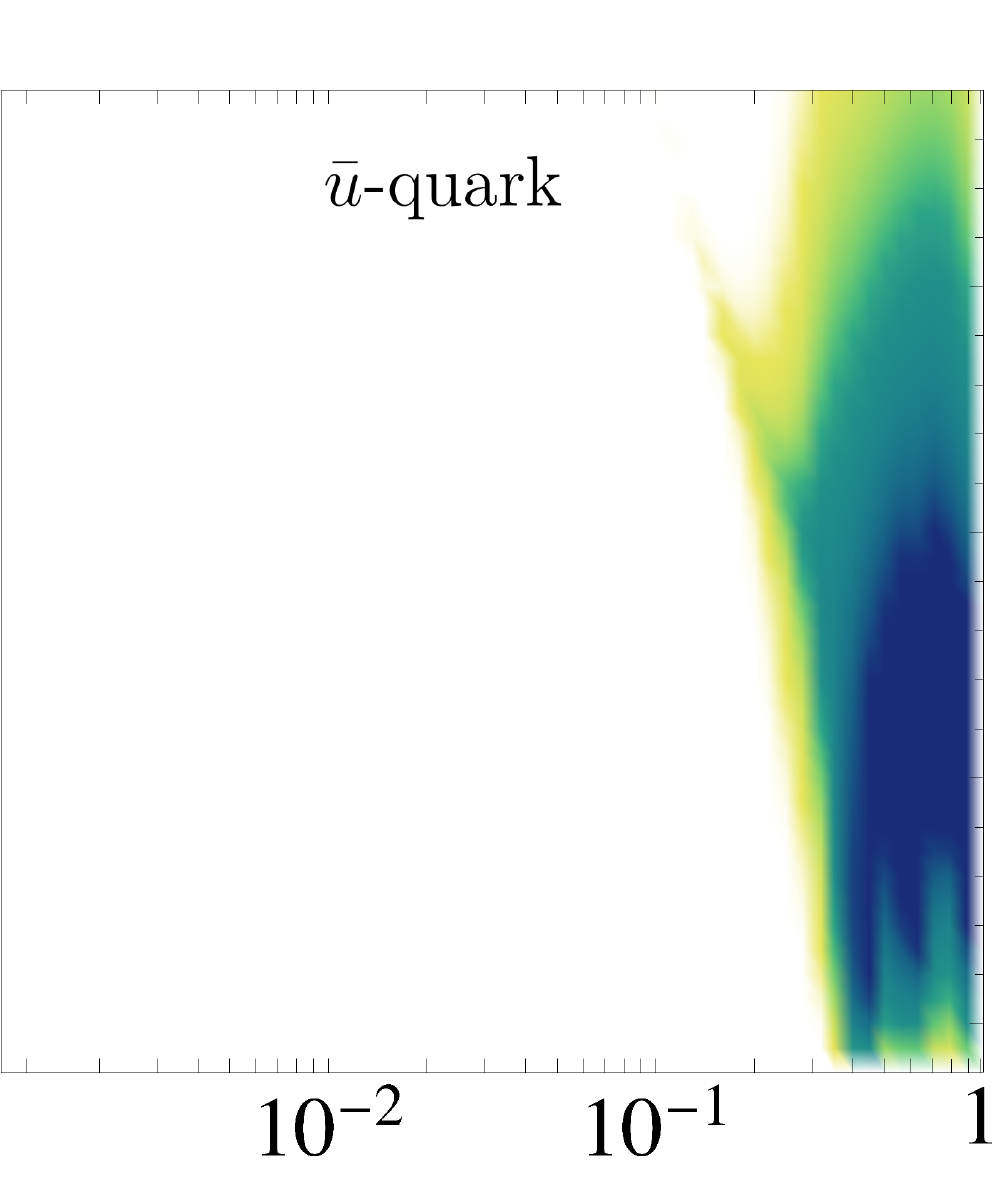}} & 
\hspace{-4.5mm}\raisebox{-.5\height}{\includegraphics[width=3.5cm]{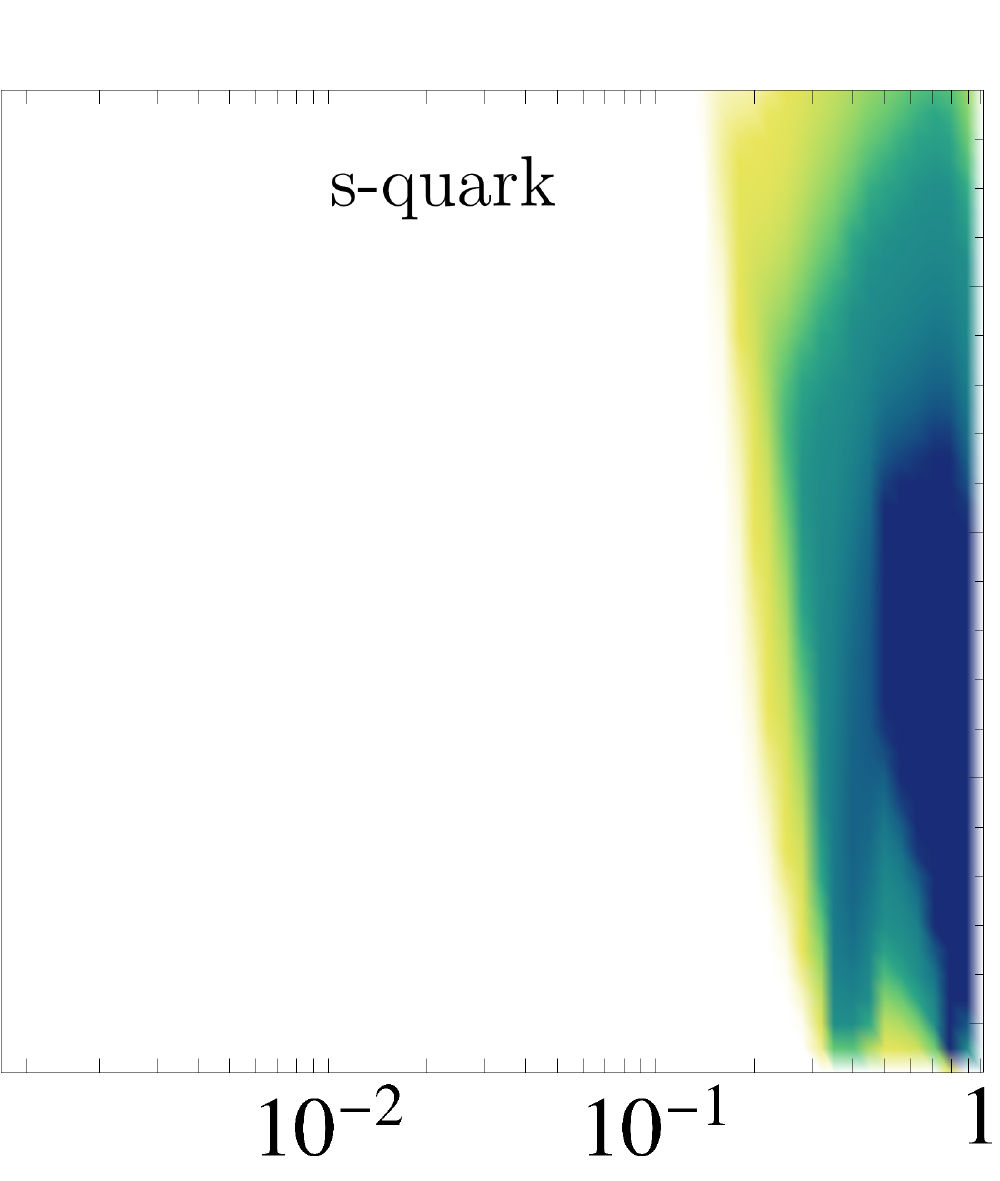}} \\
		& $x$ & $x$ & $x$   & $x$\\	
\end{tabular}
\vspace{-3mm}
\caption{\label{fig:positivity} The function $pos(x,k_T,\mu)$ defined in Eq.~(\ref{def:pos}) at $\mu=2$ (GeV) for $u$ quark, $d$ quark, $\bar{u}$ quark, $s$ quark. The positivity constraint (\ref{def:positivity}) is violated in the yellow-to-blue shaded region.}
\end{figure}

In Fig.~\ref{fig:positivity} we plot the function
\begin{eqnarray}\label{def:pos}
pos(x,k_T,\mu)=1-\frac{k_T^2}{M^2}\left(\frac{f_{1T}^\perp(x,k_T;\mu,\mu^2)}{f_1(x,k_T;\mu,\mu^2)}\right)^2,
\end{eqnarray}
as the function of $x$ and $k_T$ at $\mu=2$ GeV. One has $pos>0$ ($pos<0$) for the regions where Eq.~(\ref{def:positivity}) is (not) satisfied in the absence of $g_{1T}$ contribution. For the values of the Sivers function we take the largest boundary of 68\%CI. We observe that the positivity constraint is satisfied everywhere except for the unmeasured large-$x$ region. If we consider the lowest boundary of 68\%CI the region $pos>0$ is much larger, in particular, $u$ quark satisfies Eq.~(\ref{def:positivity}) in the full range of $(x,k_T)$. Also the picture depends on the scale, and improves (in the sense that the the region $pos>0$ becomes wider) for larger scales. We conclude that our extraction does not contradict the positivity constraint in the regions reached by the experimental data used in this analysis.

\subsection{3D tomography of the nucleon and the Sivers function}
\begin{figure}[htb]
\begin{center}
\includegraphics[width=0.45\textwidth]{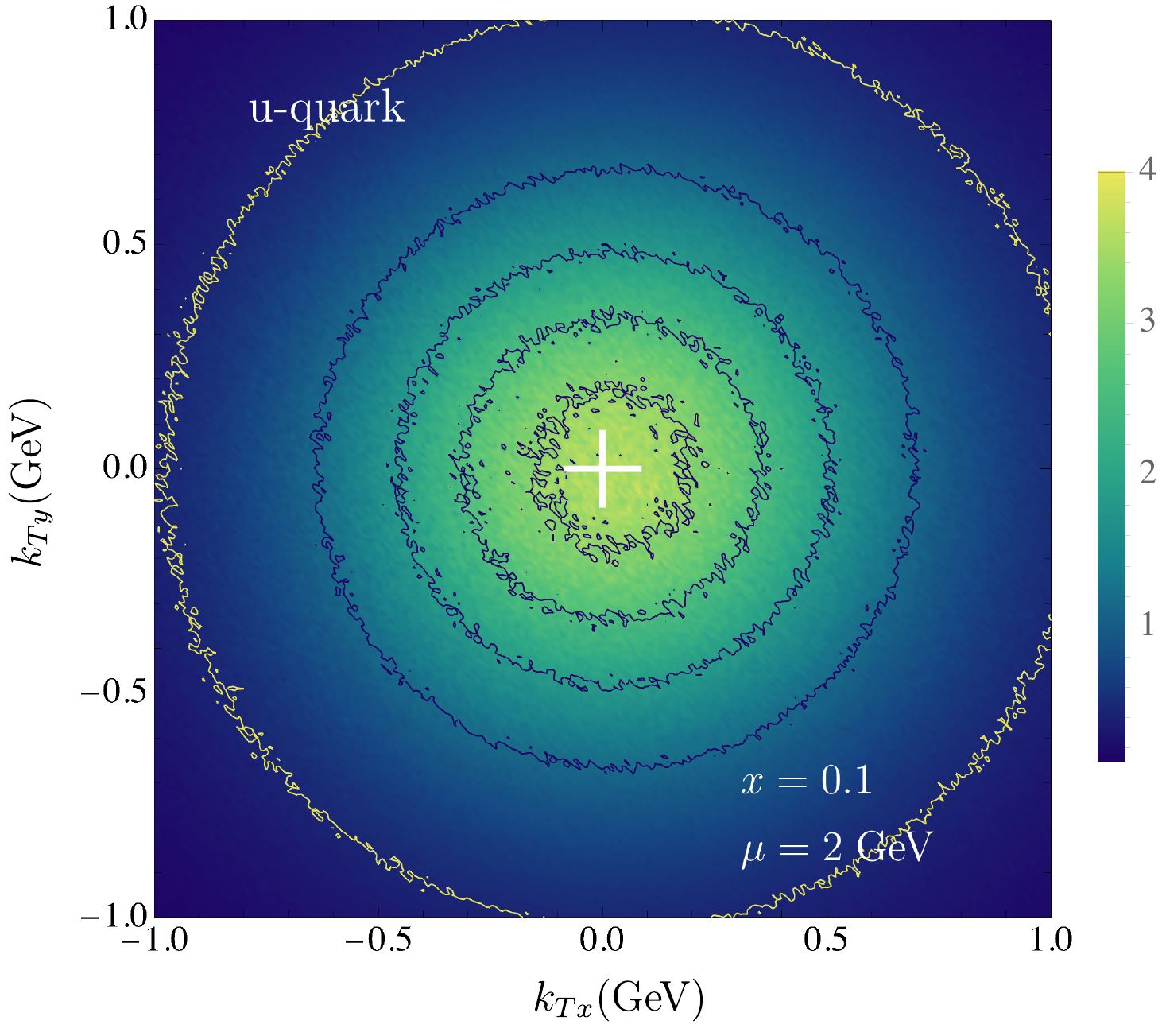}(a)
\includegraphics[width=0.45\textwidth]{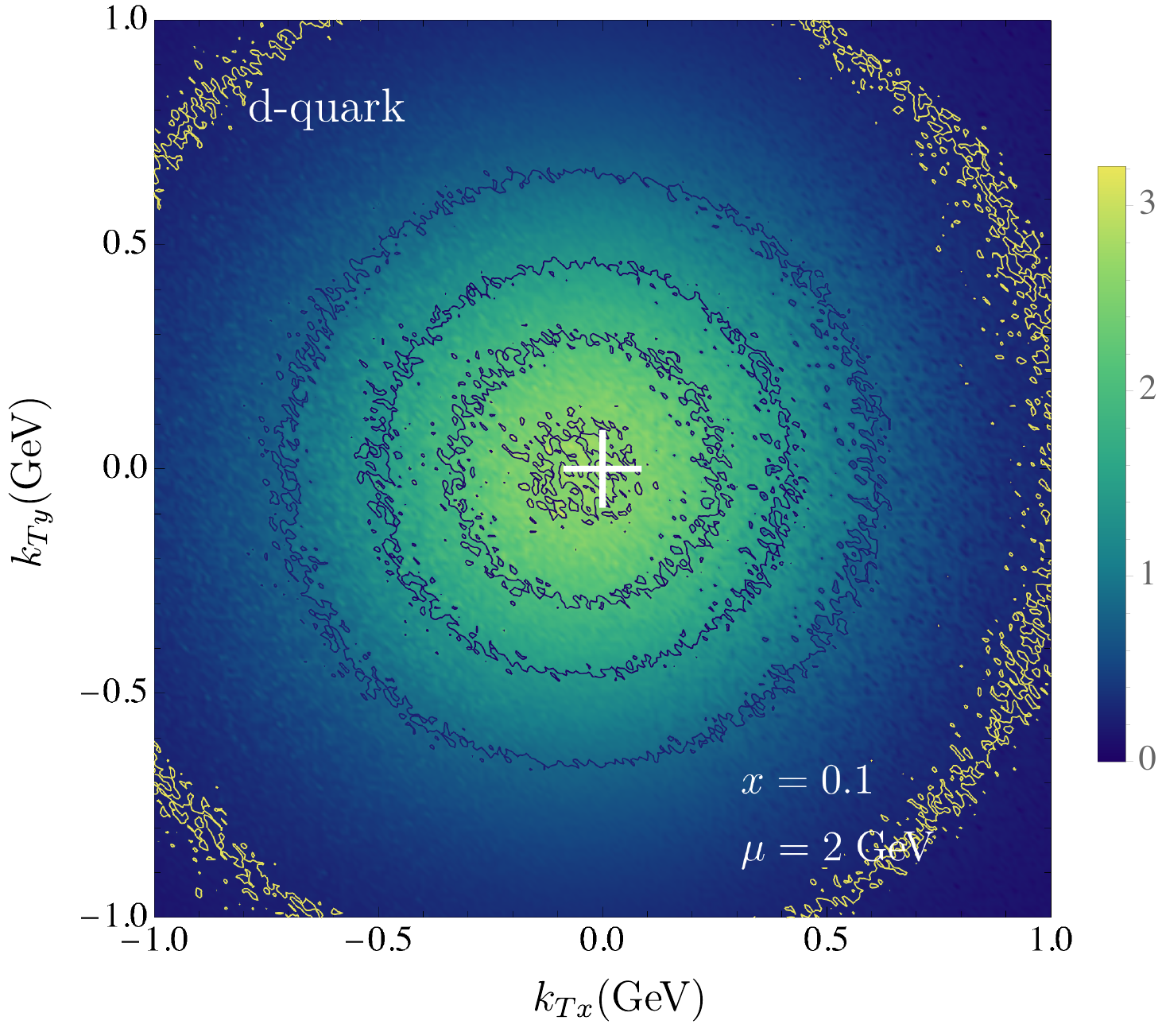}(b)
\includegraphics[width=0.465\textwidth]{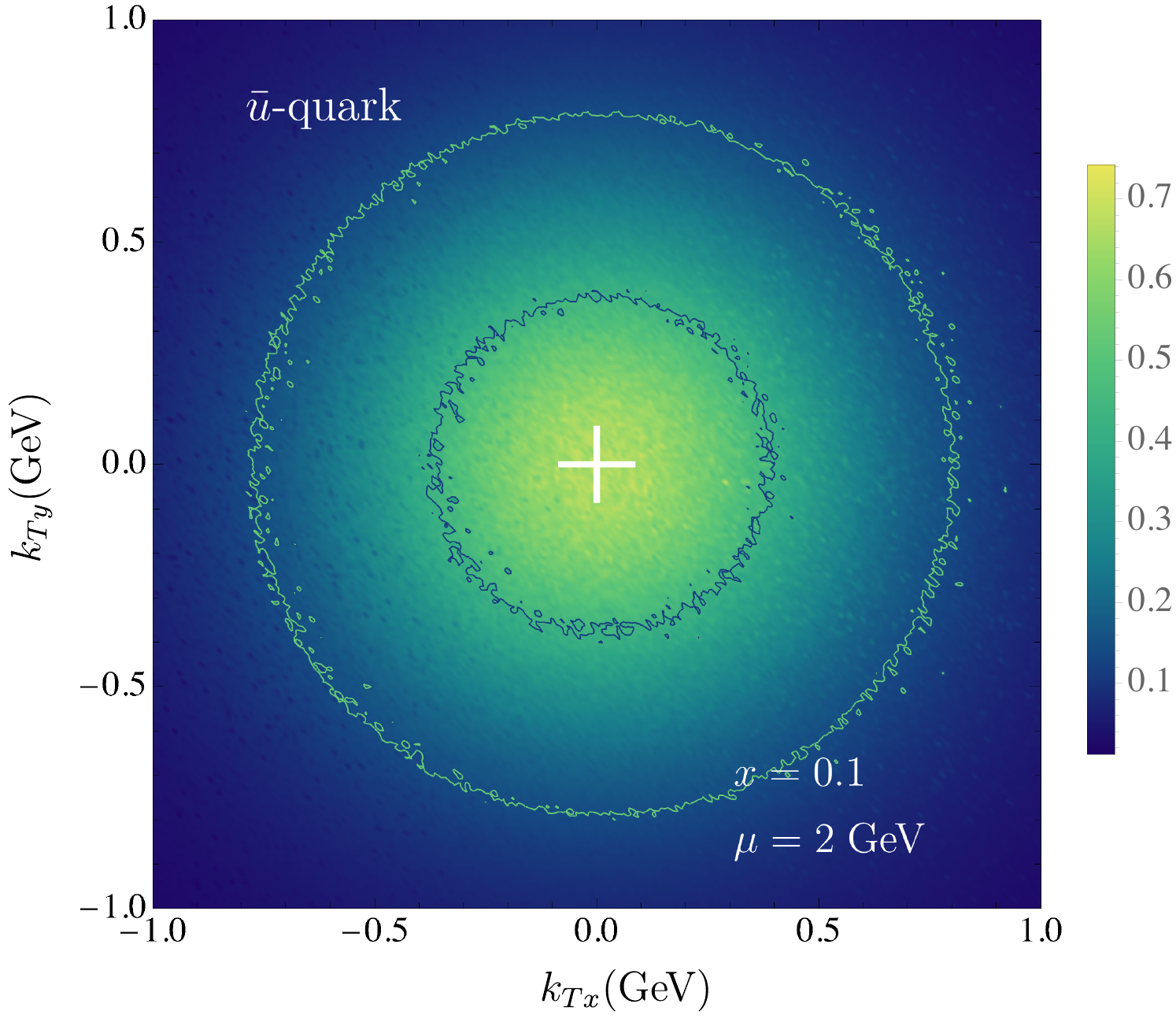}(c)
\includegraphics[width=0.465\textwidth]{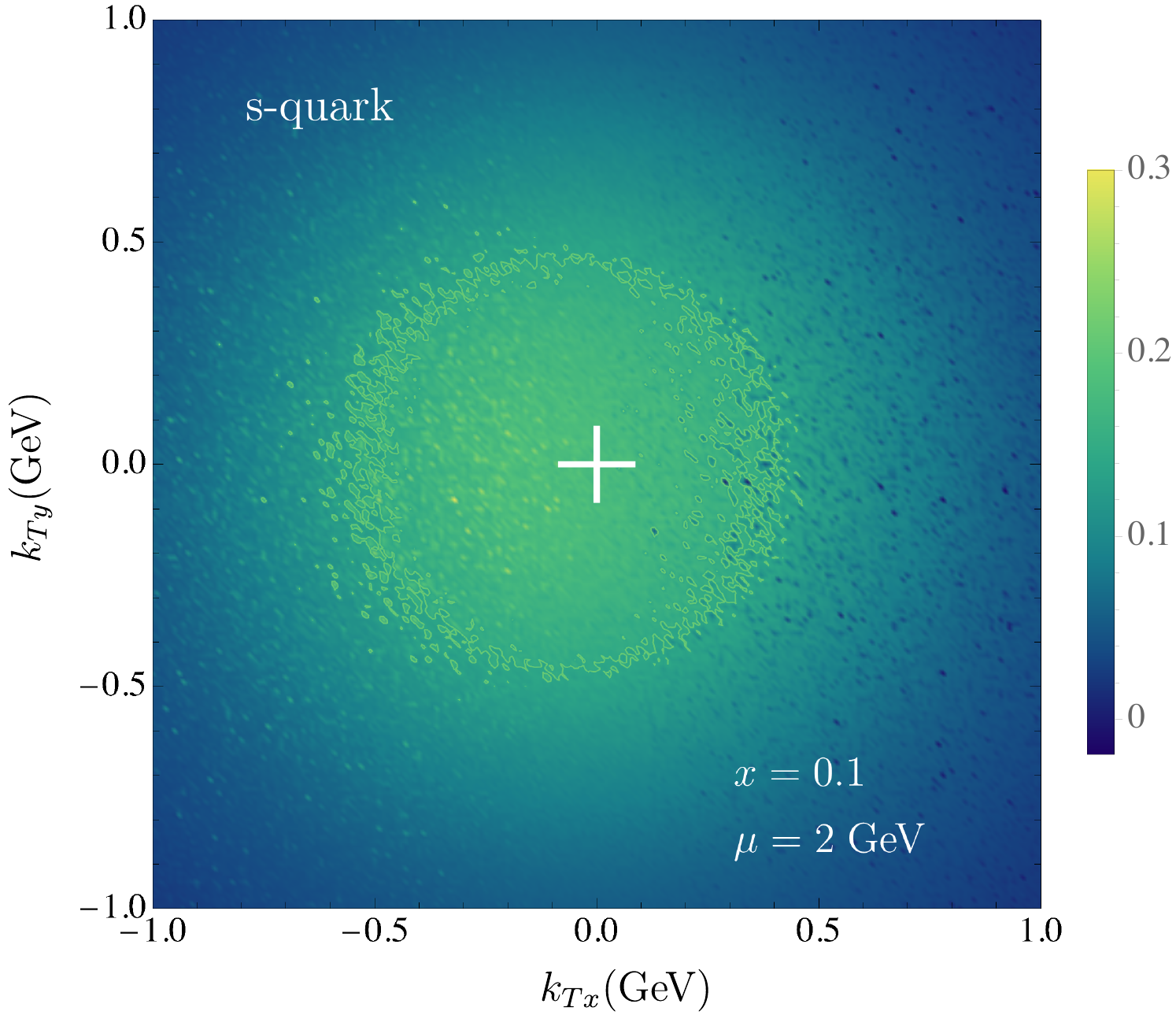}(d)
\end{center}
\caption{\label{fig:tomography} Tomographic scan of the nucleon via the momentum space quark density function $\rho_{1;q\ot h^\uparrow}(x,\vec k_T,\vec S_T,\mu)$ defined in Eq.~(\ref{eq:tomography}) at $x=0.1$ and $\mu=2$ GeV.   Panel (a) is for $u$ quarks, panel (b) is for $d$ quark, panel (c) is for $\bar u$ quark, and panel (d) is for $s$ quark. The variation of color in the plot is due to variation of replicas and  illustrates the uncertainty of the extraction. The nucleon polarization vector is along $\hat y$-direction.  White cross indicates the position of the origin $(0,0)$ in order to highlight the shift of the distributions along $\hat x$-direction due to the Sivers function.}
\end{figure}
The magnitude of the Sivers function extracted in our fit is generally much smaller than the unpolarized TMD PDF. To present the distortion effect on the unpolarized quarks driven by the hadron polarization, we introduce the momentum space quark density function
\begin{eqnarray}
\rho_{1;q\ot h^\uparrow}(x,\vec k_T,\vec S_T, \mu)=f_{1;q\ot h}(x,k_T;\mu,\mu^2)-\frac{k_{Tx}}{M} f_{1T;q\ot h}^\perp(x,k_T;\mu,\mu^2),
\label{eq:tomography}
\end{eqnarray}
where $\vec k_T$ is a two-dimensional vector $(k_{Tx},k_{Ty})$. This function reflects the TMD density of unpolarized quark $q$ in the spin-$1/2$ hadron totally polarized in $\hat y$-direction, $\vec S_T = (S_x, S_y)$, where $S_x =0$, $S_y =1$, compare to Eq.~(\ref{eq:momspace}). In Fig.~\ref{fig:tomography} we plot $\rho$ at $x=0.1$ and $\mu=2$ GeV. To present the uncertainty in unpolarized and Sivers function, we randomly select one replica for each point of a figure. Thus, the color fluctuation roughly reflects the uncertainty band of our extraction. The presented pictures have a shift of the maximum in $k_{Tx}$, which is the influence of Sivers function that introduces a dipole modulation of the momentum space quark densities. This shift corresponds to the correlation of the Orbital Angular Momentum (OAM) of quarks and the nucleon's spin. One can see from Fig.~\ref{fig:tomography} that $u$ quark has a negative correlation and $d$ quark has a positive correlation. Without OAM of quarks, such a correlation and the Sivers function are zero, and thus we can observe in Fig.~\ref{fig:tomography} the evidence of the presence of OAM of $u$ and $d$ quarks in the wave function of the nucleon.

\begin{figure}[htb]
\begin{center}
\includegraphics[width=0.45\textwidth]{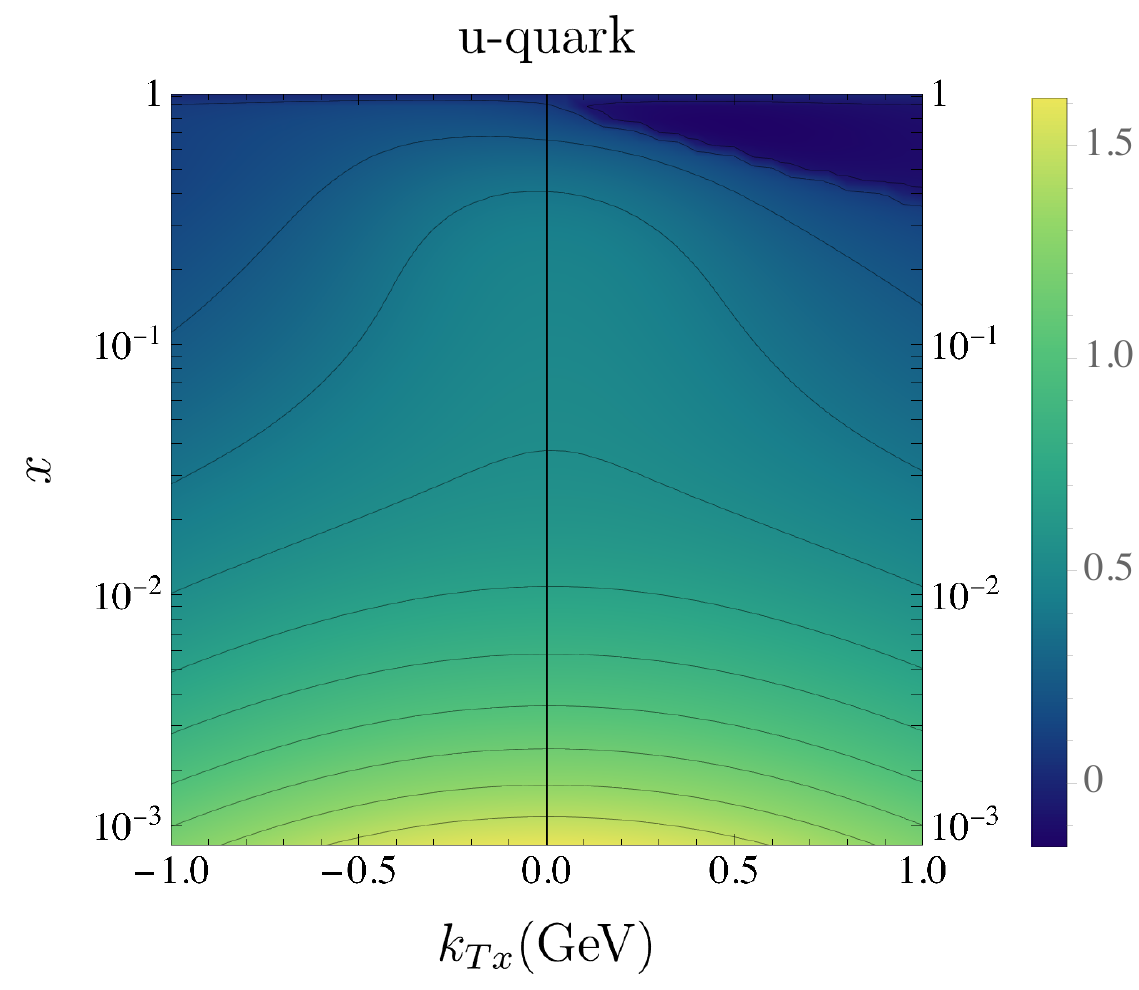}(a)
\includegraphics[width=0.45\textwidth]{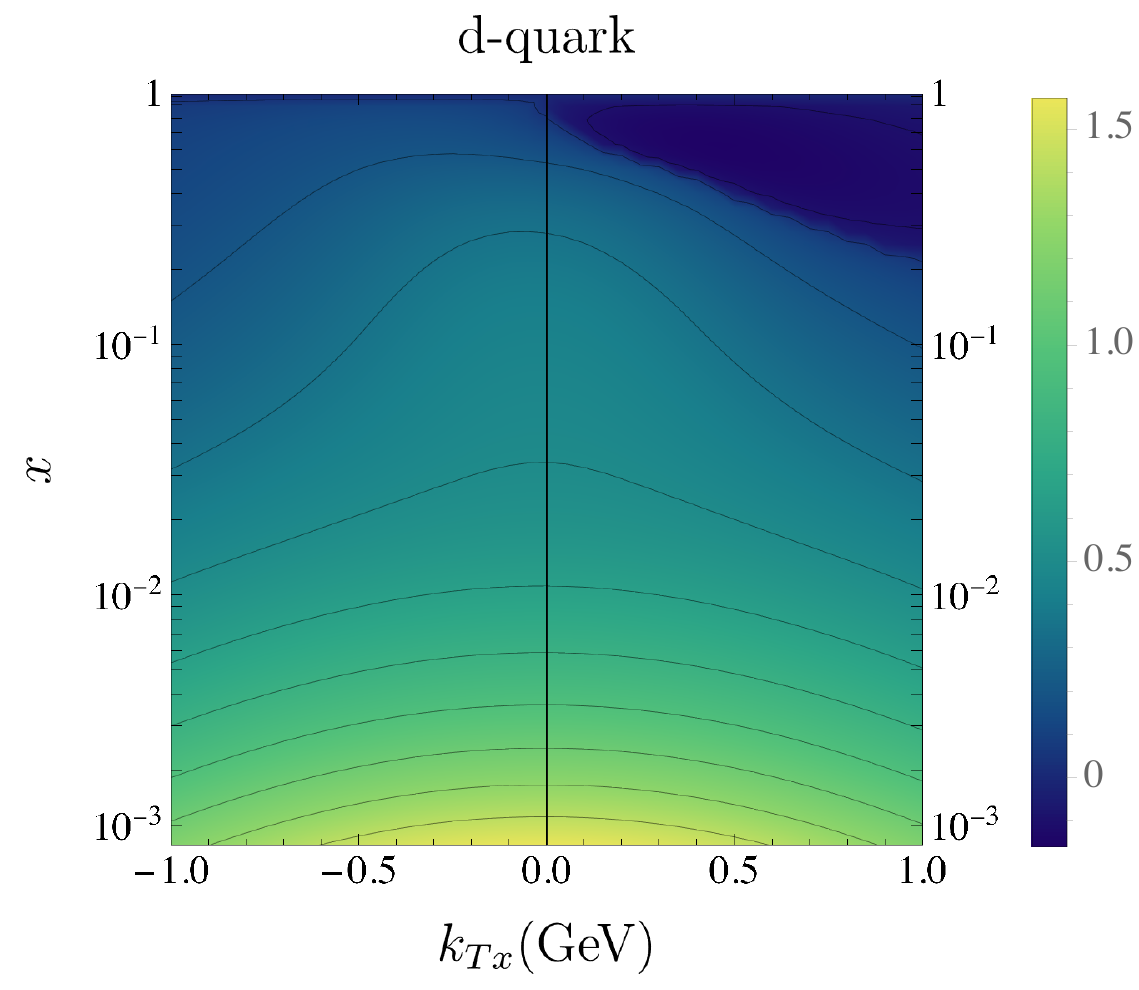}(b)
\includegraphics[width=0.465\textwidth]{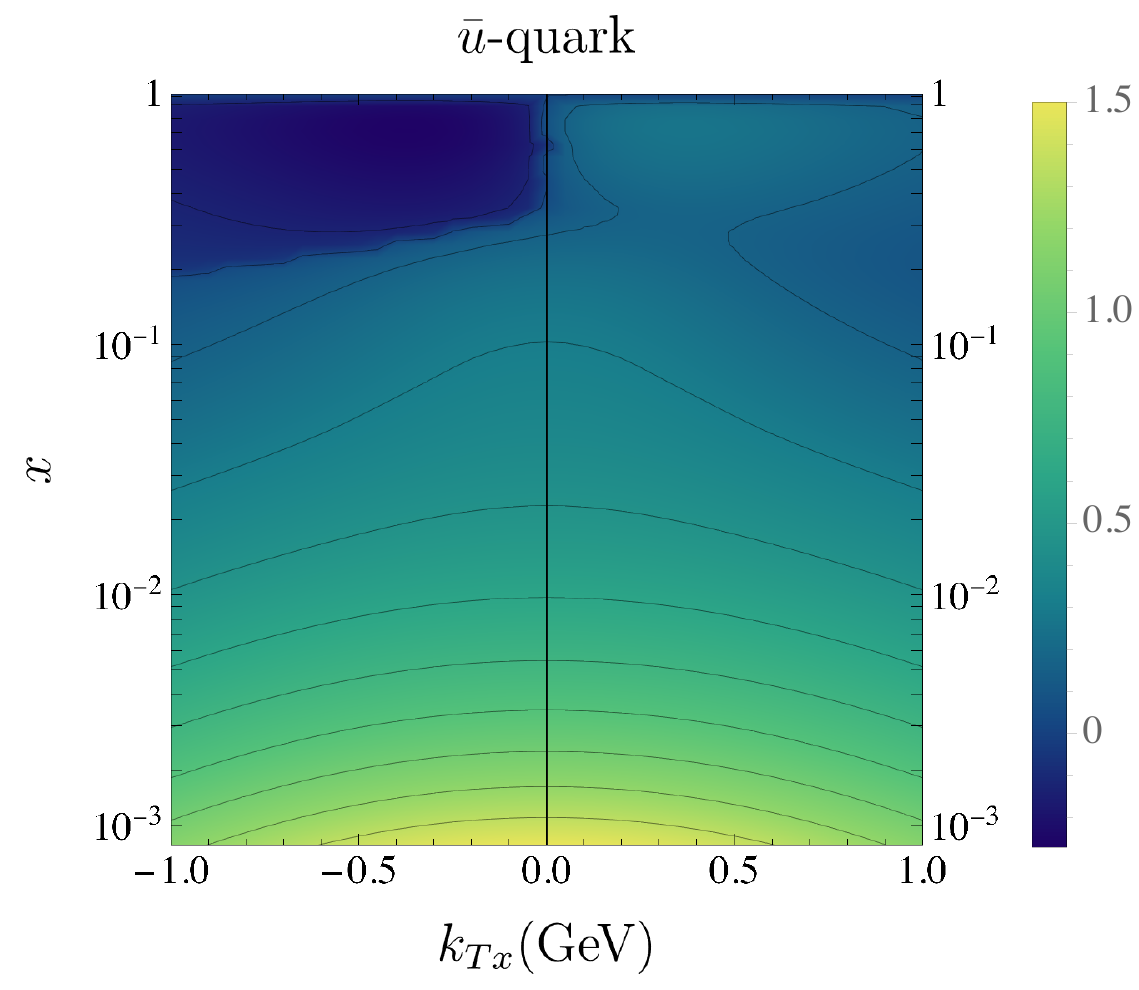}(c)
\includegraphics[width=0.465\textwidth]{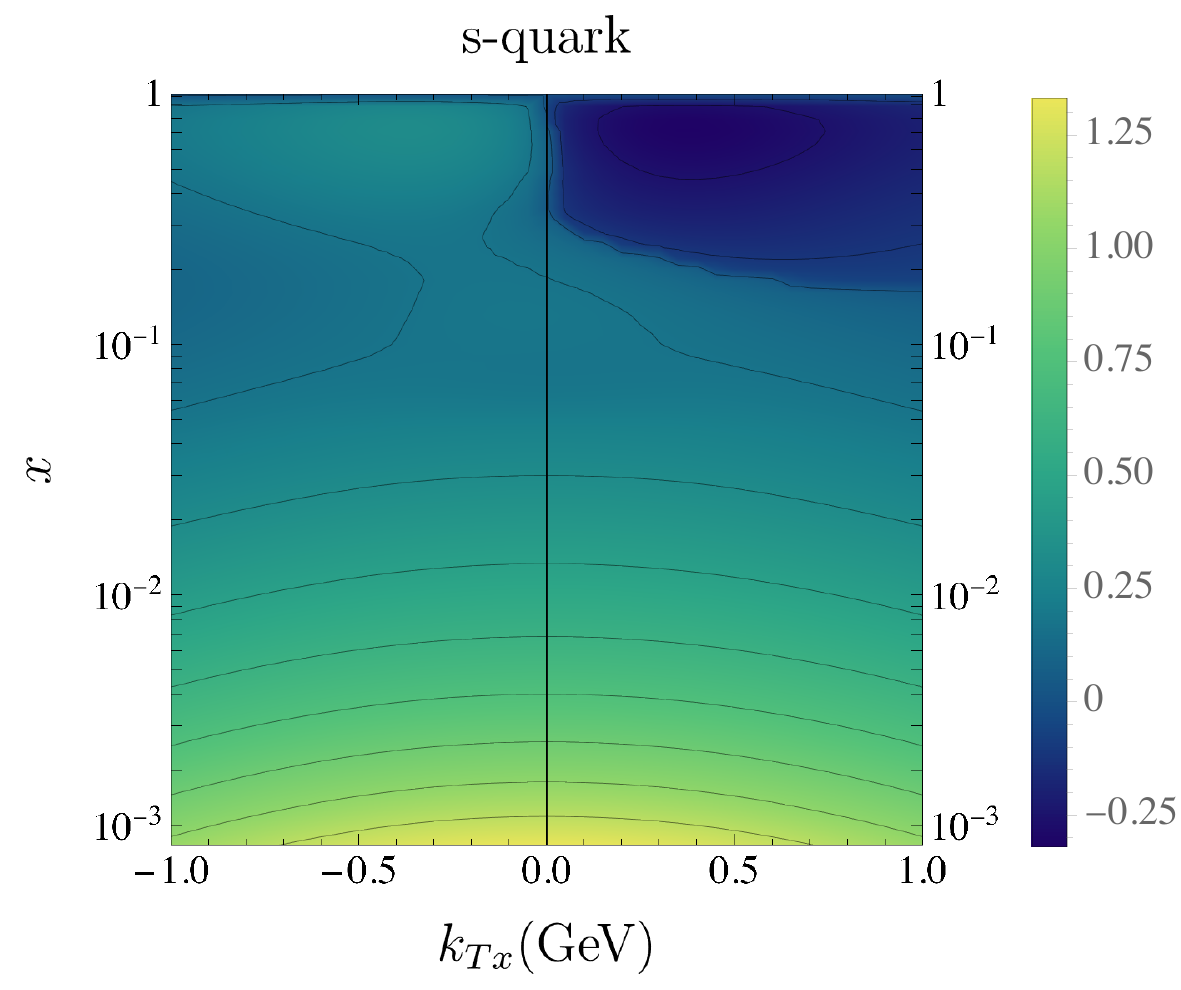}(d)
\end{center}
\caption{\label{fig:tomography1} Tomographic scan of the nucleon in $(x,k_T)$ via the momentum space quark density function $\rho_{1;q\ot h^\uparrow}(x,\vec k_T,\vec S_T,\mu)$ defined in Eq.~(\ref{eq:tomography}) at $\mu=2$ GeV.   Panel (a) is for $u$-quarks, panel (b) is for $d$-quark, panel (c) is for $\bar u$-quark, and panel (d) is for $s$-quark. The color scheme is defined as explained in the text. }
\end{figure}

Let us also discuss the tomographic scan of the nucleon both in $x$ and $k_T$. We plot in Fig.~\ref{fig:tomography1} the momentum space quark density function $\rho_{1;q\ot h^\uparrow}(x,\vec k_T,\vec S_T,\mu)$ from Eq.~(\ref{eq:tomography}) as function of both $x$ and $k_{Tx}$ in order to assess the region in which the Sivers effect has the most influence. The color scheme is chosen to be proportional to the function elevated to power $1/3$ in order not to underestimate the region where the function is not big. The asymmetry in color and contours between negative and positive $k_{Tx}$ indicates the asymmetry of the distribution and the important influence of the Sivers function. From Fig.~\ref{fig:tomography1} one can see that the existing data indicate that most of the correlation between the spin and the motion of the partons happens in the region of large to moderate $x$. In the low-$x$ region, the momentum space quark density becomes almost symmetric, and it indicates that the Sivers effect becomes smaller and corresponding experimentally observed asymmetry is small. Of course, one has to consider that there is no experimental data in the low-$x$ region available yet, so our findings must be corroborated by the future Electron-Ion Collider data. At the same time, it is crucial to explore the large-$x$ region where the effect is the largest, and the future Jefferson Lab 12 GeV data will be important for the exploration of this region.

\subsection{Determination of the Qiu-Sterman function}
\label{sec:QS}

At small-$b$ the  Sivers function $f_{1T}^\perp(x,b)$ can be expressed via the operator product expansion (OPE) through the collinear twist-3 distributions  \cite{Aybat:2011ge,Kang:2011mr,Scimemi:2019gge,Sun:2013hua,Dai:2014ala}. In our determination we do not use this relation as twist-3 functions are largely unknown. Instead, we will use the opposite strategy and determine the collinear twist-3 component from the extracted Sivers function. Such a determination has a limited power, and allows to extract only Qiu-Sterman (QS) function with certain systematic uncertainty. Nonetheless, such an extraction is meaningful, especially because the information on twist-3 distributions is very limited. Moreover, the extraction of QS function given here is much less theoretically biased in comparison to other extractions, such as those made in in Refs.~\cite{Echevarria:2020hpy,Echevarria:2014xaa}, where QS function is parametrized via twist-2 distributions and expected to have DGLAP-type evolution equation. 

The complete expression for matching of the Sivers function to collinear twist-3 distributions at NLO was derived in Ref.~\cite{Scimemi:2019gge}. In the $\zeta$-prescription, this expression reads
\begin{eqnarray}\label{th:f->QS}
&&f_{1T,q\ot h}^\perp(x,b)=-\pi\Bigg\{T_{q}(-x,0,x;\mu)+a_s(\mu)\Big[-2 \mathbf{L}_\mu P\otimes T-C_F\frac{\pi^2}{6}T(-x,0,x;\mu)+
\\\nn &&\qquad
\int_{-1}^1 d\xi \int_0^1 dy \delta(x-y\xi)\left(-\frac{\bar y}{N_c}T_q(-\xi,0,\xi;\mu)+\frac{3y\bar y}{2\xi}G^{(+)}(-\xi,0,\xi;\mu)\right)+\mathcal{O}(a_s^2)\Big]+\mathcal{O}(b^2)\Bigg\},
\end{eqnarray}
where $\bar y=1-y$, $N_c=3$ is the number of colors, $C_F=(N_c^2-1)/(2N_c)=4/3$,  $a_s=g^2/(4\pi)^2$ is the strong coupling constant, and $\mathbf{L}_\mu=\ln(\mu^2 b^2 e^{2\gamma_E}/4)$. The function $T$ is the twist-3 collinear distribution defined by the matrix element
\begin{eqnarray}
&& \langle p,s|g \bar q(z_1n)[z_1n,z_2n]\not n F_{\mu+}(z_2n)[z_2n,z_3n]q(z_3n)|p,s\rangle 
\\\nn &&\qquad\qquad
=
2\epsilon_T^{\mu\nu}s_\nu (np)^2M\int_{-1}^1 dx_1dx_2dx_3 \delta(x_1+x_2+x_3) e^{-i(np)(x_1z_1+x_2z_2+x_3z_3)} T_q(x_1,x_2,x_3),
\end{eqnarray}
where $F_{\mu\nu}$ is the gluon-strength tensor, $n$ is a light-cone vector. The function $G^{(+)}$ is a similar matrix element with three $F_{\mu+}$'s. Its explicit form is not important for the present discussion and can be found in Ref.~\cite{Scimemi:2019gge}. The notation $P\otimes T$ refers to the leading order evolution kernel for $T_q(-x,0,x)$. It has the form of a complicated integral convolution that involves function $T_q$, $\Delta T_q$ (the analog of $T$ with $\gamma^\mu\to \gamma^\mu\gamma^5$) and $G^{(\pm)}$. The expression for this kernel can be found in Refs.~\cite{Scimemi:2019gge,Braun:2009mi}. It is crucial that the evolution term involves twist-3 function for a generic argument $(x_1,x_2,x_3)$, but not just $(-x,0,x)$ as for QS matrix element. Moreover, the dominant contribution to this convolution is given by the integral along $(-x,x-\xi,\xi)$-line with $\xi\in[x,1]$, whereas the contribution from the QS-component $(-\xi,0,\xi)$ is suppressed by almost two orders of magnitude \cite{Braun:2009mi,Braun:2011aw}. The scale $\mu$ in (\ref{th:f->QS}) is the scale of OPE, and present only on the right-hand-side of Eq.~(\ref{th:f->QS}). The sum of all terms becomes $\mu$ independent, so that the left-hand-side, corresponding to the optimal Sivers function, does not depend on $\mu$.

The right-hand side of Eq.~(\ref{th:f->QS}) depends on four nonperturbative functions, each of which is a function of two variables $(x_1,x_2,-x_1-x_2)$. To reduce the number of unknowns we set
\begin{eqnarray}
\mu=\mu_b=2e^{-\gamma_E}/b,
\end{eqnarray}
such that $\mathbf{L}_\mu=0$. This choice essentially reduces number of functions and parameteric freedom since the remaining functions are only $T_q(-x,0,x)$ and $G^{(+)}(-x,0,x)$, i.e. QS-functions for the quark and the gluon. The resulting expression can be inverted by means of the perturbation theory
\begin{eqnarray}\label{th:QS=f}
T_q(-x,0,x;\mu_b)&=&-\frac{1}{\pi}\left(1+C_F a_s(\mu_b)\frac{\pi^2}{6}\right)f_{1T;q\ot h}^\perp(x,b)-\frac{a_s(\mu_b)}{\pi}\int_{-1}^1 d\xi \int_0^1 dy \delta(x-y\xi)
\nn \\ &&
\times\Big(\frac{\bar y}{\pi N_c}f_{1T,q\ot h}^\perp(\xi,b)+\frac{3y\bar y}{2\xi}G^{(+)}(-\xi,0,\xi;\mu_b)\Big)+\mathcal{O}(a_s^2)+\mathcal{O}(b^2).
\end{eqnarray}
This expression can be written as 
\begin{eqnarray}\label{th:QS=f1}
&&T_q(-x,0,x;\mu_b)=-\frac{1}{\pi}\left(1+C_F a_s(\mu_b)\frac{\pi^2}{6}\right)f_{1T;q\ot h}^\perp(x,b)
\nn \\  && -\frac{a_s(\mu_b)}{\pi} \int\limits_{x}^{1} \frac{dy}{y} 
\Big[
\frac{\bar y}{N_c}f_{1T;q\ot h}^\perp\left(\frac{x}{y},b\right)+
 \frac{3y^2\bar{y}}{2x}G^{(+)}\left(-\frac{x}{y},0,\frac{x}{y};\mu_b\right)\Big]
+\mathcal{O}(a_s^2)+\mathcal{O}(b^2)\; .
\end{eqnarray}

\begin{figure}[htb]
\begin{center}
\includegraphics[width=0.45\textwidth]{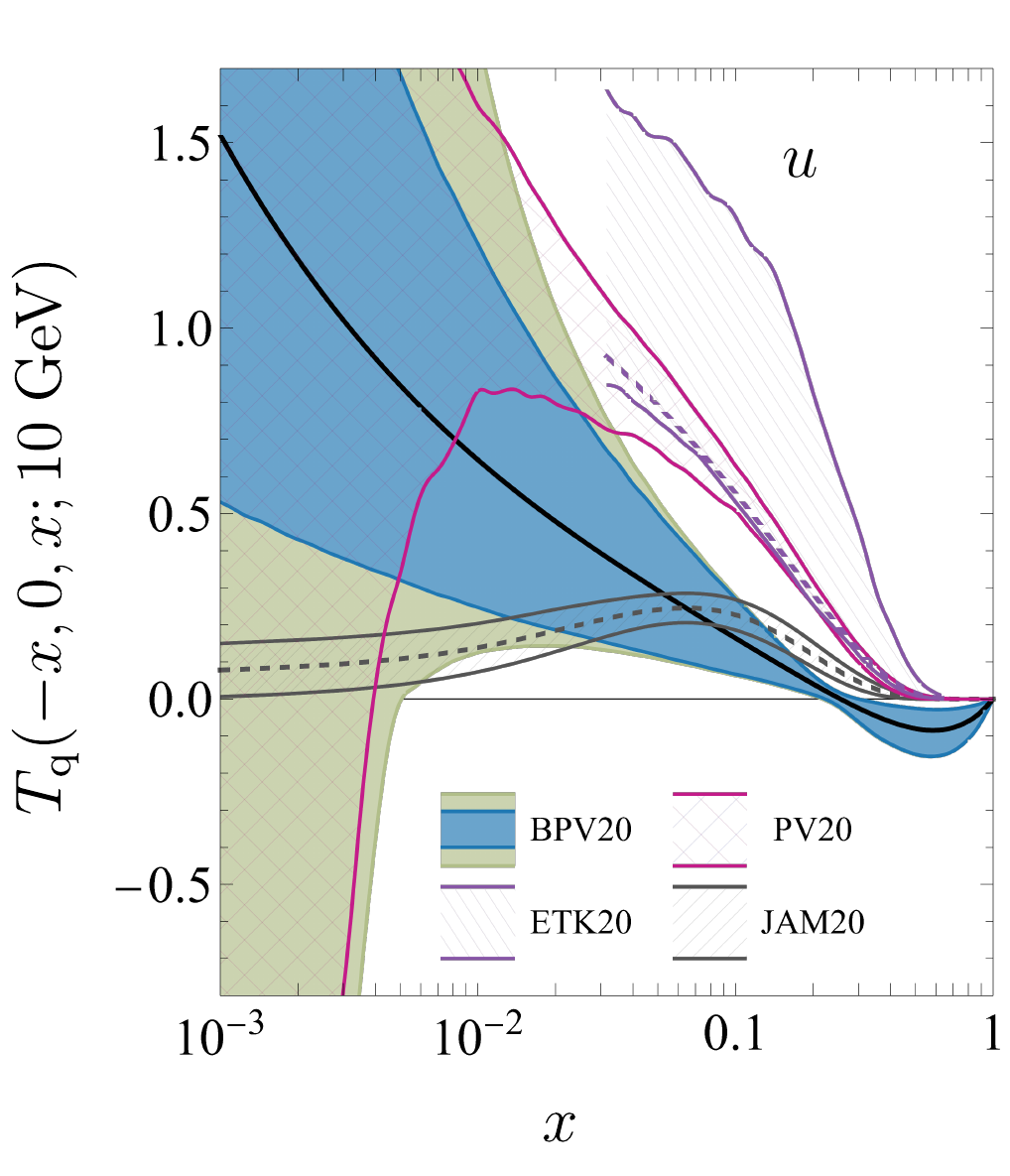}(a)
\includegraphics[width=0.45\textwidth]{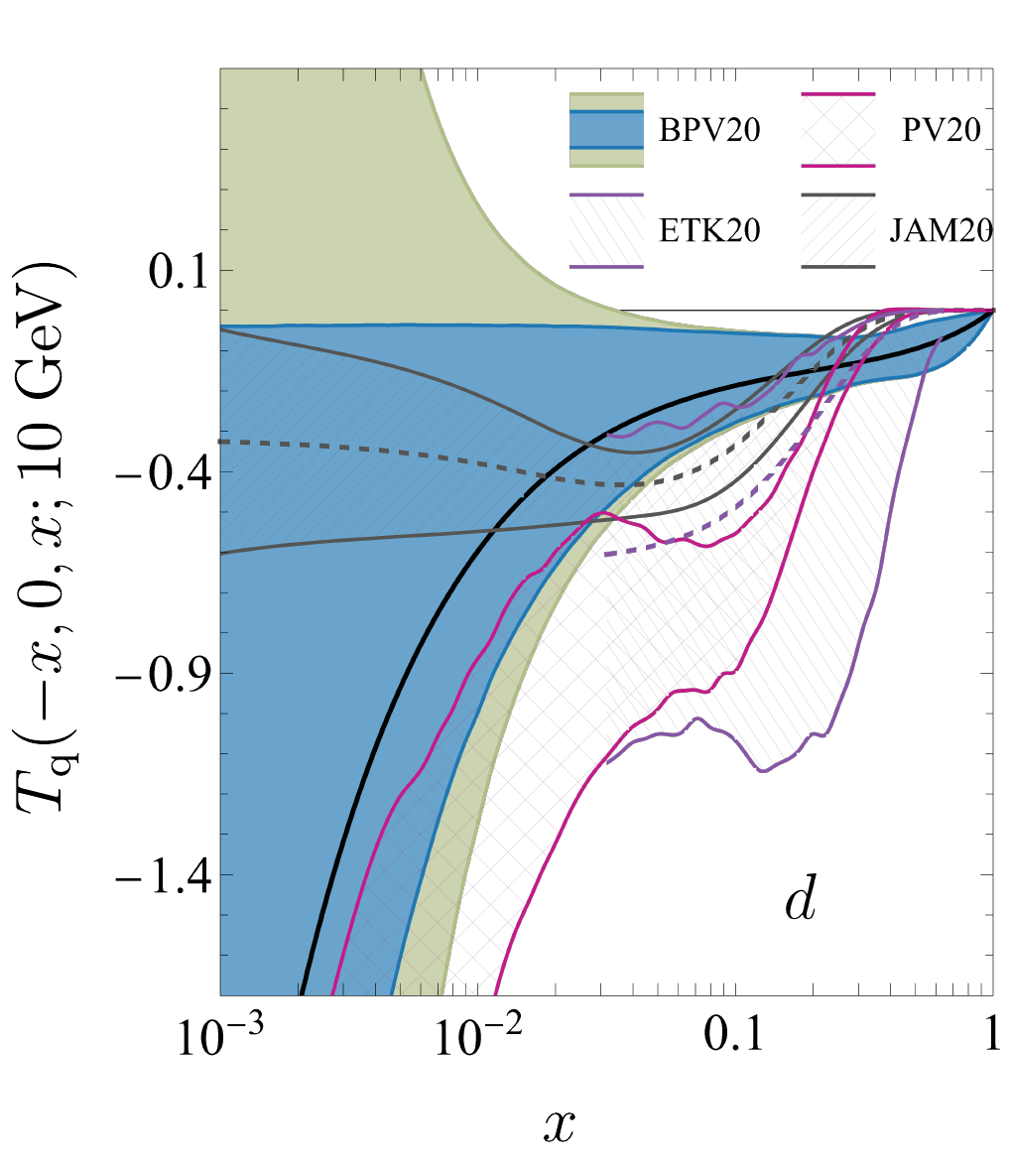}(b)
\includegraphics[width=0.45\textwidth]{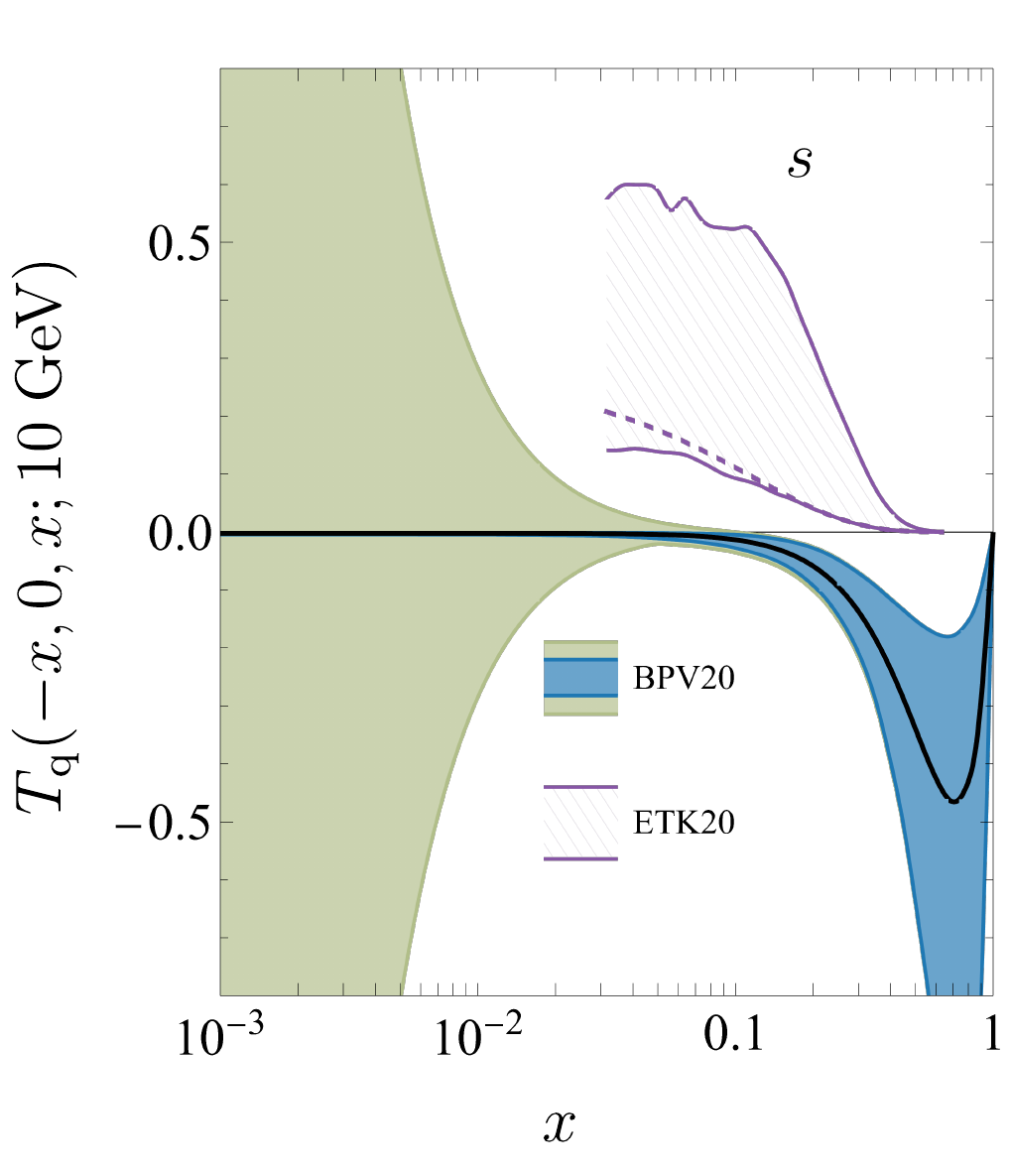}(c)
\includegraphics[width=0.45\textwidth]{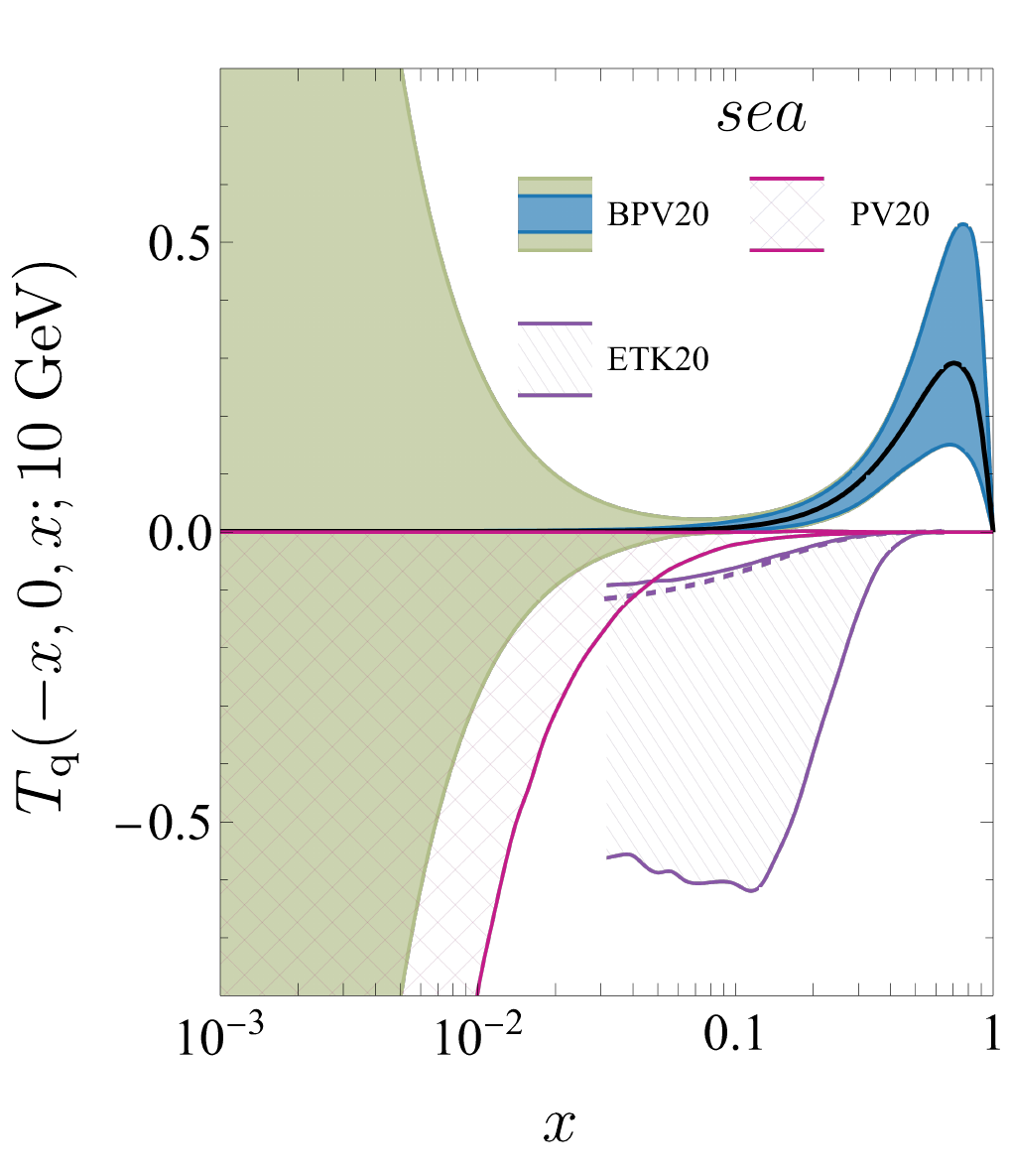}(d)
\end{center}
\caption{\label{fig:QS} Qiu-Sterman function at $\mu=10$GeV for different quark flavors, derived from the Sivers function (\ref{th:QS=f}). Our results are labeled as BPV20. The black line shows the CF value. Blue band shows 68\%CI without gluon contribution added. The green band shows the band obtained by adding the gluon contribution estimated to be $G^{(+)}=\pm |T_d+T_u|$ as described in the text.  Our results are compared to JAM20~\cite{Cammarota:2020qcw} (gray dashed line with the error corridor hatched), PV20~\cite{Bacchetta:2020gko} (magenta hatched region), ETK20~\cite{Echevarria:2020hpy} (violet hatched region, dashed line).
}
\end{figure}

To use this expression, we should select a reasonably small value of $b$, such that power corrections are negligible. Simultaneously, $b$ could not be too close to $0$ because this region corresponds to a very high-energy and thus unreliable in the current extraction. The reasonable compromise is $b\simeq 0.11$ GeV$^{-1}$ such that $\mu_b=10$ GeV. In this case, we could estimate the introduced systematic uncertainty due to omitted power corrections as $O(M^2 b^2)\sim 1\%$, which is smaller than perturbative uncertainties at this scale. Extraction of the QS function at lower scales, $\mu\sim2$ GeV, is not reliable in this approach as the corresponding value of $b\sim 0.5$ GeV$^{-1}$ is relatively large, and the power corrections become to be not negligible. The gluon function $G^{(+)}$ is also unknown, so we set it to be zero. The resulting QS-functions are shown in Fig.~\ref{fig:QS} by the black line, with 68\%CI (blue band). To estimate the uncertainty due to the unknown gluon contribution we approximate $G^{(+)}(-x,0,x)=\pm |T_d(-x,0,x)+T_u(-x,0,x)|$. The resulting band for CF value is shown in black and in green for 68\%CI. The effects of gluons are not negligible for $x\lesssim0.2$. The extracted QS-function is in general agreement with the model computations made in the light-cone wave function model in Ref.~\cite{Braun:2011aw}. We also compare our results to other extractions of the QS functions. These are the extraction from Ref.~\cite{Cammarota:2020qcw} made in the parton model approximation with SIDIS, DY, $pp$ and $A_N$ asymmetries; the NLL extraction from  Ref.~\cite{Bacchetta:2020gko} from SIDIS data; and the NLO/NNLL extraction from Ref.~\cite{Echevarria:2020hpy} from SIDIS (and DY) data. One can see that our results confirm the signs of the QS functions for $u$ and $d$ quarks found in Refs.~\cite{Cammarota:2020qcw,Bacchetta:2020gko,Echevarria:2020hpy}.

We obtain non-negligible functions for $s$ and sea quarks, and our extraction shows bigger functions in relatively large-$x$ regions and disagrees with the signs obtained in Ref.~\cite{Echevarria:2020hpy}. 
The reason partially because our QS functions for $u$ quarks that changes sign in the large-$x$ region. Another reason is that 
Ref.~\cite{Echevarria:2020hpy} and Ref.~\cite{Bacchetta:2020gko} use collinear unpolarized distributions to parametrize the QS functions and therefore cannot obtain sizable functions for sea quarks in the large-$x$ region. The QS function belongs to a different type of function, and we believe that parametrizations of twist-3 functions that utilize collinear twist-2 functions are not optimal and may bias the results of the extraction.

We have studied the functional shape of the Sivers functions, and in particular we constrained all $\epsilon>0$ to remove the nodes from the Sivers function. It turns out that a good description of the data with $\chi^2/N_{pt} < 1$ is still possible. Another study that we performed was dedicated to the large-$x$ behavior of the Sivers functions, namely, we added an extra factor $(1-x)$ to our ansatz (this choice is inspired by the model calculations made in Refs.~\cite{Brodsky:2002cx,Brodsky:2002rv}, where it was found that the Sivers functions behave as $(1-x)^2$ in the large-$x$ region). The resulting fit is also good with $\chi^2/N_{pt} < 1$, and, in particular, for the sea quarks and $s$-quarks the resulting functions become much smaller in the large-$x$ region.  

We conclude that the current data do not constrain the large-$x$ (and small-$x$) behavior of the Sivers functions and they exhibit large uncertainties in the region of $x>0.3$ and $x<0.01$. Future data from EIC and JLab 12 will be very important for exploration of both small-$x$ and large-$x$ behavior of the Sivers functions.
 
\subsection{Analysis of the sign change}
\label{sec:sign}

The sign-change of the Sivers function (\ref{th:sign-change}) is one of the principal predictions of the TMD factorization theorem. It follows from the nontrivial shape of the gauge-link contour within TMD operators (\ref{def:TMDPDF}) and would be absent in the case of a straight gauge link. Here, we attempt to estimate the significance of the sign-change.

\begin{figure}[t]
\centering
\includegraphics[width=0.55\textwidth]{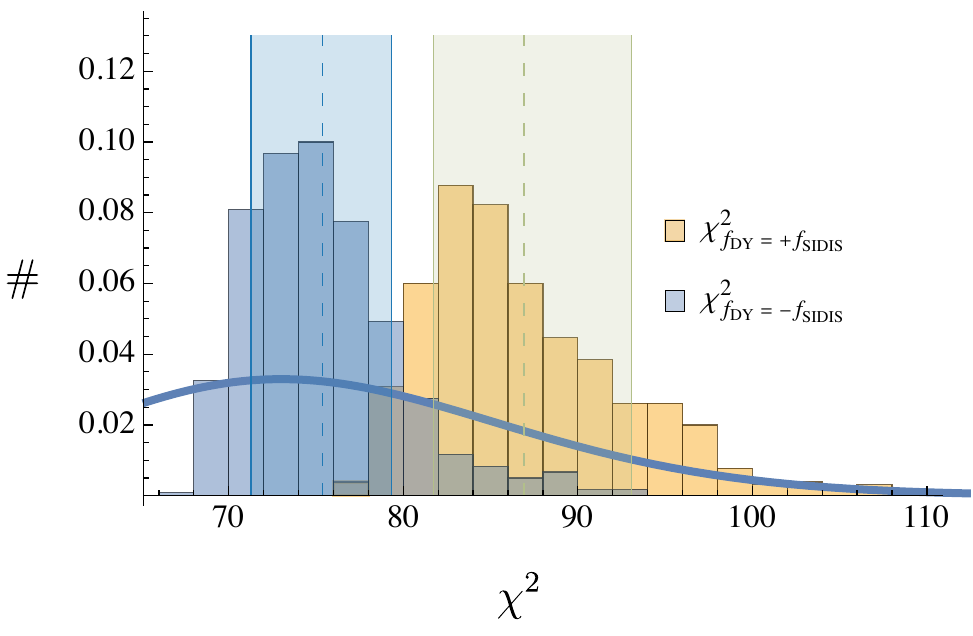}
\caption{Normalized distribution of replica's $\chi^2$ for $f_{1T\,[DY]}^{\perp}=+f_{1T\,[SIDIS]}^{\perp}$ (yellow) and $f_{1T\,[DY]}^{\perp}=-f_{1T\,[SIDIS]}^{\perp}$ (blue) cases.  The bands show the 68\%CI intervals for $\chi^2$ values. The continuous blue line is the $\chi^2$-distribution with 75 d.o.f.~.
\label{fig:chi2dist_sgnchange}}
\end{figure}
\begin{table}[htb]
\begin{center}
\begin{tabular}{|c|c|c|}
\hline 
 & $f_{1T\,[DY]}^{\perp}=-f_{1T\,[SIDIS]}^{\perp}$ & $f_{1T\,[DY]}^{\perp}=+f_{1T\,[SIDIS]}^{\perp}$\tabularnewline
\hline 
\hline 
$\chi^{2}/N_{pt}$ & $0.88_{+0.06}^{+0.16}$ & $1.00_{+0.08}^{+0.22}$\tabularnewline
\hline 
$p$-value (CF) & 0.74 & 0.44\tabularnewline
\hline
$p$-value 68\%CI & [0.60,~0.34] &[0.28,~0.08]\tabularnewline
\hline
$p$-value 68\%CI (SIDIS) & [0.67,~0.42] &[0.53,~0.11]\tabularnewline
\hline 
$p$-value 68\%CI (DY) & [0.56,~0.17] &[0.68,~0.02]\tabularnewline
\hline 
\end{tabular}
\end{center}
\caption{\label{tab:p-values} Comparison of $\chi^2$ and $p$-values between the fit with and without sign-change for Sivers function.}
\end{table}

To make a test of the sign change, we performed an independent fit of SIDIS and DY data with $f_{1T[SIDIS]}^\perp=+f_{1T[DY]}^\perp$, i.e., assuming the Sivers function does not change the sign. The fit is performed at N$^3$LO. The comparison of fits with and without sign-change is presented in Table~\ref{tab:p-values}. The CV fit demonstrates good values of $\chi^2/N_{pt}=1.00$, with the 68\%CI being [1.08, 1.22]. The (normalized) histograms of $\chi^{2}$ replicas for same- and opposite-sign fits are shown in Fig.~\ref{fig:chi2dist_sgnchange}, together with $\chi^2$ distribution for $N_{pt}-1$=75 degrees of freedom. The $p$-values of different cases are calculated as areas under the sampling distribution in $[\chi_{tot}^{2},\infty)$ interval, and given in Table~\ref{tab:p-values}. The case $f_{1T[SIDIS]}^\perp=+f_{1T[DY]}^\perp$ has somewhat higher $\chi^2$, and consequently lower $p$-value. Nonetheless, the difference is not large, and 68\%CI almost overlap. Therefore, we conclude that one cannot strictly discriminate with the current experimental data the possibility of the Sivers function having the same sign in DY and SIDIS. 

The fit with $f_{1T[SIDIS]}^\perp=+f_{1T[DY]}^\perp$ demonstrates very different features in comparison to the fit with the sign-change. In particular, the distribution of $\chi^2$ for SIDIS and DY independently is broader. So, 68\% CI of $\chi^2/N_{pt}$ for SIDIS data is [0.96, 1.21]  and for DY data is [0.80, 1.88] (compare to [0.90, 1.00] and [0.81, 1.27] in the case of the sign-change, correspondingly). Simultaneously,  the 68\%CI for the total $\chi^2$ is broader and located at higher values. This indicates a tension with the data in the same-sign approach, namely, the Sivers function that provides a better description for SIDIS gives a worse description for DY and vice-versa. 

It is also instructive to compare Sivers functions extracted in both fits. We have found that the parameters extracted in both cases  agree within 68\%CI's, except for $N_{sea}$-parameter, which flips the sign. It shows that $u$, $d$, and $s$ components are mainly constrained by the SIDIS data, where the dominant contribution comes from $q+\gamma^*\to q$ sub-process. In the DY process, the anti-quarks play a more significant role since the dominant sub-process is $q+\bar q \to \gamma^*$. Given that the unpolarized TMD for anti-quarks is much smaller than for quarks, the sign for anti-quark Sivers function almost exclusively defines the sign of the asymmetry of $W^\pm/Z$ production  in polarized $p+p$ collision.

\section{Conclusions}
\label{sec:conclusions}

We extract Sivers function from the global fit of SIDIS, pion-induced Drell-Yan and $W^\pm/Z$-bozon production experimental data. For the first time, using TMD evolution, we demonstrate the universality of TMD factorization description for SIDIS and DY transverse spin asymmetries. Our analysis is done in the $\zeta$-prescription with the unpolarized TMD distributions and nonperturbative CS-kernel extracted in \cite{Scimemi:2019cmh} (SV19), together with NNLO and N$^3$LO TMD evolution. Our results compare well in  magnitude with the existing extractions~\cite{Efremov:2004tp,Vogelsang:2005cs,Anselmino:2005ea,Anselmino:2008sga,Kang:2009bp,Aybat:2011ta,Gamberg:2013kla,Sun:2013dya,Echevarria:2014xaa,Anselmino:2016uie,Bacchetta:2020gko,Cammarota:2020qcw,Echevarria:2020hpy,Boglione:2021aha} and confirm the sign of Sivers function for $u$ and $d$ quarks while we also obtain a non-negligible Sivers function for $s$ quark and anti-quarks. The analysis was done with \texttt{artemide} package \cite{artemide}. The fitting codes  and the results of the extraction (in the form of replica-distribution for model parameters) are publicly available at \cite{dataProcessor} and \cite{artemide_sivers}.

To demonstrate the Sivers function's universality, we perform an independent fit of Sivers function from the SIDIS data only and confirm that it also describes well the DY data without any need for re-fitting. It is the first explicit check of universality for Sivers function with TMD evolution to our best knowledge. The previous successful attempt was made in the parton model approximation in Refs.~\cite{Anselmino:2016uie, Cammarota:2020qcw}.  Moreover, it is the first time SIDIS and DY data on transverse-spin asymmetries are consistently described together with a good $\chi^2$. The previous attempt to make a joined fit of Sivers function with TMD evolution \cite{Echevarria:2020hpy} faced a problem due to the difficulty in describing the large values of asymmetries in $W^\pm/Z$-production data measured by RHIC. In our analysis, we do not observe any difficulties with this data set. Although our approach is based on the same general theoretical ground as that of \cite{Echevarria:2020hpy}, our approach has a number of improvements with respect to others, and each of them could be deciding. One of such improvements is the usage of the $\zeta$-prescription. The central feature of the $\zeta$-prescription is the separation of perturbative and nonperturbative elements of the TMD factorization. So, we non-controversially use the NNLO or N$^3$LO TMD evolution, NNLO small-$b$ matching for unpolarized ingredients (TMD PDF, TMD FF, and CS-kernel), without specification of the collinear limit for the Sivers function. On the one hand, we use the best possible perturbative input and unpolarized nonperturbative parts fitted to the global data. On the other, the Sivers function is extracted as an entirely nonperturbative function of $x$ and $b$, and such a parametrization allows for sizable contributions from sea quarks in the large-$x$ region, and this may be a decisive difference with respect to the analysis of Ref.~\cite{Echevarria:2020hpy}. 

In turn, the extracted Sivers function was used to determine the QS-function, with the NLO matching relation. To our best knowledge, it is the first unbiased determination of the QS-function since all previous extractions made certain assumptions on its evolution. 

Another important point is the conservative selection of the data. The TMD factorization is valid at small values of $\delta=q_T/Q$ for DY, and $\delta=P_{hT}/(zQ)$ for SIDIS. In our analysis we used only the data with $\delta<0.3$, which is much more strict compared to other fits \cite{Echevarria:2020hpy,Bacchetta:2020gko,Echevarria:2020hpy}. It resulted in a relatively smaller data pool (76 points in total), which is guaranteed to belong to the TMD factorization domain. Additionally, we performed the test of limits for $\delta$, and found that one can raise $\delta$ to $0.4$ in the case of the transverse single-spin asymmetry measured in SIDIS.

We have also performed a test of the sign-change relation between SIDIS and DY definitions of the Sivers function. We found that the fit without sign-flip converges to values of $\chi^2$ only slightly worse than the fit with the predicted sign-flip. Therefore, we cannot statistically disregard this possibility. We have observed that the sign of the DY asymmetry is strongly correlated to the sign of the Sivers function for sea quarks, which is also apparent from the partonic channel consideration. Therefore, to clearly distinguish sign-flip/non-sign-flip scenarios, one needs the data with more substantial restrictions on the sea contribution, such as DY and kaon-production in SIDIS. Indeed, the on-going analysis of DY production by STAR, COMPASS, and the future Electron-Ion Collider will constraint the sea quark Sivers function. 

We present  in Figs.~\ref{fig:tomography} and \ref{fig:tomography1} the momentum space tomographic slices of the transversely polarized nucleon. These slices are representations of the three-dimensional (3D) nucleon structure encoded in TMD PDFs. The future and existing facilities such as the Electron-Ion Collider and Jefferson Lab 12 GeV Upgrade physics programs aim at sharpening our understanding of the 3D structure of the nucleon. We will study the impact of JLab and EIC data on the Sivers function's knowledge in the forthcoming publication.

Our results set a new benchmark and the standard of precision for studies of TMD polarized functions. They will be important for theoretical, phenomenological, and experimental studies of the 3D nucleon structure and the planning of experimental programs of existing and future facilities, such as Jefferson Lab 12 GeV Upgrade, Electron-Ion Collider, and others~\cite{Boer:2011fh,Accardi:2012qut,Dudek:2012vr,Aschenauer:2015eha,Gautheron:2010wva,Bradamante:2018ick,Chen:2019hhx,Brown:2014sea}.

\section*{Acknowledgments}
Authors are thankful to Gunar Schnell and Bakur Parsamyan for clarifications regarding data treatment, and to Zhongbo Kang for valuable discussions. The work was partially supported by   DFG FOR 2926 ``Next Generation pQCD for Hadron Structure:  Preparing  for  the  EIC'', project number 430824754 (M.B and A.V), and  by the National Science Foundation under the Contract  No.~PHY-2012002 (A.P.), and by the US Department of Energy under contract No.~DE-AC05-06OR23177 (A.P.) under which JSA, LLC operates Jefferson Lab, and within the framework of the TMD Topical Collaboration (A.P.).

\bibliography{biblio}

\providecommand{\href}[2]{#2}\begingroup\raggedright\begin{thebibliography}{100}

\bibitem{Collins:1981uk}
J.~C. Collins and D.~E. Soper, \emph{{Back-To-Back Jets in QCD}},
  \href{http://dx.doi.org/10.1016/0550-3213(81)90339-4}{\emph{Nucl. Phys. B}
  {\bf 193} (1981) 381}.

\bibitem{Collins:1984kg}
J.~C. Collins, D.~E. Soper and G.~F. Sterman, \emph{{Transverse Momentum
  Distribution in Drell-Yan Pair and W and Z Boson Production}},
  \href{http://dx.doi.org/10.1016/0550-3213(85)90479-1}{\emph{Nucl. Phys. B}
  {\bf 250} (1985) 199--224}.

\bibitem{Meng:1991da}
R.-b. Meng, F.~I. Olness and D.~E. Soper, \emph{{Semiinclusive deeply inelastic
  scattering at electron - proton colliders}},
  \href{http://dx.doi.org/10.1016/0550-3213(92)90230-9}{\emph{Nucl. Phys. B}
  {\bf 371} (1992) 79--110}.

\bibitem{Collins:2011zzd}
J.~Collins, \emph{{Foundations of perturbative QCD}}, {\emph{Camb. Monogr.
  Part. Phys. Nucl. Phys. Cosmol.} {\bf 32} (2011) 1--624}.

\bibitem{Ji:2004wu}
X.-d. Ji, J.-p. Ma and F.~Yuan, \emph{{QCD factorization for semi-inclusive
  deep-inelastic scattering at low transverse momentum}},
  \href{http://dx.doi.org/10.1103/PhysRevD.71.034005}{\emph{Phys. Rev.} {\bf
  D71} (2005) 034005}, [\href{https://arxiv.org/abs/hep-ph/0404183}{{\tt
  hep-ph/0404183}}].

\bibitem{Ji:2005nu}
X.-d. Ji, J.-P. Ma and F.~Yuan, \emph{Transverse-momentum-dependent gluon
  distributions and semi- inclusive processes at hadron colliders},
  {\emph{JHEP} {\bf 07} (2005) 020},
  [\href{https://arxiv.org/abs/hep-ph/0503015}{{\tt hep-ph/0503015}}].

\bibitem{GarciaEchevarria:2011rb}
M.~G. Echevarria, A.~Idilbi and I.~Scimemi, \emph{{Factorization Theorem For
  Drell-Yan At Low $q_T$ And Transverse Momentum Distributions
  On-The-Light-Cone}},
  \href{http://dx.doi.org/10.1007/JHEP07(2012)002}{\emph{JHEP} {\bf 07} (2012)
  002}, [\href{https://arxiv.org/abs/1111.4996}{{\tt 1111.4996}}].

\bibitem{Sivers:1989cc}
D.~W. Sivers, \emph{Single spin production asymmetries from the hard scattering
  of point - like constituents}, {\emph{Phys.~Rev.} {\bf D41} (1990) 83}.

\bibitem{Sivers:1990fh}
D.~W. Sivers, \emph{Hard scattering scaling laws for single spin production
  asymmetries}, {\emph{Phys.~Rev.} {\bf D43} (1991) 261--263}.

\bibitem{Bury:2020vhj}
M.~Bury, A.~Prokudin and A.~Vladimirov, \emph{{N$^3$LO extraction of the Sivers
  function from SIDIS, Drell-Yan, and $W^\pm/Z$ data}},
  \href{https://arxiv.org/abs/2012.05135}{{\tt 2012.05135}}.

\bibitem{Vladimirov:2017ksc}
A.~Vladimirov, \emph{{Structure of rapidity divergences in multi-parton
  scattering soft factors}},
  \href{http://dx.doi.org/10.1007/JHEP04(2018)045}{\emph{JHEP} {\bf 04} (2018)
  045}, [\href{https://arxiv.org/abs/1707.07606}{{\tt 1707.07606}}].

\bibitem{Boer:1997nt}
D.~Boer and P.~J. Mulders, \emph{{Time reversal odd distribution functions in
  leptoproduction}},
  \href{http://dx.doi.org/10.1103/PhysRevD.57.5780}{\emph{Phys. Rev. D} {\bf
  57} (1998) 5780--5786}, [\href{https://arxiv.org/abs/hep-ph/9711485}{{\tt
  hep-ph/9711485}}].

\bibitem{Bacchetta:2006tn}
A.~Bacchetta, M.~Diehl, K.~Goeke, A.~Metz, P.~J. Mulders et~al.,
  \emph{{Semi-inclusive deep inelastic scattering at small transverse
  momentum}},
  \href{http://dx.doi.org/10.1088/1126-6708/2007/02/093}{\emph{JHEP} {\bf 0702}
  (2007) 093}, [\href{https://arxiv.org/abs/hep-ph/0611265}{{\tt
  hep-ph/0611265}}].

\bibitem{Ji:2004xq}
X.-d. Ji, J.-P. Ma and F.~Yuan, \emph{Qcd factorization for spin-dependent
  cross sections in dis and drell-yan processes at low transverse momentum},
  {\emph{Phys. Lett.} {\bf B597} (2004) 299--308},
  [\href{https://arxiv.org/abs/hep-ph/0405085}{{\tt hep-ph/0405085}}].

\bibitem{Arnold:2008kf}
S.~Arnold, A.~Metz and M.~Schlegel, \emph{{Dilepton production from polarized
  hadron hadron collisions}},
  \href{http://dx.doi.org/10.1103/PhysRevD.79.034005}{\emph{Phys. Rev.} {\bf
  D79} (2009) 034005}, [\href{https://arxiv.org/abs/0809.2262}{{\tt
  0809.2262}}].

\bibitem{Boer:1997mf}
D.~Boer, R.~Jakob and P.~J. Mulders, \emph{{Asymmetries in polarized hadron
  production in e+ e- annihilation up to order 1/Q}},
  \href{http://dx.doi.org/10.1016/S0550-3213(97)00456-2}{\emph{Nucl. Phys. B}
  {\bf 504} (1997) 345--380}, [\href{https://arxiv.org/abs/hep-ph/9702281}{{\tt
  hep-ph/9702281}}].

\bibitem{Bacchetta:2017gcc}
A.~Bacchetta, F.~Delcarro, C.~Pisano, M.~Radici and A.~Signori,
  \emph{{Extraction of partonic transverse momentum distributions from
  semi-inclusive deep-inelastic scattering, Drell-Yan and Z-boson production}},
  \href{http://dx.doi.org/10.1007/JHEP06(2017)081}{\emph{JHEP} {\bf 06} (2017)
  081}, [\href{https://arxiv.org/abs/1703.10157}{{\tt 1703.10157}}].

\bibitem{Scimemi:2019cmh}
I.~Scimemi and A.~Vladimirov, \emph{{Non-perturbative structure of
  semi-inclusive deep-inelastic and Drell-Yan scattering at small transverse
  momentum}}, \href{http://dx.doi.org/10.1007/JHEP06(2020)137}{\emph{JHEP} {\bf
  06} (2020) 137}, [\href{https://arxiv.org/abs/1912.06532}{{\tt 1912.06532}}].

\bibitem{Efremov:2004tp}
A.~Efremov, K.~Goeke, S.~Menzel, A.~Metz and P.~Schweitzer, \emph{{Sivers
  effect in semi-inclusive DIS and in the Drell-Yan process}},
  \href{http://dx.doi.org/10.1016/j.physletb.2005.03.010}{\emph{Phys. Lett. B}
  {\bf 612} (2005) 233--244}, [\href{https://arxiv.org/abs/hep-ph/0412353}{{\tt
  hep-ph/0412353}}].

\bibitem{Vogelsang:2005cs}
W.~Vogelsang and F.~Yuan, \emph{{Single-transverse spin asymmetries: From DIS
  to hadronic collisions}},
  \href{http://dx.doi.org/10.1103/PhysRevD.72.054028}{\emph{Phys. Rev. D} {\bf
  72} (2005) 054028}, [\href{https://arxiv.org/abs/hep-ph/0507266}{{\tt
  hep-ph/0507266}}].

\bibitem{Anselmino:2005ea}
M.~Anselmino, M.~Boglione, U.~D'Alesio, A.~Kotzinian, F.~Murgia et~al.,
  \emph{{Extracting the Sivers function from polarized SIDIS data and making
  predictions}}, \href{http://dx.doi.org/10.1103/PhysRevD.72.094007,
  10.1103/PhysRevD.72.099903}{\emph{Phys.~Rev.} {\bf D72} (2005) 094007},
  [\href{https://arxiv.org/abs/hep-ph/0507181}{{\tt hep-ph/0507181}}].

\bibitem{Anselmino:2008sga}
M.~Anselmino, M.~Boglione, U.~D'Alesio, A.~Kotzinian, S.~Melis et~al.,
  \emph{{Sivers Effect for Pion and Kaon Production in Semi-Inclusive Deep
  Inelastic Scattering}},
  \href{http://dx.doi.org/10.1140/epja/i2008-10697-y}{\emph{Eur.~Phys.~J.} {\bf
  A39} (2009) 89--100}, [\href{https://arxiv.org/abs/0805.2677}{{\tt
  0805.2677}}].

\bibitem{Kang:2009bp}
Z.-B. Kang and J.-W. Qiu, \emph{{Testing the Time-Reversal Modified
  Universality of the Sivers Function}},
  \href{http://dx.doi.org/10.1103/PhysRevLett.103.172001}{\emph{Phys. Rev.
  Lett.} {\bf 103} (2009) 172001}, [\href{https://arxiv.org/abs/0903.3629}{{\tt
  0903.3629}}].

\bibitem{Aybat:2011ta}
S.~M. Aybat, A.~Prokudin and T.~C. Rogers, \emph{{Calculation of TMD Evolution
  for Transverse Single Spin Asymmetry Measurements}},
  \href{https://arxiv.org/abs/1112.4423}{{\tt 1112.4423}}.

\bibitem{Gamberg:2013kla}
L.~Gamberg, Z.-B. Kang and A.~Prokudin, \emph{{Indication on the
  process-dependence of the Sivers effect}}, {\emph{Phys.Rev.Lett.} {\bf 110}
  (2013) 232301}, [\href{https://arxiv.org/abs/1302.3218}{{\tt 1302.3218}}].

\bibitem{Sun:2013dya}
P.~Sun and F.~Yuan, \emph{{Energy Evolution for the Sivers Asymmetries in Hard
  Processes}}, \href{http://dx.doi.org/10.1103/PhysRevD.88.034016}{\emph{Phys.
  Rev.} {\bf D88} (2013) 034016}, [\href{https://arxiv.org/abs/1304.5037}{{\tt
  1304.5037}}].

\bibitem{Echevarria:2014xaa}
M.~G. Echevarria, A.~Idilbi, Z.-B. Kang and I.~Vitev, \emph{{QCD Evolution of
  the Sivers Asymmetry}},
  \href{http://dx.doi.org/10.1103/PhysRevD.89.074013}{\emph{Phys.~Rev.} {\bf
  D89} (2014) 074013}, [\href{https://arxiv.org/abs/1401.5078}{{\tt
  1401.5078}}].

\bibitem{Anselmino:2016uie}
M.~Anselmino, M.~Boglione, U.~D'Alesio, F.~Murgia and A.~Prokudin, \emph{{Study
  of the sign change of the Sivers function from STAR Collaboration W/Z
  production data}},
  \href{http://dx.doi.org/10.1007/JHEP04(2017)046}{\emph{JHEP} {\bf 04} (2017)
  046}, [\href{https://arxiv.org/abs/1612.06413}{{\tt 1612.06413}}].

\bibitem{Bacchetta:2020gko}
A.~Bacchetta, F.~Delcarro, C.~Pisano and M.~Radici, \emph{{The
  three-dimensional distribution of quarks in momentum space}},
  \href{https://arxiv.org/abs/2004.14278}{{\tt 2004.14278}}.

\bibitem{Cammarota:2020qcw}
{\scshape Jefferson Lab Angular Momentum} collaboration, J.~Cammarota,
  L.~Gamberg, Z.-B. Kang, J.~A. Miller, D.~Pitonyak, A.~Prokudin et~al.,
  \emph{{Origin of single transverse-spin asymmetries in high-energy
  collisions}},
  \href{http://dx.doi.org/10.1103/PhysRevD.102.054002}{\emph{Phys. Rev. D} {\bf
  102} (2020) 054002}, [\href{https://arxiv.org/abs/2002.08384}{{\tt
  2002.08384}}].

\bibitem{Echevarria:2020hpy}
M.~G. Echevarria, Z.-B. Kang and J.~Terry, \emph{{Global analysis of the Sivers
  functions at NLO+NNLL in QCD}},
  \href{http://dx.doi.org/10.1007/JHEP01(2021)126}{\emph{JHEP} {\bf 01} (2021)
  126}, [\href{https://arxiv.org/abs/2009.10710}{{\tt 2009.10710}}].

\bibitem{Boglione:2021aha}
M.~Boglione, U.~D'Alesio, C.~Flore, J.~O. Gonzalez-Hernandez, F.~Murgia and
  A.~Prokudin, \emph{{Reweighting the Sivers function with jet data from
  STAR}},  \href{https://arxiv.org/abs/2101.03955}{{\tt 2101.03955}}.

\bibitem{Anselmino:1994tv}
M.~Anselmino, M.~Boglione and F.~Murgia, \emph{{Single spin asymmetry for p
  (polarized) p ---\ensuremath{>} pi X in perturbative QCD}},
  \href{http://dx.doi.org/10.1016/0370-2693(95)01168-P}{\emph{Phys. Lett. B}
  {\bf 362} (1995) 164--172}, [\href{https://arxiv.org/abs/hep-ph/9503290}{{\tt
  hep-ph/9503290}}].

\bibitem{Airapetian:2009ae}
{\scshape HERMES} collaboration, A.~Airapetian. et~al., \emph{{Observation of
  the Naive-T-odd Sivers Effect in Deep- Inelastic Scattering}}, {\emph{Phys.
  Rev. Lett.} {\bf 103} (2009) 152002},
  [\href{https://arxiv.org/abs/0906.3918}{{\tt 0906.3918}}].

\bibitem{Airapetian:2020zzo}
{\scshape HERMES} collaboration, A.~Airapetian et~al., \emph{{Azimuthal single-
  and double-spin asymmetries in semi-inclusive deep-inelastic lepton
  scattering by transversely polarized protons}},
  \href{https://arxiv.org/abs/2007.07755}{{\tt 2007.07755}}.

\bibitem{Alekseev:2008aa}
{\scshape COMPASS} collaboration, M.~Alekseev et~al., \emph{{Collins and Sivers
  asymmetries for pions and kaons in muon-deuteron DIS}},
  \href{http://dx.doi.org/10.1016/j.physletb.2009.01.060}{\emph{Phys.Lett.}
  {\bf B673} (2009) 127--135}, [\href{https://arxiv.org/abs/0802.2160}{{\tt
  0802.2160}}].

\bibitem{Adolph:2014zba}
{\scshape COMPASS} collaboration, C.~Adolph et~al., \emph{{Collins and Sivers
  asymmetries in muonproduction of pions and kaons off transversely polarised
  protons}},
  \href{http://dx.doi.org/10.1016/j.physletb.2015.03.056}{\emph{Phys. Lett.}
  {\bf B744} (2015) 250--259}, [\href{https://arxiv.org/abs/1408.4405}{{\tt
  1408.4405}}].

\bibitem{Adolph:2012sp}
{\scshape COMPASS} collaboration, C.~Adolph et~al., \emph{{II Ð Experimental
  investigation of transverse spin asymmetries in $\mu$-$p$ SIDIS processes:
  Sivers asymmetries}},
  \href{http://dx.doi.org/10.1016/j.physletb.2012.09.056}{\emph{Phys. Lett.}
  {\bf B717} (2012) 383--389}, [\href{https://arxiv.org/abs/1205.5122}{{\tt
  1205.5122}}].

\bibitem{Adolph:2016dvl}
{\scshape COMPASS} collaboration, C.~Adolph et~al., \emph{{Sivers asymmetry
  extracted in SIDIS at the hard scales of the Drell–Yan process at
  COMPASS}},
  \href{http://dx.doi.org/10.1016/j.physletb.2017.04.042}{\emph{Phys. Lett.}
  {\bf B770} (2017) 138--145}, [\href{https://arxiv.org/abs/1609.07374}{{\tt
  1609.07374}}].

\bibitem{Aghasyan:2017jop}
{\scshape COMPASS} collaboration, M.~Aghasyan et~al., \emph{{First measurement
  of transverse-spin-dependent azimuthal asymmetries in the Drell-Yan
  process}},
  \href{http://dx.doi.org/10.1103/PhysRevLett.119.112002}{\emph{Phys. Rev.
  Lett.} {\bf 119} (2017) 112002},
  [\href{https://arxiv.org/abs/1704.00488}{{\tt 1704.00488}}].

\bibitem{Qian:2011py}
{\scshape The Jefferson Lab Hall A} collaboration, X.~Qian et~al.,
  \emph{{Single Spin Asymmetries in Charged Pion Production from Semi-Inclusive
  Deep Inelastic Scattering on a Transversely Polarized $^3$He Target}},
  \href{http://dx.doi.org/10.1103/PhysRevLett.107.072003}{\emph{Phys.Rev.Lett.}
  {\bf 107} (2011) 072003}, [\href{https://arxiv.org/abs/1106.0363}{{\tt
  1106.0363}}].

\bibitem{Zhao:2014qvx}
{\scshape Jefferson Lab Hall A} collaboration, Y.~X. Zhao et~al., \emph{{Single
  spin asymmetries in charged kaon production from semi-inclusive deep
  inelastic scattering on a transversely polarized $^3He$ target}},
  \href{http://dx.doi.org/10.1103/PhysRevC.90.055201}{\emph{Phys. Rev.} {\bf
  C90} (2014) 055201}, [\href{https://arxiv.org/abs/1404.7204}{{\tt
  1404.7204}}].

\bibitem{Adamczyk:2015gyk}
{\scshape STAR} collaboration, L.~Adamczyk et~al., \emph{{Measurement of the
  transverse single-spin asymmetry in $p^\uparrow+p \to W^{\pm}/Z^0$ at RHIC}},
  \href{http://dx.doi.org/10.1103/PhysRevLett.116.132301}{\emph{Phys. Rev.
  Lett.} {\bf 116} (2016) 132301},
  [\href{https://arxiv.org/abs/1511.06003}{{\tt 1511.06003}}].

\bibitem{Dudek:2012vr}
J.~Dudek, R.~Ent, R.~Essig, K.~Kumar, C.~Meyer et~al., \emph{{Physics
  Opportunities with the 12 GeV Upgrade at Jefferson Lab}},
  \href{http://dx.doi.org/10.1140/epja/i2012-12187-1}{\emph{Eur.~Phys.~J.} {\bf
  A48} (2012) 187}, [\href{https://arxiv.org/abs/1208.1244}{{\tt 1208.1244}}].

\bibitem{Boer:2011fh}
D.~Boer, M.~Diehl, R.~Milner, R.~Venugopalan, W.~Vogelsang et~al.,
  \emph{{Gluons and the quark sea at high energies: Distributions,
  polarization, tomography}},  \href{https://arxiv.org/abs/1108.1713}{{\tt
  1108.1713}}.

\bibitem{Accardi:2012qut}
A.~Accardi et~al., \emph{{Electron Ion Collider: The Next QCD Frontier}},
  \href{http://dx.doi.org/10.1140/epja/i2016-16268-9}{\emph{Eur. Phys. J.} {\bf
  A52} (2016) 268}, [\href{https://arxiv.org/abs/1212.1701}{{\tt 1212.1701}}].

\bibitem{Aschenauer:2015eha}
E.-C. Aschenauer et~al., \emph{{The RHIC SPIN Program: Achievements and Future
  Opportunities}},  \href{https://arxiv.org/abs/1501.01220}{{\tt 1501.01220}}.

\bibitem{Gautheron:2010wva}
{\scshape COMPASS} collaboration, F.~Gautheron et~al., \emph{{COMPASS-II
  Proposal}}, .

\bibitem{Bradamante:2018ick}
{\scshape COMPASS} collaboration, F.~Bradamante, \emph{{The future SIDIS
  measurement on transversely polarized deuterons by the COMPASS
  Collaboration}}, \href{http://dx.doi.org/10.22323/1.346.0045}{\emph{PoS} {\bf
  SPIN2018} (2018) 045}, [\href{https://arxiv.org/abs/1812.07281}{{\tt
  1812.07281}}].

\bibitem{Chen:2019hhx}
{\scshape SeaQuest} collaboration, A.~Chen et~al., \emph{{Probing
  nucleon\textquoteright{}s spin structures with polarized Drell-Yan in the
  Fermilab SpinQuest experiment}},
  \href{http://dx.doi.org/10.22323/1.346.0164}{\emph{PoS} {\bf SPIN2018} (2019)
  164}, [\href{https://arxiv.org/abs/1901.09994}{{\tt 1901.09994}}].

\bibitem{Brown:2014sea}
C.~Brown et~al., \emph{{Letter of Intent for a Drell-Yan Experiment with a
  Polarized Proton Target}}, .

\bibitem{Belitsky:2002sm}
A.~V. Belitsky, X.~Ji and F.~Yuan, \emph{Final state interactions and gauge
  invariant parton distributions}, {\emph{Nucl. Phys.} {\bf B656} (2003)
  165--198}, [\href{https://arxiv.org/abs/hep-ph/0208038}{{\tt
  hep-ph/0208038}}].

\bibitem{Brodsky:2002rv}
S.~J. Brodsky, D.~S. Hwang and I.~Schmidt, \emph{Initial-state interactions and
  single-spin asymmetries in drell-yan processes}, {\emph{Nucl. Phys.} {\bf
  B642} (2002) 344--356}, [\href{https://arxiv.org/abs/hep-ph/0206259}{{\tt
  hep-ph/0206259}}].

\bibitem{Brodsky:2002cx}
S.~J. Brodsky, D.~S. Hwang and I.~Schmidt, \emph{Final-state interactions and
  single-spin asymmetries in semi-inclusive deep inelastic scattering},
  {\emph{Phys. Lett.} {\bf B530} (2002) 99--107},
  [\href{https://arxiv.org/abs/hep-ph/0201296}{{\tt hep-ph/0201296}}].

\bibitem{Collins:2002kn}
J.~C. Collins, \emph{Leading-twist single-transverse-spin asymmetries:
  Drell-yan and deep-inelastic scattering}, {\emph{Phys.~Lett.} {\bf B536}
  (2002) 43--48}, [\href{https://arxiv.org/abs/hep-ph/0204004}{{\tt
  hep-ph/0204004}}].

\bibitem{Ji:2006ub}
X.~Ji, J.-W. Qiu, W.~Vogelsang and F.~Yuan, \emph{{A unified picture for single
  transverse-spin asymmetries in hard processes}},
  \href{http://dx.doi.org/10.1103/PhysRevLett.97.082002}{\emph{Phys. Rev.
  Lett.} {\bf 97} (2006) 082002},
  [\href{https://arxiv.org/abs/hep-ph/0602239}{{\tt hep-ph/0602239}}].

\bibitem{Boer:2003cm}
D.~Boer, P.~J. Mulders and F.~Pijlman, \emph{Universality of t-odd effects in
  single spin and azimuthal asymmetries}, {\emph{Nucl.~Phys.} {\bf B667} (2003)
  201--241}, [\href{https://arxiv.org/abs/hep-ph/0303034}{{\tt
  hep-ph/0303034}}].

\bibitem{Koike:2007dg}
Y.~Koike, W.~Vogelsang and F.~Yuan, \emph{{On the Relation Between Mechanisms
  for Single-Transverse-Spin Asymmetries}},
  \href{http://dx.doi.org/10.1016/j.physletb.2007.11.096}{\emph{Phys.Lett.}
  {\bf B659} (2008) 878--884}, [\href{https://arxiv.org/abs/0711.0636}{{\tt
  0711.0636}}].

\bibitem{Kang:2011mr}
Z.-B. Kang, B.-W. Xiao and F.~Yuan, \emph{{QCD Resummation for Single Spin
  Asymmetries}},
  \href{http://dx.doi.org/10.1103/PhysRevLett.107.152002}{\emph{Phys. Rev.
  Lett.} {\bf 107} (2011) 152002}, [\href{https://arxiv.org/abs/1106.0266}{{\tt
  1106.0266}}].

\bibitem{Scimemi:2018mmi}
I.~Scimemi and A.~Vladimirov, \emph{{Matching of transverse momentum dependent
  distributions at twist-3}},
  \href{http://dx.doi.org/10.1140/epjc/s10052-018-6263-5}{\emph{Eur. Phys. J.
  C} {\bf 78} (2018) 802}, [\href{https://arxiv.org/abs/1804.08148}{{\tt
  1804.08148}}].

\bibitem{Efremov:1981sh}
A.~V. Efremov and O.~V. Teryaev, \emph{On spin effects in quantum
  chromodynamics}, {\emph{Sov.~J.~Nucl.~Phys.} {\bf 36} (1982) 140}.

\bibitem{Efremov:1984ip}
A.~Efremov and O.~Teryaev, \emph{{QCD Asymmetry and Polarized Hadron Structure
  Functions}},
  \href{http://dx.doi.org/10.1016/0370-2693(85)90999-2}{\emph{Phys.Lett.} {\bf
  B150} (1985) 383}.

\bibitem{Qiu:1991pp}
J.-W. Qiu and G.~Sterman, \emph{Single transverse spin asymmetries},
  {\emph{Phys.~Rev.~Lett.} {\bf 67} (1991) 2264--2267}.

\bibitem{Qiu:1991wg}
J.-W. Qiu and G.~Sterman, \emph{Single transverse spin asymmetries in direct
  photon production}, {\emph{Nucl.~Phys.} {\bf B378} (1992) 52--78}.

\bibitem{Scimemi:2017etj}
I.~Scimemi and A.~Vladimirov, \emph{{Analysis of vector boson production within
  TMD factorization}},
  \href{http://dx.doi.org/10.1140/epjc/s10052-018-5557-y}{\emph{Eur. Phys. J.}
  {\bf C78} (2018) 89}, [\href{https://arxiv.org/abs/1706.01473}{{\tt
  1706.01473}}].

\bibitem{Bertone:2019nxa}
V.~Bertone, I.~Scimemi and A.~Vladimirov, \emph{{Extraction of unpolarized
  quark transverse momentum dependent parton distributions from
  Drell-Yan/Z-boson production}},
  \href{http://dx.doi.org/10.1007/JHEP06(2019)028}{\emph{JHEP} {\bf 06} (2019)
  028}, [\href{https://arxiv.org/abs/1902.08474}{{\tt 1902.08474}}].

\bibitem{Bacchetta:2019sam}
A.~Bacchetta, V.~Bertone, C.~Bissolotti, G.~Bozzi, F.~Delcarro, F.~Piacenza
  et~al., \emph{{Transverse-momentum-dependent parton distributions up to
  N$^{3}$LL from Drell-Yan data}},
  \href{http://dx.doi.org/10.1007/JHEP07(2020)117}{\emph{JHEP} {\bf 07} (2020)
  117}, [\href{https://arxiv.org/abs/1912.07550}{{\tt 1912.07550}}].

\bibitem{Scimemi:2018xaf}
I.~Scimemi and A.~Vladimirov, \emph{{Systematic analysis of double-scale
  evolution}}, \href{http://dx.doi.org/10.1007/JHEP08(2018)003}{\emph{JHEP}
  {\bf 08} (2018) 003}, [\href{https://arxiv.org/abs/1803.11089}{{\tt
  1803.11089}}].

\bibitem{artemide}
\emph{\texttt{artemide} repository, https://github.com/ vladimirovalexey/
  artemide-public},  2020.

\bibitem{dataProcessor}
``Data processing library for \texttt{artemide},
  https://github.com/vladimirovalexey/artemide-dataprocessor.''

\bibitem{Boer:2011xd}
D.~Boer, L.~Gamberg, B.~Musch and A.~Prokudin, \emph{{Bessel-Weighted
  Asymmetries in Semi Inclusive Deep Inelastic Scattering}},
  \href{http://dx.doi.org/10.1007/JHEP10(2011)021}{\emph{JHEP} {\bf 1110}
  (2011) 021}, [\href{https://arxiv.org/abs/1107.5294}{{\tt 1107.5294}}].

\bibitem{Gamberg:2011my}
L.~Gamberg, D.~Boer, B.~Musch and A.~Prokudin, \emph{{Semi-Inclusive Deep
  Inelastic Scattering and Bessel-Weighted Asymmetries}},
  \href{http://dx.doi.org/10.1063/1.3667306}{\emph{AIP Conf.Proc.} {\bf 1418}
  (2011) 72--78}, [\href{https://arxiv.org/abs/1111.0603}{{\tt 1111.0603}}].

\bibitem{Scimemi:2019gge}
I.~Scimemi, A.~Tarasov and A.~Vladimirov, \emph{{Collinear matching for Sivers
  function at next-to-leading order}},
  \href{http://dx.doi.org/10.1007/JHEP05(2019)125}{\emph{JHEP} {\bf 05} (2019)
  125}, [\href{https://arxiv.org/abs/1901.04519}{{\tt 1901.04519}}].

\bibitem{Chiu:2011qc}
J.-y. Chiu, A.~Jain, D.~Neill and I.~Z. Rothstein, \emph{{The Rapidity
  Renormalization Group}},
  \href{http://dx.doi.org/10.1103/PhysRevLett.108.151601}{\emph{Phys. Rev.
  Lett.} {\bf 108} (2012) 151601}, [\href{https://arxiv.org/abs/1104.0881}{{\tt
  1104.0881}}].

\bibitem{Vladimirov:2020umg}
A.~A. Vladimirov, \emph{{Self-contained definition of Collins-Soper kernel}},
  \href{https://arxiv.org/abs/2003.02288}{{\tt 2003.02288}}.

\bibitem{Gehrmann:2010ue}
T.~Gehrmann, E.~Glover, T.~Huber, N.~Ikizlerli and C.~Studerus,
  \emph{{Calculation of the quark and gluon form factors to three loops in
  QCD}}, \href{http://dx.doi.org/10.1007/JHEP06(2010)094}{\emph{JHEP} {\bf 06}
  (2010) 094}, [\href{https://arxiv.org/abs/1004.3653}{{\tt 1004.3653}}].

\bibitem{Grozin:2014hna}
A.~Grozin, J.~M. Henn, G.~P. Korchemsky and P.~Marquard, \emph{{Three Loop Cusp
  Anomalous Dimension in QCD}},
  \href{http://dx.doi.org/10.1103/PhysRevLett.114.062006}{\emph{Phys. Rev.
  Lett.} {\bf 114} (2015) 062006}, [\href{https://arxiv.org/abs/1409.0023}{{\tt
  1409.0023}}].

\bibitem{Li:2016ctv}
Y.~Li and H.~X. Zhu, \emph{{Bootstrapping Rapidity Anomalous Dimensions for
  Transverse-Momentum Resummation}},
  \href{http://dx.doi.org/10.1103/PhysRevLett.118.022004}{\emph{Phys. Rev.
  Lett.} {\bf 118} (2017) 022004},
  [\href{https://arxiv.org/abs/1604.01404}{{\tt 1604.01404}}].

\bibitem{Vladimirov:2016dll}
A.~A. Vladimirov, \emph{{Correspondence between Soft and Rapidity Anomalous
  Dimensions}},
  \href{http://dx.doi.org/10.1103/PhysRevLett.118.062001}{\emph{Phys. Rev.
  Lett.} {\bf 118} (2017) 062001},
  [\href{https://arxiv.org/abs/1610.05791}{{\tt 1610.05791}}].

\bibitem{Vladimirov:2019bfa}
A.~Vladimirov, \emph{{Pion-induced Drell-Yan processes within TMD
  factorization}}, \href{http://dx.doi.org/10.1007/JHEP10(2019)090}{\emph{JHEP}
  {\bf 10} (2019) 090}, [\href{https://arxiv.org/abs/1907.10356}{{\tt
  1907.10356}}].

\bibitem{Gourdin:1973qx}
M.~Gourdin, \emph{{Semiinclusive reactions induced by leptons}},
  \href{http://dx.doi.org/10.1016/0550-3213(72)90615-3}{\emph{Nucl. Phys.} {\bf
  B49} (1972) 501}.

\bibitem{Kotzinian:1994dv}
A.~Kotzinian, \emph{{New quark distributions and semiinclusive
  electroproduction on the polarized nucleons}},
  \href{http://dx.doi.org/10.1016/0550-3213(95)00098-D}{\emph{Nucl. Phys.} {\bf
  B441} (1995) 234}, [\href{https://arxiv.org/abs/hep-ph/9412283}{{\tt
  hep-ph/9412283}}].

\bibitem{Diehl:2005pc}
M.~Diehl and S.~Sapeta, \emph{{On the analysis of lepton scattering on
  longitudinally or transversely polarized protons}},
  \href{http://dx.doi.org/10.1140/epjc/s2005-02242-9}{\emph{Eur. Phys. J.} {\bf
  C41} (2005) 515}, [\href{https://arxiv.org/abs/hep-ph/0503023}{{\tt
  hep-ph/0503023}}].

\bibitem{Bacchetta:2004jz}
A.~Bacchetta, U.~D'Alesio, M.~Diehl and C.~A. Miller, \emph{Single-spin
  asymmetries: The trento conventions}, {\emph{Phys.~Rev.} {\bf D70} (2004)
  117504}, [\href{https://arxiv.org/abs/hep-ph/0410050}{{\tt hep-ph/0410050}}].

\bibitem{Aybat:2011zv}
S.~Aybat and T.~C. Rogers, \emph{{TMD Parton Distribution and Fragmentation
  Functions with QCD Evolution}},
  \href{http://dx.doi.org/10.1103/PhysRevD.83.114042}{\emph{Phys.Rev.} {\bf
  D83} (2011) 114042}, [\href{https://arxiv.org/abs/1101.5057}{{\tt
  1101.5057}}].

\bibitem{Aybat:2011ge}
S.~M. Aybat, J.~C. Collins, J.-W. Qiu and T.~C. Rogers, \emph{{The QCD
  Evolution of the Sivers Function}}, {\emph{Phys.~Rev.} {\bf D85} (2012)
  034043}, [\href{https://arxiv.org/abs/1110.6428}{{\tt 1110.6428}}].

\bibitem{Echevarria:2012pw}
M.~G. Echevarria, A.~Idilbi, A.~Sch\"afer and I.~Scimemi,
  \emph{{Model-Independent Evolution of Transverse Momentum Dependent
  Distribution Functions (TMDs) at NNLL}},
  \href{http://dx.doi.org/10.1140/epjc/s10052-013-2636-y}{\emph{Eur. Phys. J.
  C} {\bf 73} (2013) 2636}, [\href{https://arxiv.org/abs/1208.1281}{{\tt
  1208.1281}}].

\bibitem{Idilbi:2004vb}
A.~Idilbi, X.-d. Ji, J.-P. Ma and F.~Yuan, \emph{{Collins-Soper equation for
  the energy evolution of transverse-momentum and spin dependent parton
  distributions}},
  \href{http://dx.doi.org/10.1103/PhysRevD.70.074021}{\emph{Phys. Rev.} {\bf
  D70} (2004) 074021}, [\href{https://arxiv.org/abs/hep-ph/0406302}{{\tt
  hep-ph/0406302}}].

\bibitem{vonManteuffel:2020vjv}
A.~von Manteuffel, E.~Panzer and R.~M. Schabinger, \emph{{Cusp and collinear
  anomalous dimensions in four-loop QCD from form factors}},
  \href{http://dx.doi.org/10.1103/PhysRevLett.124.162001}{\emph{Phys. Rev.
  Lett.} {\bf 124} (2020) 162001},
  [\href{https://arxiv.org/abs/2002.04617}{{\tt 2002.04617}}].

\bibitem{Kang:2009sm}
Z.-B. Kang and J.-W. Qiu, \emph{{Single transverse spin asymmetry of dilepton
  production near Z0 pole}},
  \href{http://dx.doi.org/10.1103/PhysRevD.81.054020}{\emph{Phys. Rev. D} {\bf
  81} (2010) 054020}, [\href{https://arxiv.org/abs/0912.1319}{{\tt
  0912.1319}}].

\bibitem{Anselmino:2009st}
M.~Anselmino, M.~Boglione, U.~D'Alesio, S.~Melis, F.~Murgia and A.~Prokudin,
  \emph{{Sivers effect in Drell-Yan processes}},
  \href{http://dx.doi.org/10.1103/PhysRevD.79.054010}{\emph{Phys. Rev. D} {\bf
  79} (2009) 054010}, [\href{https://arxiv.org/abs/0901.3078}{{\tt
  0901.3078}}].

\bibitem{Boglione:2019nwk}
M.~Boglione, A.~Dotson, L.~Gamberg, S.~Gordon, J.~O. Gonzalez-Hernandez,
  A.~Prokudin et~al., \emph{{Mapping the Kinematical Regimes of Semi-Inclusive
  Deep Inelastic Scattering}},
  \href{http://dx.doi.org/10.1007/JHEP10(2019)122}{\emph{JHEP} {\bf 10} (2019)
  122}, [\href{https://arxiv.org/abs/1904.12882}{{\tt 1904.12882}}].

\bibitem{Echevarria:2016scs}
M.~G. Echevarria, I.~Scimemi and A.~Vladimirov, \emph{{Unpolarized Transverse
  Momentum Dependent Parton Distribution and Fragmentation Functions at
  next-to-next-to-leading order}},
  \href{http://dx.doi.org/10.1007/JHEP09(2016)004}{\emph{JHEP} {\bf 09} (2016)
  004}, [\href{https://arxiv.org/abs/1604.07869}{{\tt 1604.07869}}].

\bibitem{Collins:2014jpa}
J.~Collins and T.~Rogers, \emph{{Understanding the large-distance behavior of
  transverse-momentum-dependent parton densities and the Collins-Soper
  evolution kernel}},
  \href{http://dx.doi.org/10.1103/PhysRevD.91.074020}{\emph{Phys. Rev. D} {\bf
  91} (2015) 074020}, [\href{https://arxiv.org/abs/1412.3820}{{\tt
  1412.3820}}].

\bibitem{Efremov:1983eb}
A.~V. Efremov and O.~V. Teryaev, \emph{{THE TRANSVERSAL POLARIZATION IN QUANTUM
  CHROMODYNAMICS}}, {\emph{Sov. J. Nucl. Phys.} {\bf 39} (1984) 962}.

\bibitem{Qiu:1998ia}
J.-W. Qiu and G.~Sterman, \emph{{Single transverse-spin asymmetries in hadronic
  pion production}}, {\emph{Phys.~Rev.} {\bf D59} (1998) 014004},
  [\href{https://arxiv.org/abs/hep-ph/9806356}{{\tt hep-ph/9806356}}].

\bibitem{Braun:2009mi}
V.~Braun, A.~Manashov and B.~Pirnay, \emph{{On the scale dependence of
  twist-three contributions to single spin asymmetries}},
  \href{http://dx.doi.org/10.1103/PhysRevD.80.114002}{\emph{Phys.~Rev.} {\bf
  D80} (2009) 114002}, [\href{https://arxiv.org/abs/0909.3410}{{\tt
  0909.3410}}].

\bibitem{Echevarria:2015byo}
M.~G. Echevarria, I.~Scimemi and A.~Vladimirov, \emph{{Universal transverse
  momentum dependent soft function at NNLO}},
  \href{http://dx.doi.org/10.1103/PhysRevD.93.054004}{\emph{Phys. Rev. D} {\bf
  93} (2016) 054004}, [\href{https://arxiv.org/abs/1511.05590}{{\tt
  1511.05590}}].

\bibitem{iminuit}
``\texttt{iminuit}, https://doi.org/10.5281/zenodo.3949207.''

\bibitem{Ball:2008by}
{\scshape NNPDF} collaboration, R.~D. Ball, L.~Del~Debbio, S.~Forte,
  A.~Guffanti, J.~I. Latorre, A.~Piccione et~al., \emph{{A Determination of
  parton distributions with faithful uncertainty estimation}},
  \href{http://dx.doi.org/10.1016/j.nuclphysb.2008.09.037}{\emph{Nucl. Phys. B}
  {\bf 809} (2009) 1--63}, [\href{https://arxiv.org/abs/0808.1231}{{\tt
  0808.1231}}].

\bibitem{bootstrapbook}
A.~Davison and D.~Hinkley, \emph{{Bootstrap Methods and their Application}},
  {\emph{Cambridge University Press: Cambridge} (1997) }.

\bibitem{artemide_sivers}
\emph{\texttt{artemide} repository with sivers extraction,
  https://github.com/vladimirovalexey/artemide-public/tree/master/models/sivers20},
  2020.

\bibitem{Boer:2011fx}
D.~Boer, \emph{{On a possible node in the Sivers and Qiu-Sterman functions}},
  \href{http://dx.doi.org/10.1016/j.physletb.2011.07.006}{\emph{Phys. Lett. B}
  {\bf 702} (2011) 242--245}, [\href{https://arxiv.org/abs/1105.2543}{{\tt
  1105.2543}}].

\bibitem{Kang:2011hk}
Z.-B. Kang, J.-W. Qiu, W.~Vogelsang and F.~Yuan, \emph{{An Observation
  Concerning the Process Dependence of the Sivers Functions}},
  \href{http://dx.doi.org/10.1103/PhysRevD.83.094001}{\emph{Phys. Rev. D} {\bf
  83} (2011) 094001}, [\href{https://arxiv.org/abs/1103.1591}{{\tt
  1103.1591}}].

\bibitem{Lu:2004au}
Z.~Lu and B.-Q. Ma, \emph{{Sivers function in light-cone quark model and
  azimuthal spin asymmetries in pion electroproduction}},
  \href{http://dx.doi.org/10.1016/j.nuclphysa.2004.06.006}{\emph{Nucl. Phys. A}
  {\bf 741} (2004) 200--214}, [\href{https://arxiv.org/abs/hep-ph/0406171}{{\tt
  hep-ph/0406171}}].

\bibitem{Courtoy:2008vi}
A.~Courtoy, F.~Fratini, S.~Scopetta and V.~Vento, \emph{{A Quark model analysis
  of the Sivers function}},
  \href{http://dx.doi.org/10.1103/PhysRevD.78.034002}{\emph{Phys. Rev. D} {\bf
  78} (2008) 034002}, [\href{https://arxiv.org/abs/0801.4347}{{\tt
  0801.4347}}].

\bibitem{Bacchetta:2008af}
A.~Bacchetta, F.~Conti and M.~Radici, \emph{{Transverse-momentum distributions
  in a diquark spectator model}},
  \href{http://dx.doi.org/10.1103/PhysRevD.78.074010}{\emph{Phys. Rev. D} {\bf
  78} (2008) 074010}, [\href{https://arxiv.org/abs/0807.0323}{{\tt
  0807.0323}}].

\bibitem{Bacchetta:1999kz}
A.~Bacchetta, M.~Boglione, A.~Henneman and P.~J. Mulders, \emph{Bounds on
  transverse momentum dependent distribution and fragmentation functions},
  {\emph{Phys. Rev. Lett.} {\bf 85} (2000) 712--715},
  [\href{https://arxiv.org/abs/hep-ph/9912490}{{\tt hep-ph/9912490}}].

\bibitem{Tangerman:1994eh}
R.~D. Tangerman and P.~J. Mulders, \emph{{Intrinsic transverse momentum and the
  polarized Drell-Yan process}},
  \href{http://dx.doi.org/10.1103/PhysRevD.51.3357}{\emph{Phys. Rev. D} {\bf
  51} (1995) 3357--3372}, [\href{https://arxiv.org/abs/hep-ph/9403227}{{\tt
  hep-ph/9403227}}].

\bibitem{Kotzinian:1995cz}
A.~M. Kotzinian and P.~J. Mulders, \emph{{Longitudinal quark polarization in
  transversely polarized nucleons}},
  \href{http://dx.doi.org/10.1103/PhysRevD.54.1229}{\emph{Phys. Rev. D} {\bf
  54} (1996) 1229--1232}, [\href{https://arxiv.org/abs/hep-ph/9511420}{{\tt
  hep-ph/9511420}}].

\bibitem{Gutierrez-Reyes:2019rug}
D.~Gutierrez-Reyes, S.~Leal-Gomez, I.~Scimemi and A.~Vladimirov,
  \emph{{Linearly polarized gluons at next-to-next-to leading order and the
  Higgs transverse momentum distribution}},
  \href{http://dx.doi.org/10.1007/JHEP11(2019)121}{\emph{JHEP} {\bf 11} (2019)
  121}, [\href{https://arxiv.org/abs/1907.03780}{{\tt 1907.03780}}].

\bibitem{Sun:2013hua}
P.~Sun and F.~Yuan, \emph{{TMD Evolution: Matching SIDIS to Drell-Yan and W/Z
  Boson Production}},
  \href{http://dx.doi.org/10.1103/PhysRevD.88.114012}{\emph{Phys.~Rev.} {\bf
  D88} (2013) 114012}, [\href{https://arxiv.org/abs/1308.5003}{{\tt
  1308.5003}}].

\bibitem{Dai:2014ala}
L.-Y. Dai, Z.-B. Kang, A.~Prokudin and I.~Vitev, \emph{{Next-to-leading order
  transverse momentum-weighted Sivers asymmetry in semi-inclusive deep
  inelastic scattering: the role of the three-gluon correlator}},
  \href{http://dx.doi.org/10.1103/PhysRevD.92.114024}{\emph{Phys. Rev.} {\bf
  D92} (2015) 114024}, [\href{https://arxiv.org/abs/1409.5851}{{\tt
  1409.5851}}].

\bibitem{Braun:2011aw}
V.~Braun, T.~Lautenschlager, A.~Manashov and B.~Pirnay, \emph{{Higher twist
  parton distributions from light-cone wave functions}},
  \href{http://dx.doi.org/10.1103/PhysRevD.83.094023}{\emph{Phys. Rev. D} {\bf
  83} (2011) 094023}, [\href{https://arxiv.org/abs/1103.1269}{{\tt
  1103.1269}}].

\end{thebibliography}\endgroup

\end{document}